\documentclass[prd,twocolumn,aps,amsmath,nofootinbib,superscriptaddress]{revtex4}

\usepackage{amsmath}
\usepackage{graphicx}
\usepackage{bm}
\usepackage{xspace}
\usepackage{epsfig}
\usepackage{color}
\usepackage{xcolor}
\usepackage[hypertex]{hyperref}

  \newcount\hour \newcount\minute
  \hour=\time \divide \hour by 60
  \minute=\time
  \newcommand{\mydate}{\ \today \ - \number\hour :\ifnum \minute<10 0\fi 
\number\minute}

\def\SppP{{\cal {P\!\!\!\!\hspace{0.04cm}\slash}}_\perp}

\def\nslash{n\!\!\!\slash}
\def\bnslash{\bar n\!\!\!\slash}
\def\pslash{p\!\!\!\slash}
\def\qslash{q\!\!\!\slash}

\def\OMIT#1{}

\newcommand{\nn}{\nonumber} 

\newcommand{\bn}{{\bar n}}

\newcommand{\bnP}{\bar {\cal P}}

\newcommand{\cP}{{\cal P}}
\newcommand{\cPslash}{ {\cal P}\!\!\!\!\slash}

\newcommand{\mcdot}{\!\cdot\!}

\newcommand{\plus}{\ensuremath{\! + \!}}
\newcommand{\minus}{\ensuremath{\! - \!}}
\newcommand{\mpm}{\ensuremath{\! \pm \!}}

\newcommand{\SCETa}{\ensuremath{{\rm SCET}_{\rm I}}\xspace}
\newcommand{\SCETb}{\ensuremath{{\rm SCET}_{\rm II}}\xspace}

\begin{document}


\preprint{ \hbox{MIT-CTP 3828} \hbox{CMU-HEP-07-xx} 
 \hbox{hep-ph/xxxxxx}  }

\title{\boldmath
  Penguin Loops for Nonleptonic $B$-Decays in the Standard Model:\\
  Is There a Penguin Puzzle?
}

\author{Ambar Jain\vspace{0.3cm}}
\affiliation{Center for Theoretical Physics, Massachusetts Institute of
  Technology, Cambridge, MA 02139}

\author{Ira Z.~Rothstein}
\affiliation{Department of Physics, Carnegie Mellon University,
    Pittsburgh, PA 15213  \vspace{0.3cm}}
\author{Iain W.~Stewart\vspace{0.3cm}}
\affiliation{Center for Theoretical Physics, Massachusetts Institute of
  Technology, Cambridge, MA 02139}
\begin{abstract}
  
  We compute standard model penguin amplitudes in nonleptonic $B$-decays to
  light charmless mesons using tree amplitude data to fix hadronic parameters.
  The leading calculation is carried out for the $\alpha_s(m_b)$ penguin
  contributions from charm quark, up quark, and magnetic penguin loops in the
  NDR and HV renormalization schemes.  Power suppressed penguins that are
  proportional to the chiral condensate are also computed using a new
  factorization formula for these terms, which is derived working to all orders
  in $\alpha_s(\sqrt{m_b\Lambda})$. We demonstrate using \SCETa that this
  formula exhibits only small perturbative phases and does not have endpoint
  singularities.  Due to our use of data to fix hadronic parameters we obtain
  significantly more accurate predictions for the short-distance standard model
  penguin amplitudes than have been found in the past.  Analyzing data in
  $B\to\pi\pi$, $B\to K\pi$, and $B\to \rho\rho$ for the penguin amplitudes we
  find that standard model short-distance imaginary parts are an order of
  magnitude smaller than current measurements, while real parts are up to a
  factor of two smaller with the correct sign.  This difference is most likely a
  consequence of long-distance charm contributions or new physics. Constraints
  on the type of new physics that could help explain the data are derived, and
  used to show that current data favors sizeable long-distance strong phases.

\end{abstract}

\maketitle
 
\section{Introduction}

$B$-physics experiments have made considerable progress in improving our
understanding of standard model CP violation~\cite{HFAG}. Several analyses have
fairly small theoretical uncertainty and yield precise results, such as
$\sin(2\beta)$ from $B\to J/\Psi K_s$ or $B\to \eta' K_s$ type-decays.  However,
for a large number of observables, extracting short-distance information depends
upon our ability to handle QCD effects.  Many of these observables are sensitive
to new physics, and thus considerable effort has gone into understanding how to
calculate strong decay amplitudes with controlled approximations~\cite{xeg}.
Examples of the type of observables are the magnitude and relative strong phase
of penguin contributions in charmless non-leptonic $B$-decays, $B\to \pi\pi,
K\pi, \rho\rho$ etc, which have significant contributions from loop-dominated
penguin amplitudes.

In this paper we classify standard model contributions to penguin amplitudes
using the SCET factorization theorem for nonleptonic decays from
Ref.~\cite{Bauer:2004tj}, and compute a missing set of ${\cal O}(\alpha_s(m_b))$
short-distance perturbative corrections.  (These missing corrections were also
recently computed in Ref.~\cite{Beneke:2006mk}, and we compare results at the
end.)  In principle these corrections have the potential of making up for an
observed shortfall in explaining the penguin amplitude data with leading order
strong phases. However, we find that these contributions to the amplitudes are
quite small.  We also derive a new factorization theorem for ``chiraly
enhanced'' penguin amplitudes, which are suppressed by $1/m_b$ but enhanced by
the chiral condensate. Our result involves a new generalized form factor
$\zeta_\chi^{BM}(z)$ and a single twist-3 meson distribution $\phi_{pp}^M(u)$,
and it does not suffer from endpoint divergences.  We find that these
contributions also have small imaginary contributions.  Indeed, all known
imaginary short-distance corrections to the penguin amplitudes are small,
roughly an order of magnitude below the experimental values in $B\to\pi\pi$
decays and $B\to K\pi$ decays.  Explanations for this discrepancy from
long-distance standard model contributions are critiqued and weighed against a
beyond the standard model explanation.

In the standard model the amplitude for a channel $\bar B\to M_1 M_2$ may be
written as
\begin{eqnarray} \label{A}
  A^{M_1 M_2} = \lambda_u^{(f)} T^{M_1 M_2} + \lambda_c^{(f)} P^{M_1 M_2} 
   \,,
\end{eqnarray}
with $\lambda_p^{(f)}=V_{pb}^{\phantom{*}} V_{pf}^*$, and where we use CKM
unitarity to remove $V_{tb}^{\phantom{*}}V_{tf}^*$ ($f=d$ or $f=s$).  In this
paper $T^{M_1M_2}$ and $P^{M_1M_2}$ will be called tree and penguin amplitudes
respectively.  We derive amplitudes for all two-body pseudoscalar and vector
modes that do not involve isosinglets in the final state. In comparing with
experimental penguin amplitudes extracted from data, we focus on the
$B\to\pi\pi$, $B\to K\pi$, and $B\to\rho\rho$ channels.

With the latest data one may extract values for the penguin amplitudes in the
$B\to\pi\pi$ and $B\to\rho\rho$ channels using isospin symmetry. Isospin implies
that $P^{\pi^+\pi^-}$ also appears in the $\pi^0\pi^0$ channel, and is absent for
$\pi^0\pi^-$ (up to small electroweak penguin
terms~\cite{Gronau:2005pq,Gardner:1998gz}). The same is true for $P^{\rho^+\rho^-}$
(using the fact that the $\rho$'s are measured to be primarily longitudinal,
$f_L^{\rho^+\rho^-}\simeq 98\%$ and $f_L^{\rho^+\rho^0}\simeq
91\%$~\cite{rhorhodata}, and neglecting interference due to the large rho
width~\cite{Falk:2003uq}). To quote experimental values for the penguin
amplitudes one must pick a phase convention.  We take $T^{\pi^+\pi^-}$ and
$T^{\rho^+\rho^-}$ to be real and positive, and quote other phases relative to
this. For the penguins we will quote results for
\begin{align} \label{Psign}
 P^{\pi\pi} \equiv - P^{\pi^+\pi^-} \!\!, \:\:
 P^{K\pi} \equiv - P^{\pi^+ K^-} \!\!, \:\:
 P^{\rho\rho} \equiv - P^{\rho^+\rho^-} \!\!, 
\end{align}
etc.  In addition, we must also fix the value of the well-determined weak phase
$\beta=21.2^{\circ}$~\cite{HFAG} and the less well determined weak phase $\gamma$.  The
latest global CKM fits give~\cite{CKMfitter,UTfit}
\begin{align} \label{gammaGLOBAL}
 \gamma^{\rm CKMfitter}_{\rm global} &= 59^\circ {}^{+9.2^\circ}_{-3.7^\circ}\,,
 & \gamma^{\rm UTfitter}_{\rm global}   & =64.6^\circ \pm 4.2^\circ \,.
\end{align}

An alternative method to obtain $\gamma$ is to use $B\to\pi\pi$ or $B\to \rho\rho$ data
alone.  In principle for $B\to \pi\pi$ this is possible using only isospin
~\cite{GL}, and for $B\to\rho\rho$ it is possible using isospin, the
polarization data, and neglecting the $\rho$ width.  However the current
experimental uncertainties need information beyond isospin, such as an expansion
in $\Lambda_{\rm QCD}/m_b$, necessary to obtain results competitive with
Eq.~(\ref{gammaGLOBAL}).  An approach with small
uncertainties~\cite{Bauer:2004dg}, which we label the BRS method, augments the
isospin analysis by using the factorization theorem for nonleptonic decays in a
specific limited way, namely to use ${\rm Im}(T^{\pi^0\pi^0}/T^{\pi^+\pi^-})
\sim {\cal O}[\Lambda/E, {\cal O}(\alpha_s(m_b)]$. Data on ${\rm Br}(\bar B^0\to
\pi^+\pi^-)$, ${\rm Br}(B^-\to \pi^-\pi^0)$, ${\rm Br}(\bar B^0\to \pi^0\pi^0)$,
$S_{\pi^+\pi^-}$, and $C_{\pi^+\pi^-}$ or the analogs for $B\to \rho\rho$ are
then used to determine $\gamma$. With the latest non-leptonic
data~\cite{pipidata,rhorhodata} as summarized by HFAG~\cite{HFAG}, this gives
\begin{align} \label{gammaBPRS}
  \gamma_{\pi\pi}^{\rm BRS} &= 73.9^\circ
  {}^{+7.5^\circ}_{-10.3^\circ}\Big|_{\rm exp}
  {}^{+1.0^\circ}_{-2.5^\circ}\Big|_{\rm thy}  \,, \\
    \gamma_{\rho\rho}^{\rm BRS} &= 77.3^\circ
  {}^{+7.6^\circ}_{-32^\circ} \Big|_{\rm exp}
  {}^{+1.0^\circ}_{-4.6^\circ}\Big|_{\rm thy}
  \nn \,,
\end{align}
where we quote the experimental and theory errors separately. With these values
of $\gamma$, factorization in SCET exactly reproduces the observed $\bar
B^0\to\pi^0\pi^0$ and $\bar B^0\to \rho^0\rho^0$ branching fractions. There is
also a second solution
\begin{align} \label{gammaBPRS2}
  (\gamma_{\pi\pi}^{\rm BRS })^{\rm 2nd} &= 27.7^\circ
  {}^{+9.9^\circ}_{-7.3^\circ}\Big|_{\rm exp}
  {}^{+10^\circ}_{-4.5^\circ}\Big|_{\rm thy}   \,, \\
  (\gamma_{\rho\rho}^{\rm BRS})^{\rm 2nd} &= 52.8^\circ 
  {}^{+32^\circ}_{-7.7^\circ}\Big|_{\rm exp}
  {}^{+6.7^\circ}_{-4.1^\circ}\Big|_{\rm thy}   \,,\nn
\end{align}
that is, however, disfavored by the additional piece of information that the
form factor parameter $\zeta_J>0$. The value of $\gamma$ from $B\to \rho\rho$ in
Eq.~(\ref{gammaBPRS}) has not been quoted earlier in the literature, but the
analysis is identical to that for $B\to\pi\pi$ in Ref.~\cite{Bauer:2004dg}.
Currently the global fit values in Eq.~(\ref{gammaGLOBAL}) and BRS values in
Eq.~(\ref{gammaBPRS}) are consistent with each other at the $1$--$\sigma$ level.
Suppression of ${\rm Im}(T^{\pi^0\pi^0}/T^{\pi^+\pi^-})$ can also be studied in
a convention where $\lambda_c^{(f)}$ is eliminated from
Eq.~(\ref{A})~\cite{Grossman:2005jb}, however in this case charm penguins
contribute to the tree amplitudes which can induce contamination by
long-distance contributions . We will quote numerical results for the penguin
amplitudes using $\gamma=59^\circ$ and $\gamma=74^\circ$ to give some indication
of the spread of possible values.

With the latest $B\to\pi\pi$ and $B\to \rho\rho$
data~\cite{pipidata,rhorhodata}, the isospin formula quoted below in
Eq.~(\ref{z0}) gives the penguin amplitudes for $\gamma=59^\circ$,
\begin{align} \label{Ppipi} 
%
 10^3\, \hat P^{\pi\pi} &= \phantom{-}(1.77 \pm 0.73) - i (2.91 \pm 0.58) \,,\nn\\
 10^3 \, \hat P^{\rho\rho} &= (-2.91\pm 2.63 ) - i (0.78 \pm 1.82) \,,
\end{align}
while for $\gamma=74^\circ$ we find
\begin{eqnarray} \label{Ppipi2}
 10^3\, \hat P^{\pi\pi} &= (4.41 \pm 0.61) - i (2.91 \pm 0.58) \,,\nn\\
 10^3 \, \hat P^{\rho\rho} &= (3.81\pm 2.34 ) - i (0.78 \pm 1.82) \,.
\end{eqnarray}
Here for convenience we pulled out a prefactor to quote a dimensionless penguin
amplitude $\hat P$, using 
\begin{align}
  \frac{ P^{M_1 M_2} }{(1\,{\rm GeV})} \equiv
   \frac{G_F m_B^2 }{\sqrt{2}}\: \hat P^{M_1 M_2} 
  \,.
\end{align}
Note that for fixed CP-asymmetries $C$ and $S$, the extracted real part of $\hat
P^{\pi\pi}$ and $\hat P^{\rho\rho}$ depends fairly strongly on the value of
$\gamma$, but the imaginary part is completely independent of the choice for
$\gamma$. (This is demonstrated explicitly in Eq.~(\ref{z0}) below.) Neither
result depends on the error in $|V_{ub}|$. The experimental errors here have
decreased noticeably from early penguin extractions~\cite{pipianalysis}. The
challenge for standard model predictions is to reproduce or rule out the values
in Eqs.~(\ref{Ppipi}-\ref{Ppipi2}).

The extraction of both the real and imaginary part of penguin amplitudes in the
$K\pi$ system currently requires further theoretical input. In $B\to K\pi$
decays the penguin amplitudes dominate the tree amplitudes due to CKM
suppression, making a precise comparison of their values even more
interesting. Both types of amplitudes are important in CP asymmetries. Using a
$\Lambda/m_b$ expansion, the tree amplitude for $\bar B^0\to K^-\pi^+$ at LO
depend only on hadronic parameters $\zeta^{B\pi}$ and $\zeta_J^{B\pi}$ that are
fully determined by the tree amplitudes in the $B\to \pi\pi$ channels, plus the
$\phi_K$ twist-2 distribution function~\cite{Bauer:2005kd}.  This allows the
phase of the penguin amplitude $P^{K^+\pi^-}$ to be extracted from the data
using only factorization for the tree amplitudes (which we will refer to by
adding a subscript TF).  The tree amplitudes are reliable since a proof of
factorization to all orders in $\alpha_s$ was given in
Ref.~\cite{Bauer:2004tj,chay}, extending the original proposal and one-loop
analysis in Ref.~\cite{BBNS}.  Although factorization has also been demonstrated
for light-quark penguin loops ($u,d,s$), a complete analysis for charm-loops is
still lacking.  Using the phase convention where $T^{K^-\pi^+}$ is real we find
for $\gamma = 59^\circ$
\begin{eqnarray} \label{PKpi}
 10^3\, \hat P_{\rm TF}^{K^-\pi^+} = \left\{
\begin{array}{l}
   \phantom{-}(4.87 \pm 0.39) - i (2.22\pm 0.77)\\
    -(4.22\pm 0.36) - i (2.22\pm 0.34)
\end{array} \right.
  ,
\end{eqnarray}
while for $\gamma = 74^\circ$
\begin{align} \label{PKpi2}
  10^3\, \hat P_{\rm TF}^{K^-\pi^+} = \left\{
 \begin{array}{l}
   \phantom{-}(4.73\pm 0.36)-i(2.16 \pm 0.73) \\ 
    -(4.41\pm 0.34)-i(2.16\pm 0.68)
 \end{array} \right.
  .
\end{align}
The only $B\to K\pi$ data used here was $Br(K^-\pi^+)$ and $A_{\rm
  CP}(K^-\pi^+)$, and there are two solutions for each $\gamma$. Alternatives
to the above analysis extract the $K\pi$ penguin amplitudes using a SU(3) based
analysis with the $\pi\pi$ data~\cite{SU3buras} or by a global SU(3) based
fit~\cite{Chiang:2004nm}, and these yield similar conclusions for the size of the
penguin amplitudes. Again the data gives $K\pi$-penguin amplitudes with large
imaginary components which require explanation in the standard model.

To determine penguin amplitudes for charmless $B$-decays in the
standard model it is convenient to organize the relevant mass scales
as an expansion in $\Lambda/m_b$ and $\Lambda/m_c$~\cite{BBNS}. This
can be done from first principles using the effective field theory
SCET~\cite{SCET}. In this expansion certain contributions to these
amplitudes factorize allowing them to be parameterized by well defined
universal hadronic matrix elements.  Since we are interested in the
standard model prediction, we also organize the amplitude according to
large ($C_{1,2,8g}$) and small ($C_{3-10}$) Wilson coefficients.  To
explain which terms will be computed in this paper we schematically
give a result $\hat P_0$ for channels $B\to MM'$ with pseudoscalars,
$MM' = PP$, with pseudoscalars and vectors $MM'=PV_0$, and with two
longitudinal vectors $MM'=V_0 V_0$. For completeness we also quote an
analogous result $\hat P_T$ for transverse polarizations $MM'=V_T V_T$:
\begin{widetext}
\begin{eqnarray} 
\label{hatP}
  \hat P_{0} &\sim& 
\Big(C_{3,4} \!+\! \frac{\alpha_s(m_b) C_{1,2,8g}}{\pi}\Big) 
   \zeta^{BM} \phi^{M'}
  +
   \Big(C_{3,4} \!+\! \frac{\alpha_s(m_b) C_{1,2,8g}}{\pi}\Big) 
   \zeta_J^{BM} \phi^{M'}
 \\[5pt]
&& 
+  \Big( C_{5,6} \!+\! \frac{\alpha_s(m_b) C_{1,2,8g}}{\pi} \Big) 
 \Big[  \frac{\mu_{M'}}{m_b}\,  \zeta^{BM} \phi_{pp}^{M'}
 + \frac{\mu_{M'}}{m_b}\, \zeta_J^{BM} \phi_{pp}^{M'} \Big]
 +\! \Big( C_{3,4} \!+\! \frac{\alpha_s(m_b) C_{1,2,8g}}{\pi} \Big)
   \frac{\mu_M}{m_b}\zeta_{\chi}^{BM} \phi^{M'}\!
\nn\\[5pt]
&& 
+
\, C_{1,2}\, \alpha_s(2m_c) v \hat A_{c\bar c}^{BMM'}
+ 
  \frac{\alpha_s(m_b)}{m_b}\left( C_{3,4}  f_B\phi^M \phi^{M'}
+ 
 C_{5,6} f_B\phi_B^+ \phi^{3M} \phi^{M'} 
\right)
+
 C_{5,6} \frac{\alpha_s(m_b)\mu_{M} }{m_b^2}  f_B  \phi^{M}_{pp} \phi^{M'}
\,, \nn \\
 \label{hatPT}
 \hat P_{T} &\sim & \alpha_{\rm em} \frac{m_b}{\Lambda} \hat A^{BVV}_{\gamma-\rho}  \\
  && 
+  \Big( C_{3,4} \!+\! \frac{\alpha_s(m_b) C_{1,2,8g}}{\pi} \Big) 
  \Big[ \frac{1}{m_b}\, \zeta_\perp^{BV} \phi_{pp\perp}^{V'} +
    \frac{1}{m_b}\, \zeta_{J\perp}^{BV} \phi_{pp\perp}^{V'} \Big]
+ \! \Big( C_{5,6} \!+\! \frac{\alpha_s(m_b) C_{1,2,8g}}{\pi} \Big) 
  \frac{1}{m_b}\, \zeta_{K \perp}^{BV} \phi_{\perp}^{V'}
  \nn \\
&& 
+
\, C_{1,2}\, \alpha_s(2m_c) v\, \hat A_{c\bar c}^{BVV'} 
  +
 C_{5,6} \frac{\alpha_s(m_b)}{m_b^2}\: \hat A^{BVV'}_{(2ann-\chi)} 
  \,. \nn
\end{eqnarray}
\end{widetext}
When coefficients $C_{3,4,5,6}$ appear here we leave implicit the fact
that the electroweak penguin coefficients $C_{7,8,9,10}$ can also
appear. 

The terms on the first line of Eq.~(\ref{hatP}) for $\hat P_0$ are
leading in the $\Lambda/m_{b,c}$ expansion. Working to all orders in
$\alpha_s(\mu_i)$ at the intermediate scale $\mu_i\simeq
\sqrt{m_b\Lambda}$, they involve a so-called soft form factor
$\zeta^{BM}$ and a hard form factor $\zeta_J^{BM}$. (Though it should
be emphasized these names are somewhat misleading, since both form
factors involve hard-collinear exchange, and thus the same length
scales.) The terms on the second line of $\hat P_0$ are the
``chiraly-enhanced'' power corrections suppressed by
$\mu_{M'}/m_b$. Here $\mu_{M'} \sim {\cal O}(\Lambda)$ is a ratio of
the squared meson mass to a sum of quark masses, and is important for
nonleptonic decays because it is numerically enhanced $\mu_{M'}\simeq
2\,{\rm GeV}$~\cite{BBNS}. In Eq.~(\ref{hatP}) we display a new result
that we will derive below, namely that to all orders in $\alpha_s$ the
chiraly enhanced terms in $\hat P_0$ are determined by one new form
factor, $\zeta_{\chi}^{BM}$ and one twist-3 distribution,
$\phi_{pp}^M$ for longitudinal polarizations/pseudoscalars.

On the third line of Eq.~(\ref{hatP}) for $\hat P_0$ we have a term
$\hat A_{c\bar c}^{BMM'}$. This is the so-called ``charming penguin''
due to long-distance charm loop effects~\cite{Ciuchini,Colangelo},
whose leading contribution is expected to come from the charm
threshold region~\cite{Bauer:2004tj,Bauer:2005wb}. It is
parametrically down by a power of the non-relativistic velocity $v$
relative to the leading-power result.

The remaining terms on the third line of Eq.~(\ref{hatP}) are due to
annihilation and are suppressed by one or more powers of $\Lambda/m_b$. Terms at
order $1/m_b$ contribute only to longitudinal polarizations and include: $f_B
\phi^M\phi^{M'}$ which was studied in Refs.~\cite{Keum,Lu,BBNS2}, and the
remaining leading annihilation amplitude $f_B\phi_B^+\phi^{3M}\phi^{M'}$, which
was computed recently in Ref.~\cite{Arnesen:2006dc}. The $\mu_m f_B \phi_{pp}^M
\phi^{M'}$ terms are chiraly enhanced annihilation studied in
Refs.~\cite{BBNS2,Arnesen:2006vb}. At lowest order in the $\alpha_s$
expansion the annihilation terms shown in Eq.~(\ref{hatP}) are
real~\cite{Arnesen:2006vb} (using the zero-bin
procedure~\cite{Manohar:2006nz}).  A nonperturbative complex
annihilation amplitude involving soft exchange occurs at order
$\alpha_s^2(\mu_i)/m_b$.

The terms shown in the $\hat P_T$ amplitude follow a similar notation to that
for $\hat P_0$, but there are no analogs of the LO terms on the first line for
$\hat P_0$~\cite{Kagan:2004uw}. The term shown on the first line of $\hat P_T$
comes from $\gamma$-$\rho$ conversion~\cite{Beneke:2005we}, and is suppressed by
$\alpha_{em}$, but enhanced by $m_b/\Lambda$. $\hat P_T$ does not contain
chiraly enhanced terms, so we show on the second line the terms from the
analogous operators, which generate three form factors $\zeta_\perp^{BV}$,
$\zeta_{J\perp}^{BV}$, and $\zeta_{K\perp}^{BV}$.  In Ref.~\cite{Kagan:2004uw}  analogs of the terms on the second line of Eq.~(\ref{hatPT}) were computed with an expansion in $\alpha_s(\mu_i)$, and the penguin annihilation term $\hat A_{2ann-\chi}^{B V V'}$ was also treated. Following
Ref.~\cite{Bauer:2005wb} we included a term $\hat A_{c\bar c}^{BVV'}$ in the
amplitude to produce transverse vector mesons.  This is the most conservative
approach given that so little is known about the factorization properties of
$\hat A_{c\bar c}$.

We remark that the second and third lines of Eqs.~(\ref{hatP}) and (\ref{hatPT})
do not contain the complete set of $1/m_b$ or $1/m_b^2$ power corrections, but
rather a collection of terms that are believed to be important due to numerical
enhancement.\footnote{Also note that to our knowledge it has not been
  demonstrated that the chiraly enhanced terms shown in Eq.~(\ref{hatP}) give
  the complete set of such enhanced contributions at this order in the power
  expansion. In particular it remains undetermined whether time-ordered product
  terms appearing at this order are or are not chiraly enhanced.} The additional
uncertainty from missing $1/m_b$ corrections will be taken into account in our
final error estimates.  Also the $C_{3,4}$ and $C_{5,6}$ terms always come
together with an $\alpha_s C_{1,2}$ term, which is the matrix element
responsible for canceling the largest scheme dependence in these coefficients at
next-to-leading-log (NLL) order.  Since tree level $C_{3-10}$ terms
could compete numerically with $\alpha_s C_{1,2,8g}$ terms, we will require both
to be included in what we call our leading order penguin amplitude.  In this
paper we will neglect $\alpha_s(m_b) C_{3,4}$ terms relative to
$\alpha_s(m_b)C_{1,2,8g}$ since numerically they are 6-30 times smaller. This is
the same strategy that was adopted for the ``NNLL computations'' in $B\to
X_s\ell^+\ell^-$~\cite{bseeReview}.

Eq.~(\ref{hatP}) is schematic because we have not yet displayed the
precise coefficients in front of each term.  The coefficients for the
$\zeta^{BM}$ terms on the first and second lines were computed in
Ref.~\cite{BBNS}.  To obtain Eq.~(\ref{hatP}), we expanded in
$\Lambda/m$ and $\alpha_s(m_b)$, but avoided the additional
uncertainties from expanding in $\alpha_s(\mu_i)$, where the
intermediate scale $\mu_i\simeq \sqrt{E\Lambda}\sim 1.3\,{\rm GeV}$.
This is made possible by the fact that the form-factor parameters
$\zeta_J^{BM}$ and $\zeta^{BM}$ are universal hadronic parameters when
we distinguish $m^2 \gg E\Lambda\gg
\Lambda^2$~\cite{Bauer:2004tj}.  Without expanding in $\alpha_s(\mu_i)$, the
third term on line 1 was obtained in Ref.~\cite{Bauer:2004tj}.  The third term
on line 1 and third and non-$\alpha_s(m_b)$ terms on line 2 were computed at
leading order in an expansion in $\alpha_s(\mu_i)$ in Ref.~\cite{BBNS}.

A main goal of this paper is the computation of the $\alpha_s C_{1,2,8g}
\zeta_J^{BM} \phi^{M'}$ terms on line 1 of
Eq.~(\ref{hatP}).  Note that this term is a leading-order contribution
to the penguin amplitudes due to the hierarchy in the Wilson
coefficients. Our other main goal is to derive the factorization
theorem for the terms on the second lines of $\hat P_0$, and compute
the corresponding Wilson coefficients. We derive the factorization
theorem working to all orders in $\alpha_s$, and perform tree level
matching for all contributions. We also compute the
$\alpha_s\zeta^{BM}$ terms, however the perturbative matching
computation that determines the $\alpha_s(m_b)\zeta_J^{BM}$ and
$\alpha_s(m_b)\zeta_\chi^{BM}$ terms on line 2 will not be considered
here

We would like to compute the magnitude and phase of $\hat P$ for the
$\pi^+\pi^-$, $K^+\pi^-$, and $\rho^+\rho^-$ channels from the terms in
Eq.~(\ref{hatP}) and compare with Eqs.~(\ref{Ppipi}-\ref{PKpi2}). The phase here
is that of $\hat P/\hat T$.  A schematic expression for $\hat T$ is given by
\begin{eqnarray} 
\label{hatT}
  \hat T &\sim& 
\Big(C_{1,2} \!+\! \frac{\alpha_s(m_b) C_{1,2}}{\pi}\Big) \zeta^{BM}\: \phi^{M'} \nn \\
&& +
\Big(C_{1,2} \!+\! \frac{\alpha_s(m_b) C_{1,2}}{\pi}\Big) \zeta_J^{BM}\: \phi^{M'} \nn \\
&& + \, C_1\,\frac{\alpha_s(m_b) }{m_b} \, T_{(ann)}^{BMM'}+\ldots
 \,,
\end{eqnarray}
where we suppress terms from Wilson coefficients $C_{3-10}$ and power
suppressed terms other than annihilation.  For $\hat T_{\pi^+\pi^-}$,
$\hat T_{\rho^+\rho^-}$, and $\hat T_{K^+\pi^-}$ the leading term in
$\hat T$ is real and numerically dominant.  Therefore in $\hat P/\hat
T$ the $C_{1,2}\alpha_s$ corrections from $\hat P$ are leading, while
those from $\hat T$ are higher order in $\alpha_s$. In fact it would
be inconsistent to keep the $C_{1,2}\alpha_s$ corrections in $\hat T$
without keeping the $C_{3,4}\alpha_s$ terms in $\hat P$, because both
of these terms carry $\mu$-dependence that cancels that in
$\zeta^{BM}$ and $\zeta_J^{BM}$. Thus to compute the imaginary part of
$\hat P/\hat T$ at the order we are working, we can take $\hat
T^{\pi^+\pi^-}$ and $\hat T^{K^+\pi^-}$ to be real. Numerically
Refs.~\cite{Guido,Beneke:2006mk} found that including one and some
two-loop corrections gives $10^3 \hat T^{\pi\pi} = (31{}^{+7}_{-9} -
i\, 0.07 {}^{+.7}_{-3}) +\hat T^{\pi\pi}_{( ann)}$. Thus, the
imaginary part is significantly smaller than the real part. Using the
zero-bin procedure, the annihilation contributions
$T^{M_1M_2}_{(ann)}$ are also real at leading order in an expansion of
$\alpha_s$ at the hard and intermediate scales~\cite{Arnesen:2006dc,
Arnesen:2006vb}.

Besides $\zeta_J^{BM}(z)$ and $\zeta^{BM}$, the other hadronic parameters in
Eq.~(\ref{hatP}) include the twist-2 distribution $\phi^{M'}$ and twist-3
chiraly enhanced distribution $\phi_{pp}^{M'}$ defined below in
Eq.~(\ref{phipp}). In this paper we adopt the point of view that $\zeta^{BM}$
and the normalization
\begin{eqnarray}\label{zJnorm}
  \zeta_J^{BM} \equiv \int dz\ \zeta_J^{BM}(z)
\end{eqnarray}
should be fixed using other data (tree amplitudes and/or form factors) and then
used to make predictions for the penguin amplitudes~\cite{Bauer:2004tj}.  We
will see that, relative to adopting models for all the hadronic parameters, 
fitting to tree amplitudes removes the dominant hadronic uncertainty in the
computation of the short-distance penguin amplitudes.  In a generic new physics
model, $\hat P^{\rm expt} = \hat P^{\rm SM} + \hat P^{\rm BSM}$, so to test the
data for new physics we must have control over $ \hat P^{\rm SM}$.

The plan for the paper is as follows. In section~\ref{sect:zeta} we give
formulas for determining the penguin amplitudes and the soft and hard form
factor parameters from the data. Section~\ref{sect:LO} reviews the leading
factorization formula, and section~\ref{sect:endpt} discusses the endpoint
behavior of $\zeta_J^{BM}(z)$. In section~\ref{sect:summary} we give a summary
of all ${\cal O}(\alpha_s(m_b))$ one-loop hard coefficients at LO, and then in
sections~\ref{sect:NDR} and~\ref{sect:HV} provide more details of their
calculation in the NDR and HV schemes respectively. A factorization theorem for
chiraly enhanced penguins is derived in section~\ref{sect:chiral} working to all
orders in $\alpha_s$ at the intermediate scale $\sqrt{m_b\Lambda}$. In
section~\ref{sect:long} we discuss long-distance charm contributions.  Penguin
annihilation contributions are reviewed in section~\ref{sect:ann}. Our analysis
strategy is outlined in section~\ref{sect:models} and input parameters are
summarized in section~\ref{sect:inputs}. 

Our numerical analysis for standard model penguins is taken up in
section~\ref{sect:analysis}.  This is followed by section~\ref{sect:np} where we
derive constraints on the effect of new physics contributions, and discuss what
is needed to shift the penguin amplitudes closer to the data.  Further
discussion and conclusions are given in section~\ref{sect:conclusion}. Several
calculational details are relegated to appendices.


\section{Determining Penguin Amplitudes and the $\zeta^{BM}$ and $\zeta_J^{BM}$ Form Factors with
  Nonleptonic Data}  \label{sect:zeta}

The $B\to\pi\pi$ data can be used to extract the penguin amplitude
$P^{\pi^+\pi^-}$ and the tree amplitudes $T^{\pi^+\pi^-}$ and $T^{\pi^0\pi^-}$,
including the strong phase in $P^{\pi^+\pi^-}/T^{\pi^+\pi^-}$.  Solving
equations in Ref.~\cite{Bauer:2004dg} with our phase convention the penguin
amplitude is
\begin{widetext}
\begin{align} \label{z0}
 {\rm Re}\big( \hat P^{\pi\pi}\big) &= 
    \frac{ N_{\pi^0\pi^-}}{(1\,{\rm GeV}) } \:
  \: \bigg\{ \frac{ (t_c^{\pi\pi})^2 \big[ \sin2\beta -\sin(2\beta\plus 2\gamma)\big]
    -\overline R_c\,\big(\sin2\beta+S_{\pi^+\pi^-}\big) }
    {|V_{cb}^{\phantom{*}} V_{cd}^* |\, (2t_c^{\pi\pi}\sin\gamma) \cos2\beta }
    \bigg\} \,,
 \nn\\
{\rm Im}\big( \hat P^{\pi\pi}\big) &= 
   \frac{ N_{\pi^0\pi^-}}{(1\,{\rm GeV})  } \:
   \: \frac{C_{\pi^+\pi^-}\, \overline R_c}
   {|V_{cb}^{\phantom{*}} V_{cd}^* |\, (2 t_c^{\pi\pi} \sin\gamma)} 
  \,, 
\end{align}
where the parameters on the right-hand-side are determined by nonleptonic data:
\begin{align} \label{z2}
  N_{\pi^0\pi^-} 
  & =  \bigg[ \frac{64\pi }{m_B^3 G_F^2} \frac{{\rm Br}(B^-\to
    \pi^0\pi^-)}{\tau_{B^-}}\bigg]^{1/2}
   ,
  & t_c^{\pi\pi} &
  = \frac{1}{\sin\gamma} 
  \Bigg[ \frac{\overline R_c}{2}\: 
  \Big( 1 \!+\! B_{\pi\pi}\cos 2\beta+S_{\pi^+\pi^-}\sin2\beta \Big)  
  \Bigg]^{1/2} 
  ,\nn\\
   \overline R_c &=  \frac{{\rm Br}(\bar B^0 \to \pi^+\pi^-)\tau_{B^-}}
  {2{\rm Br}(B^-\to\pi^0\pi^-)\tau_{B^0} } 
  \,,
 & B_{\pi\pi} &=\sqrt{1-C_{\pi^+\pi^-}^2 - S_{\pi^+\pi^-}^2} \: .
\end{align}
\end{widetext}
Eqs.~(\ref{z0}) and (\ref{z2}) also determine the penguin for longitudinal $B\to
\rho\rho$ decays, by simply taking all superscripts and subscripts $\pi\to
\rho$, and were used to determine the numbers quoted in
Eqs.~(\ref{Ppipi}-\ref{Ppipi2}) with $|V_{cb}|=0.0417$ and $|V_{cd}|=0.227$.
The ${\rm Im}(\hat P)$ is mainly sensitive to the direct CP-asymmetry. Since
$(t_c^{\pi\pi}\sin\gamma)$ and $(t_c^{\rho\rho}\sin\gamma)$ do not explicitly
depend on the weak phase $\gamma$, the same is true for the values extracted for
${\rm Im}(P^{\pi\pi})$ and ${\rm Im}(P^{\rho\rho})$ (demonstrating the statement
we made in the introduction).  The amplitude parameter $t_c^{\pi\pi} =
|T^{\pi^+\pi^-}|/|T^{\pi^0\pi^+}|$ gives information about the size of the
color-suppressed tree amplitude.  The results in Eq.~(\ref{z0}) and (\ref{z2})
are based on isospin symmetry and neglect small electroweak penguin
contributions in $B^-\to\pi^0\pi^-$. This analysis leaves a $\pm$ sign ambiguity
in front of the $B_{\pi\pi}$ dependence in $t_c^{\pi\pi}$, which we resolved in
Eq.~(\ref{z2}) by taking the ``$+$'' solution. This solution is preferred by the
standard model and rigorous power counting for the QCD amplitudes.  The other
experimentally allowed solution, $B_{\pi\pi}\to -B_{\pi\pi}$, has very large
penguin amplitudes, $10^3 |\hat P_{\pi\pi}|\sim 11$ and $10^3 |\hat
P_{\rho\rho}|\sim 24$, which are extremely difficult to accommodate in the
standard model.

An important question for the phenomenology of charmless nonleptonic $B$-decays is
the relative size of the form factors $\zeta^{BM}$ and $\zeta_J^{BM}$ (defined
below in Eq.~(\ref{zeta})). Here we follow Ref.~\cite{Bauer:2004tj} and organize
the expansion according to $\zeta^{BM}_J \sim \zeta^{BM}$ which is a natural
power counting when factorization is not used at the intermediate scale, and is
also the scaling used in the KLS approach~\cite{Keum}. This counting is
supported by the $B\to \pi\pi$ data and factorization with the zero-bin
procedure of Ref.~\cite{Manohar:2006nz} (which implies that both of these form
factors have an $\alpha_s(\mu_i)$ at leading order in the expansion at the
intermediate scale). In the BBNS approach~\cite{BBNS}, a hierarchy $\zeta^{BM}_J
\sim \alpha_s(\mu_i) \zeta^{BM}$ is adopted. This changes the order of terms in
the perturbative expansion which we discussed in Eq.~(\ref{hatP}) by making
certain terms higher order in $\alpha_s$.

At leading order the $B\to \pi\pi$ factorization theorem for the tree amplitudes
can be used to extract the normalization of the soft and hard form factor
parameters. Expressed in terms of observables the result is~\cite{Arnesen:2005ez}
\begin{widetext}
\begin{align} \label{z1}
 |V_{ub}| \big( \zeta^{B\pi} +\zeta_J^{B\pi} \big)
    & =   \frac{N_{\pi^0\pi^-}}{f_\pi |V_{ud}|}
   \bigg[\frac{  (C_1+C_2){ t_c^{\pi\pi}} -C_2-C_3
}{C_1^2-C_2^2 + (C_1\plus C_2)(C_4\minus C_3)} \bigg]
\bigg[ 1+ {\cal O}\Big(\alpha_s(m_b),\frac{\Lambda}{E}\Big) \bigg] 
  ,\\
 |V_{ub}| \langle x^{-1}\rangle_{\pi}\, \zeta_J^{B\pi} &=
   \frac{N_{\pi^0\pi^-}}{f_\pi |V_{ud}|}
  \bigg[   \frac{ 3C_1\plus C_2\plus C_3 \plus 3 C_4 -
  4 (C_1\plus C_2)  t_c^{\pi\pi}}
  {C_1^2 - C_2^2 + (C_1\plus C_2)(C_4\minus C_3) }
  \bigg] \label{z1a}
\bigg[ 1+ {\cal O}\Big(\alpha_s(m_b),\frac{\Lambda}{E}\Big) \bigg],
\end{align}
\end{widetext}
where $E$ is the pion energy in the CM frame of the $B$.  On the left-hand-side of
Eq.~(\ref{z1}) we have the semileptonic $B\to \pi\ell\bar\nu$ form factor at
$q^2\simeq 0$, which is given by $f_+^{B\pi}(0)=\zeta^{B\pi}+\zeta_J^{B\pi}$.
The expression for $\langle x^{-1}\rangle_{\pi} \zeta_J^{B\pi}$ in
Eq.~(\ref{z1a}) follows in a straightforward manner from results in
Ref.~\cite{Bauer:2004tj}, but to our knowledge has not been presented in this
simple closed form in the literature.  For the hadronic parameter $\langle
x^{-1}\rangle_{\pi}= \int_0^1\!  dx \: \phi_\pi(x)/x$, a fit of the
$\gamma$-$\pi$ form factor to $\gamma^*\gamma\to \pi^0$ data
gives~\cite{Bakulev:2003cs}
\begin{align} \label{invpiexp}
 \langle x^{-1}\rangle_{\pi}=2.9 \pm 0.4\,.
\end{align}

Using the latest experimental data~\cite{HFAG}, $\gamma=67\pm
10^\circ$, $|V_{ud}|=0.9738$, and LL Wilson coefficients,
Eq.~(\ref{z1}) gives
\begin{align} \label{flzJ}
 f^{B\pi}_+(0) &= \Big(0.182 \pm 0.011  \pm 0.036 \Big) 
  \frac{4.2\!\times\! 10^{-3}}{ |V_{ub}|} \,,\nn\\
  \langle x^{-1}\rangle_{\pi}\, \zeta_J^{B\pi} &= 
   \Big(0.262 \pm 0.052   \pm 0.052 \Big) 
   \frac{4.2\!\times\! 10^{-3}}{ |V_{ub}|} \,.
\end{align}
We emphasize that these results do not rely on Eq.~(\ref{invpiexp}).  In
Eq.~(\ref{flzJ}) the first errors are experimental and the second theoretical.
Theoretical errors are computed as a generic 20\% error on the central value,
both here and below in Eq.~(\ref{flzJrho}).  Setting $|V_{ub}|=4.2\times
10^{-3}$ and using $\langle x^{-1}\rangle_{\pi}=2.9$ in Eq.~(\ref{flzJ}) gives
\begin{align}
 \zeta^{B \pi} &= 0.092 \pm 0.027 ,\qquad
 \zeta_J^{B\pi} =0.090 \pm 0.018 \,.
\end{align}
These values favor $\zeta_J^{B\pi}\sim \zeta^{B\pi}$. There is a sizeable
correlation in their quoted errors, and to take this correlation into account in
our numerical analysis we will do Gaussian scans over the range of experimental
errors quoted in Eq.~(\ref{flzJ}).  A 12\% error in $|V_{ub}|$ is also included
in our final results.

The same results, Eqs.~(\ref{z0}-\ref{z1a}) apply for $B\to \rho\rho$ for
longitudinal $\rho$'s, where now one uses $N_{\rho^0\rho^-}$ and determines
$t_c^{\rho\rho}$ from the $B\to \rho\rho$ branching ratios and CP-asymmetries
$S_{\rho^+\rho^-}$ and $C_{\rho^+\rho^-}$.  Here the analog of $f_+^{B\pi}(0)$
is the longitudinal $B\to \rho\ell\bar\nu$ form factor at $q^2\simeq 0$
\begin{align}
 A_\parallel(0) &= \frac{m_B^2\: A_2(0)}{2m_V(m_B\plus m_V)}  -\frac{(m_B\plus
   m_V)}{2 m_V} A_1(0) \nn\\
  &= - \zeta^{B\rho} - \zeta_J^{B\rho}  \,.
\end{align}
Taking $\gamma=67\pm 10^\circ$ the nonleptonic data gives
\begin{align} \label{flzJrho}
  A^{B\rho}_\parallel(0) &= -\Big(0.261 \pm 0.022 \pm 0.052 
\Big) \frac{4.2\!\times\! 10^{-3}}{ |V_{ub}|} \,,\nn\\
  \langle x^{-1}\rangle_{\rho}\, \zeta_J^{B\rho} &= 
   \Big(0.06 \pm 0.11   
 \Big) 
   \frac{4.2\!\times\! 10^{-3}}{ |V_{ub}|} \, .
\end{align}
Again the first errors are experimental and the second theoretical. Due to the
large uncertainty in the central value of $\langle x^{-1}\rangle_{\rho}\,
\zeta_J^{B\rho}$ a 20\% theoretical uncertainty would not be noticeable.

It is interesting to make a comparison of the $\pi$ and $\rho$ parameters, which
from the results in Eq.~(\ref{flzJ}) and (\ref{flzJrho}) give
\begin{align} \label{ratio1}
 R_{\rho\pi}\equiv \frac{\langle x^{-1}\rangle_{\rho} \, \zeta_J^{B\rho} }
    {\langle x^{-1}\rangle_{\pi} \, \zeta_J^{B\pi} }
  = 0.23 \pm 0.42
\,.
\end{align}
This large $0.42$ experimental error is induced by our current knowledge of the
$B\to \rho\rho$ CP-asymmetries together with greater sensitivity to $\gamma$ and
$t_c$. Now in \SCETb one can derive a factorization theorem for $\zeta_J^{BM}$
(discussed in Eq.~(\ref{zetaJfactor}) below) that implies that
\begin{align}
 R_{\rho\pi} &= \frac{ f_\rho  [\langle x^{-1}\rangle_{\rho}]^2 }
    { f_\pi [\langle x^{-1}\rangle_{\pi }]^2 } 
   \bigg[ 1+ {\cal O}\Big(\alpha_s(\mu_i),\frac{\Lambda}{E}\Big) \bigg]
  \,.
\end{align}
Here the theoretical errors should be increased to $\simeq 35\%$ to account for
the additional expansion in $\alpha_s$ at the intermediate scale $\mu_i$.  Using
$f_\rho \simeq 1.6\,f_\pi$ we find that the nonleptonic data plus factorization
at the intermediate scale currently implies
\begin{align} \label{rcharmless}
  \frac{\langle x^{-1}\rangle_{\rho}}
    {\langle x^{-1}\rangle_{\pi}  } & = 0.38 \pm 0.35 \pm 0.13 \,,
\end{align} 
where the first error is experimental and the second is the $35\%$ theoretical
error.  

A result for this ratio can also be obtained from the factorization theorem for
the color-suppressed decays $\bar B^0\to D^0\rho^0$ and $\bar B^0\to D^0\pi^0$
derived in Eq.(69) of Ref.~\cite{Mantry:2003uz}, which gives
\begin{align} \label{rcharm}
   \frac{\langle x^{-1}\rangle_{\rho}}
    {\langle x^{-1}\rangle_{\pi}  }  &= \frac{f_\pi}{f_\rho}
\sqrt{\frac{{\rm Br}(\bar B^0\to D^0\rho^0)}{{\rm Br}(\bar B^0\to
D^0\pi^0)}}
  = 0.62 \pm 0.24 \,.
\end{align}
It is quite interesting that this ratio is found to be less than
unity, and that the results extracted in Eq.~(\ref{rcharmless}) from
charmless decays, and in Eq.~(\ref{rcharm}) from charmed final states,
agree within errors. The significant range allowed by the errors can
be reduced by noting that there is a rigorous lower bound on the
inverse moment for positive definite $\phi^M(x)$,
\begin{align}
  \langle x^{-1} \rangle_{M} &= \int_0^1\!\! dx \: \frac{\phi^M(x) }{x} 
   \ge  \int_0^1\!\! dx \: \phi^M(x)  = 1 \,.
\end{align} 
For $M=\pi$ and $M=\rho$ we have $\phi^M(x)= \phi^M(1-x)$ from
isospin and charge conjugation, and this bound can be strengthened:
\begin{align}
  \langle x^{-1} \rangle_{M} &= \int_0^1\!\! dx \: \frac{\phi^M(x) }{x} 
  =   \int_0^{1/2}\!\!\!\! dx \: \Big(\frac{1}{x} + \frac{1}{1\minus x}\Big)
   \phi^M(x) \nn\\
  &\ge  4 \int _0^{1/2}\!\!\!\! dx \: \phi^M(x)  = 2 \,.
\end{align} 
With the mean value in Eq.~(\ref{invpiexp}) this bound is close to the central value
in Eq.~(\ref{rcharm}).

For our analysis we take into account the bound and the data in
Eqs.~(\ref{rcharmless}) and (\ref{rcharm}), and hence use a model for
$\phi_\rho(x)$ that is constrained such $\langle
x^{-1}\rangle_{\rho}=2.2 {}^{+0.6}_{-0.2}$ which gives $\langle
x^{-1}\rangle_{\rho}/\langle x^{-1}\rangle_{\pi}\simeq 0.76$. Taking
$|V_{ub}|=4.2\times 10^{-3}$ we then find using Eq.~(\ref{flzJrho})
that
\begin{align} 
 \zeta^{B\rho}  =0.234 \pm 0.065  , \qquad 
   \zeta_{J}^{B \rho} &= 0.027  \pm 0.049
   \,.
\end{align}
Again we scan over the range of experimental errors in
Eq.~(\ref{flzJrho}) to take into account the sizeable correlations.
Note that $\zeta_{J}^{B \rho}$ is sensitive to the values of the $B\to
\rho\rho$ branching ratios, which dominate the error and favor a
smaller color suppressed amplitude than in $B\to\pi\pi$. The central
values in $B\to \rho\rho$ are consistent with $\zeta_J^{B\rho}\lesssim
\alpha_s(\mu_i) \zeta^{B\rho}$.  However, within errors the scaling
that we adopt, $\zeta_J^{B\rho}\sim
\zeta^{B\rho}$, is also consistent (given that $\sim$ still means that the
factors can differ numerically by a factor like $2$).  As the experimental
uncertainties on the nonleptonic decays decrease, we expect the combined
analysis of $B\to\pi\pi,\rho\rho$ introduced in this section to play an
important role in furthering our knowledge of hadronic parameters appearing in
the factorization theorem for charmless nonleptonic decays.


\section{Factorization at Leading Power} \label{sect:LO}

In this section we review the SCET factorization analysis at
leading order from~\cite{Bauer:2004tj} to setup our notation. The
decays $B \to M_1 M_2$ are mediated in full QCD by the weak $\Delta
B=1$ Hamiltonian, which for $\Delta S=0$ reads
\begin{align} \label{Hw}
 H_W = \frac{G_F}{\sqrt{2}} \sum_{p=u,c} \lambda_p^{(d)}
 \Big( C_1 O_1^p + C_2 O_2^p 
  +\!\!\! \sum_{i=3}^{10,7\gamma,8g}\!\! C_i O_i \Big),
\end{align}
where the CKM factor is $\lambda_p^{(f)} = V_{pb} V^*_{pf}$ with $f=d$ and at LL
order 
\begin{align}
 C_{1-10}(m_b) &= \{
  1.107\,, 
  -.249\,,
  .011\,,
 -.026\,, 
  .008\,, 
 -.031 \,, 
  \nn\\
 & \hspace{-1.3cm}
  4.2 \!\times\! 10^{-4} \,,
  4.2 \!\times\! 10^{-4} \,,
  -9.7 \!\times\! 10^{-3} \,,
  1.9 \!\times\! 10^{-3} \}.
\end{align}
The coefficients in Eq.~(\ref{Hw}) are known at NLL order~\cite{fullWilson}, and
the values we used for our main analysis are presented in
section~\ref{sect:inputs}.  The basis of operators is
\begin{align}\label{fullops}
 O_1^p  &= (\overline{p} b)_{V\!-\!A}
  (\overline{d} p)_{V\!-\!A}, \ \
 \\
 O_2^p &= (\overline{p}_{\beta} b_{\alpha})_{V\!-\!A}
  (\overline{d}_{\alpha} p_{\beta})_{V\!-\!A},
  \nonumber \\
 O_{3} &=  (\overline{d} b)_{V\!-\!A}
  (\overline{q} q)_{V\! - \!A}\,, 
 \nonumber \\
  O_{4} &=  (\overline{d}_{\beta} b_{\alpha})_{V\!-\!A}
  (\overline{q}_{\alpha} q_{\beta})_{V\! - \!A} \,, 
 \nonumber \\
 O_{5}  &=  (\overline{d} b)_{V\!-\!A}
  (\overline{q} q)_{V\! + \!A}\,, 
  \nonumber \\
 O_{6}  &=  (\overline{d}_{\beta} b_{\alpha})_{V\!-\!A}
  (\overline{q}_{\alpha} q_{\beta})_{V\! + \!A} \,,
  \nonumber \\
 O_{7}  &= \frac{3e_q}{2} (\overline{d} b)_{V\!-\!A}
  (\overline{q} q)_{V\! + \!A},
  \nonumber \\
 O_{8}  &= \frac{3e_q}{2} (\overline{d}_{\beta} b_{\alpha})_{V\!-\!A}
  (\overline{q}_{\alpha} q_{\beta})_{V\! + \!A},
  \nonumber \\
 O_{9}  &= \frac{3e_q}{2} (\overline{d} b)_{V\!-\!A}
  (\overline{q} q)_{V\! - \!A}
  ,\nonumber\\
 O_{10}  &= \frac{3e_q}{2} (\overline{d}_{\beta} b_{\alpha})_{V\!-\!A}
  (\overline{q}_{\alpha} q_{\beta})_{V\! - \!A}
  ,\nonumber\\
  O_{8g} &= -\frac{g \bar m_b}{4\pi^2} \: 
     \overline{d} \sigma_{\mu\nu} G^{\mu\nu} P_R b , \nn\\
  O_{7\gamma} &= -\frac{e \bar m_b}{4\pi^2} \: 
     \overline{d} \sigma_{\mu\nu} F^{\mu\nu} P_R b \,. \nn
\end{align}
Here $\alpha, \beta$ are color indices and $e_q$ are electric charges and the
$q$ are summed over the light quarks, $q=u,d,s,c,b$. The $\Delta S=1$ $H_W$ is
obtained by replacing $(f=d)\to (f=s)$ in Eqs.~(\ref{Hw},\ref{fullops}). The
numerical dominance of $C_1,C_2,C_{8g}$ will allow us to simplify the
calculation since we need only include the effects of $O_{1,2,8g}$ at one-loop.
Perturbative corrections due to the other operators are numerically tiny.  Our
sign for $g$ is such that the QCD fermion Feynman rule is $i g T^A \gamma^\mu$.

The matching onto SCET occurs in two stages. First one matches onto \SCETa by
integrating out fluctuations at the scale $m_b$.  One then matches onto $\SCETb$
at the scale $\sqrt{\Lambda m_b}$.  For the LO factorization theorem for
nonleptonic B-decays this second step of matching can not lead to strong phases,
as discussed in \cite{Bauer:2004tj}, and for $\zeta_{J}^{BM}$ is known at
one-loop order~\cite{Becher:2004kk,Beneke:2005gs}.  In this paper we wish to
complete the ${\cal O}(\alpha_s)$ matching for the first stage.  For tree
amplitudes the corresponding computation was carried out in
Ref.~\cite{Beneke:2005vv}. Here we consider the result for the penguin
amplitudes.  In particular we present the short-distance up and charm loop
contributions in two different regulation schemes for $\gamma_5$, as well as
corresponding contributions from the magnetic gluon operator.

At the scale $\mu\simeq m_b$ the Hamiltonian in Eq.~(\ref{Hw}) is matched onto
operators in SCET. Due to the nature of the matching onto \SCETb the first two
orders of the power expansion of the Hamiltonian in \SCETa are needed to
determine the leading-order amplitudes
\begin{align} \label{match}
 H_W \!\!&=  &\!\! \frac{2G_F}{\sqrt{2}} \sum _{n,\bn} \bigg\{ 
  \sum_i \int [d\omega_{j}]_{j=1}^{3}
       c_i^{(f)}(\omega_j)  Q_{if}^{(0)}(\omega_j) \nn\\ 
 && \hspace{-1cm}
  + \sum_i \int [d\omega_{j}]_{j=1}^{4}  b^{(f)}_i(\omega_j) 
  Q_{if}^{(1)}(\omega_j) 
  + \ldots \bigg\} \,.
\end{align}
  The operators for the $\Delta S=0$
transitions are~\cite{chay,Bauer:2004tj}
\begin{align} \label{Q0}
  Q_{1d}^{(0)} &=  \big[ \bar u_{n,\omega_1} \bnslash P_L b_v\big]
  \big[ \bar d_{\bn,\omega_2}  \nslash P_L u_{\bn,\omega_3} \big]
  \,,  \\
  Q_{2d,3d}^{(0)} &=  \big[ \bar d_{n,\omega_1} \bnslash P_L b_v \big]
  \big[ \bar u_{\bn,\omega_2} \nslash P_{L,R} u_{\bn,\omega_3} \big]
   \,,\nn \\
  Q_{4d}^{(0)} &=  \big[ \bar q_{n,\omega_1} \bnslash P_L b_v \big]
  \big[ \bar d_{\bn,\omega_2} \nslash P_{L}\, q_{\bn,\omega_3} \big]
   \,, \nn \\
  Q_{5d,6d}^{(0)} &=  
  \big[ \bar d_{n,\omega_1} \bnslash P_L b_v \big]
  \big[ \bar q_{\bn,\omega_2} \nslash P_{L,R} q_{\bn,\omega_3} \big]
  \,, \nn
\end{align}
and
\begin{align} \label{Q1}
  Q_{1d}^{(1)} &= \frac{-2}{m_b} 
     \big[ \bar u_{n,\omega_1}\, ig\,\slash\!\!\!\!{\cal B}^\perp_{n,\omega_4} 
     P_L b_v\big]
     \big[ \bar d_{\bn,\omega_2}  \nslash P_L u_{\bn,\omega_3} \big] 
     \,,  \\
  Q_{2d,3d}^{(1)} &=  \frac{-2}{m_b}  
     \big[ \bar d_{n,\omega_1} \, ig\,\slash\!\!\!\!{\cal B}^\perp_{n,\omega_4} 
     P_L b_v \big]
     \big[ \bar u_{\bn,\omega_2} \nslash P_{L,R} u_{\bn,\omega_3} \big]
      \,,\nn \\
  Q_{4d}^{(1)} &=  \frac{-2}{m_b} 
     \big[ \bar q_{n,\omega_1} \, ig\,\slash\!\!\!\!{\cal B}^\perp_{n,\omega_4} 
     P_L b_v \big]
     \big[ \bar d_{\bn,\omega_2} \nslash P_{L}\, q_{\bn,\omega_3} \big]
      \,,\nn \\
  Q_{5d,6d}^{(1)} &= \frac{-2}{m_b} 
    \big[ \bar d_{n,\omega_1} \, ig\,\slash\!\!\!\!{\cal B}^\perp_{n,\omega_4} 
     P_L b_v \big]
    \big[ \bar q_{\bn,\omega_2} \nslash P_{L,R} q_{\bn,\omega_3} \big]
      \,, \nn\\
  Q_{7d}^{(1)} &= \frac{-2}{m_b} 
   \big[ \bar u_{n,\omega_1}\, ig\,{\cal B}^{\perp\, \mu}_{n,\omega_4} 
    P_L b_v\big]
   \big[ \bar d_{\bn,\omega_2}  \nslash \gamma^\perp_\mu P_R u_{\bn,\omega_3} \big] 
      \,,\nn\\
  Q_{8d}^{(1)} &= \frac{-2}{m_b} 
   \big[ \bar q_{n,\omega_1}\, ig\,{\cal B}^{\perp\, \mu}_{n,\omega_4} 
    P_L b_v\big]
   \big[ \bar d_{\bn,\omega_2}  \nslash \gamma^\perp_\mu P_R q_{\bn,\omega_3} \big] 
      \,. \nn
\end{align}
Here $P_L=(1-\gamma_5)/2$ and $P_R=(1+\gamma_5)/2$.  At lowest order,
$ Q_{7d,8d}^{(1)}$ give a vanishing contribution to the rates. $
Q_{5d,6d}^{(0,1)}$ will not be relevant in our analysis since we will
not be considering isosinglet final states. At tree level the matching
onto $ Q_{5d,6d}^{(0,1)}$ was done in Ref.~\cite{JureAlex}. From
Eqs.~(\ref{Q0},\ref{Q1}) the $\Delta S=1$ operators $Q_{is}^{(0)}$ are
obtained by swapping $\bar d\to \bar s$.  The ``quark'' fields in
Eqs.~(\ref{Q0},\ref{Q1}) with subscripts $n$ and $\bn$ are products of
collinear quark fields and Wilson lines with large momenta $\omega_i$.
In particular we have defined
\begin{align}
  \bar u_{n,\omega} &= [ \bar\xi_n^{(u)} W_n\, \delta(\omega\!-\!
\bn\mcdot\cP^\dagger) ]\,, \\
 ig\,{\cal B}^{\perp\,\mu}_{n,\omega} &= \frac{1}{(-\omega)}\, 
 \big[ W^\dagger_n [ i\bn\mcdot D_{c,n} , i D^\mu_{n,\perp} ] W_n 
  \delta(\omega-\bnP^\dagger) \big] \nn 
\end{align}
where $\bar\xi_n^{(u)}$ creates a $n$-collinear up quark or
annihilates an antiquark.  The $b_v$ field is the standard HQET field.
For a complete basis we also need operators with octet bilinears,
$T^A\otimes T^A$, but their matrix elements vanish at LO.

The leading-order amplitude is generated by time-ordered products of
both the operators $Q^{(0)}$ and $Q^{(1)}$ with insertions of a
subleading Lagrangians~\cite{bps4,ps1}.  T-products with $Q^{(0)}$ can
be factorized as $T_1^i \tilde Q_i^\bn$ and contribute to terms with
$\zeta^{BM}$, while T-products with $Q^{(1)}$ can be written as $T_2^i
\tilde Q_i^\bn$ and contribute to terms with $\zeta_J^{BM}$. Here
\begin{align} \label{Tproducts}
  T_1[\tilde Q_i^{(0)}]   &\equiv  \mbox{\large $\int$} d^4y\, d^4y'\,T 
   \big[\tilde Q^{(0)}_i(0) \: i{\cal L}^{(1)}_{\xi_n q}(y) \: i{\cal
    L}_{\xi_n\xi_n}^{(1)}\!(y') \big] \!\nn\\
  & \ \ + \mbox{\large $\int$} d^4y\, d^4y'\,T 
   \big[\tilde Q^{(0)}_i(0) \: i{\cal L}^{(1)}_{\xi_n q}(y) \: 
   i{\cal L}_{cg}^{(1)}(y')\} \big]\nn \\
  & \ \ 
     +  \mbox{\large $\int$} d^4y\, 
    T \big[\tilde Q_i^{(0)}(0),i{\cal L}^{(1,2)}_{\xi_n q}(y) \big], \ \nn\\
  T_2[\tilde Q_i^{(1)}] &\equiv \mbox{\large $\int$} d^4y \:
    T \big[\tilde Q_i^{(1)}(0),i{\cal L}^{(1)}_{\xi_n q}(y) \big] ,
\end{align} 
and it was convenient to define
\begin{align}
  \tilde Q_i^{(0)} &= \big[ \bar q^{i}_{n,\omega_1} \bnslash P_L b_v \big] \,,
  \\
  \tilde Q_i^{(1)} &= \frac{-2}{m_b} \big[ \bar q^i_{n,\omega_1}\,
  ig\,\slash\!\!\!\!{\cal B}^\perp_{n,\omega_4} P_L b_v\big] \,,\nn\\
  \tilde Q_i^\bn &= \bigg\{
  \begin{array}{l}
   \bar q^i_{\bn,\omega_2} \nslash P_{L} q^{\prime
    i}_{\bn,\omega_3} \qquad i=1,2,4,5 \\
     \bar q^i_{\bn,\omega_2} \nslash P_{R} q^{\prime
    i}_{\bn,\omega_3} \qquad i=3,6
  \end{array} 
  \,. \nn
\end{align}
The Lagrangians in Eq.~(\ref{Tproducts}) can be found in Ref.~\cite{bps5}. Note
that only the $n$-collinear fields appear in the T-products $T_1^i$ and $T_2^i$,
which explains why the same $T_{1,2}^i$ appear for heavy-to-light form factors
at large meson energies~\cite{Bauer:2005wb}. The form factors simply do not have
the extra $\tilde Q_i^\bn$.  In addition we have operators/T-products whose
matrix elements give $A_{cc}$. We refer to section~\ref{sect:long} below for
further discussion of these contributions.

In this paper we use factorization at the scale $m_b$, where the hadronic
parameters are defined by matrix elements of $T_1$ and $T_2$ and the
$\bn$-collinear operator, namely
\begin{align} \label{zeta}
%
  \big\langle M_n \big| T_1\big[ \bar q_{n\omega_1}^L \bnslash b_v \big]
   \big| B\big\rangle 
  & =  {\cal C}_{q_L}^{BM}  
   \bar \delta_{\omega_1} \: {m_B} \: \zeta^{BM} , \\
  \big\langle M_n \big| T_2\big[ \bar q_{n\omega_1}^L \,
  ig\,\slash\!\!\!\!{\cal B}^\perp_{n\omega_4}  b_v \big] \big| B\big\rangle 
 &= - {\cal C}_{q_L}^{BM} 
   \bar \delta_{\omega_1\omega_4} \frac{m_B}{2} \zeta_J^{BM}(z)\,,
  \nn\\
   \big\langle M_\bn \big| \bar q_{\bn\omega_2}^{\,\prime L} \nslash 
       q^{L}_{\bn\omega_3} \big| 0 \big\rangle 
  &= \frac{i}{2}\:{\cal C}_{q'_L q}^M   \bar \delta_{\omega_2\omega_3}\, f_M
  \, \phi_{M}(u)  
   \nn , \\
   \big\langle M_\bn \big| \bar q_{\bn\omega_2}^{\,\prime R} \nslash 
     q^{R}_{\bn\omega_3} \big| 0 \big\rangle 
  &= \frac{i}{2}\:{\cal C}_{q'_R q}^M   \bar \delta_{\omega_2\omega_3}\, f_M
  \, \phi_{M}(u) 
   \nn , 
\end{align}
where $z=\omega_1/m_B$, $u=\omega_2/m_B$ and we have made the
momentum-conserving $\delta$-functions explicit, $\bar
\delta_{\omega}=\delta(\omega_1-m_B)$, $\bar
\delta_{\omega_1\omega_4}=\delta(\omega_1+\omega_4-m_B)$, and $\bar
\delta_{\omega_2}^{\omega_3}=\delta(\omega_2-\omega_3-m_B)$.  As pictured
in Fig.~\ref{figQCDb}, $u$ and $1-u$ are momentum fractions for the quark and
antiquark $\bn$-collinear fields, and $z$ and $1-z$ are the momentum fractions
carried by the $n$-collinear quark and gluon field in $\tilde Q_i^{(1)}$.
Finally, ${\cal C}_i^{BM}$ and ${\cal C}_{i}^M$ are Clebsch-Gordan coefficients.
We fix the following sign convention for the states
\begin{align}
&\pi^+ = +u\bar d, ~~~  \pi^0 = \frac{1}{\sqrt{2}}(d\bar d-u\bar u),
   ~~~ \pi^- = -d\bar u,  \nn \\
& \bar K^0 = s\bar d, ~~ K^- = -s \bar u, ~~ K^+ = u \bar s, 
  ~~ K^0 = d\bar s , \nn \\
&\, \bar B^0 = b\bar d, ~~ \, B^- = -b \bar u, ~~ \, B^+ = u \bar b, 
  ~~  B^0 = d\bar b\, ,
\end{align}
and take vector meson states to have a negative sign relative to the
corresponding pseudoscalar mesons. The over all phase convention is fixed so
that the Clebsch-Gordan ${\cal C}_{u_L}^{\bar B^0\pi^+}=+1$, ${\cal
  C}_{d_Lu}^{\pi^-}=+1$, ${\cal C}_{d_Ru}^{\pi^-}=-1$, ${\cal C}_{u_L}^{\bar
  B^0\rho^+}=+1$, and ${\cal C}_{d_L u}^{\rho^-}= {\cal C}_{d_R u}^{\rho^-}=+1$. One can
then compute that ${\cal C}_{d_L}^{B^-\pi^-}=+1$ and ${\cal C}_{u_ru}^{\pi^0}=-
\frac{1}{\sqrt{2}}$, etc.  Note that the signs take into account whether the
operators have left or right-handed quarks. Putting the pieces together gives the
leading order factorization theorem which integrates out hard $\sim m_b^2$
fluctuations
\begin{align}
\label{newfact}
 A^{LO} &\equiv  -i\, \langle M_1 M_2 | H_W | \bar B \rangle \\
 &= \frac{G_F m_B^2}{\sqrt{2}} f_{M_1}  \bigg[ 
  \int_0^1\!\! du dz \: T_{1 J}(u,z) \zeta_J^{B M_2}(z) \phi^{M_1}(u)
  \nn\\
 &\qquad +  \zeta^{B M_2} \int_0^1\!\! du\: T_{1\zeta}(u) \phi^{M_1}(u)  
  \bigg] + (1\leftrightarrow 2) \,
  \nn .
\end{align}
Here the hard coefficients $T_{1\zeta}$ and $T_{1 J}$ depend on
channel specific linear combinations of the matching coefficients 
\begin{align} \label{Tz}
%
 T_{1\zeta}(u) &= 
  {\cal C}_{u_L}^{BM_2} \, {\cal C}_{f_Lu}^{M_1} \, c_1^{(f)}(u)
  +   {\cal C}_{f_L}^{BM_2} \, {\cal C}_{u_Lu}^{M_1} \, c_2^{(f)}(u)
  \nn \\
 & +   {\cal C}_{f_L}^{BM_2} \, {\cal C}_{u_Ru}^{M_1} \, c_3^{(f)}(u)
  +  {\cal C}_{q_L}^{BM_2} \, {\cal C}_{f_Lq}^{M_1} \, c_4^{(f)}(u)
  ,\nn\\
 T_{1 J}(u,z) &=
   {\cal C}_{u_L}^{BM_2} \, {\cal C}_{f_Lu}^{M_1} \, b_1^{(f)}(u,z)
  +   {\cal C}_{f_L}^{BM_2} \, {\cal C}_{u_Lu}^{M_1} \, b_2^{(f)}(u,z)
   \nn\\
 &+   {\cal C}_{f_L}^{BM_2} \, {\cal C}_{u_Ru}^{M_1} \, b_3^{(f)}(u,z)
  +  {\cal C}_{q_L}^{BM_2} \, {\cal C}_{f_Lq}^{M_1} \, b_4^{(f)}(u,z)
  . 
\end{align}
Results for these $T$'s in different decay channels can be read off of
Table~I in Ref.~\cite{Bauer:2004tj}.  Power counting implies
$\zeta^{BM}\sim \zeta^{BM}_J\sim (\Lambda/m_b)^{3/2}$ while
$\phi_{M}\sim 1$.  Here the non-perturbative parameters $\zeta^{BM}$,
$\zeta^{BM}_{J}(z)$, and $\phi^M(u)$, all occur in the $B\to M$
semileptonic and rare form factors.  For a model independent analysis
they need to be determined from data. Note that in the leading order
factorization theorem all terms involve a form factor times a meson
distribution function.

Taking the terms proportional to $\lambda_c^{(f)}$ from Eq.~(\ref{newfact})
generates the penguin amplitude terms on the first line of Eq.~(\ref{hatP}).
Using Eq.~(\ref{newfact}) still requires matching the full theory $O_i$'s onto
the $Q_{if}^{(0,1)}$ to determine the Wilson coefficients $c_i^{(f)}$ and
$b_i^{(f)}$.  For the coefficients of $Q_i^{(0)}$ with $f=d,s$ we have
\begin{eqnarray} \label{ci}
  c_1^{(f)} \!\!&=&\!\! 
       \lambda_u^{(f)}\Big( C_1 \!+\! \frac{C_2}{N_c} \Big)
       -\lambda_t^{(f)} \frac32 \Big( C_{10} \!+\! \frac{C_9}{N_c}\Big)
       + \Delta c_1^{(f)}
      \,,\nn\\
  c_2^{(f)} \!\!&=&\!\! 
       \lambda_u^{(f)}\Big( C_2 \!+\! \frac{C_1}{N_c}\Big) 
       -\lambda_t^{(f)} \frac32 \Big(C_9 \!+\! \frac{C_{10}}{N_c}\Big)
       +  \Delta c_2^{(f)}
      \,,\nn \\
  c_3^{(f)} \!\!&=&\!\!
       -\lambda_t^{(f)}\frac32  \Big( C_7 + \frac{C_8}{N_c}\Big)
       +  \Delta c_3^{(f)}
     \,, \nn \\
  c_4^{(f)} \!\!&=&\!\! 
        -\lambda_t^{(f)}\Big( C_4 + \frac{C_3}{N_c}
       -\frac{C_{10}}{2} - \frac{C_9}{2N_c} \Big)
      +  \Delta c_4^{(f)}
     \,,
\end{eqnarray}
and for the $Q_i^{(1)}$ we have the coefficients
\begin{eqnarray} \label{bi}
   b_1^{(f)}  \!\!&=&\!\!
    \lambda_u^{(f)}\Big[ C_1 + \Big(1 \!+\!\frac{1}{\bar u} \Big)
      \frac{C_2}{N_c} \Big] 
      \\
&& 
    - \lambda_t^{(f)}\Big[ \frac{3}{2} C_{10} + 
    \Big(1 \!+\!\frac{1}{\bar u} \Big)
      \frac{3 C_9}{2 N_c} \Big]
     + \Delta b_1^{(f)}
     \,,\nn
     \\
   b_2^{(f)}  \!\!&=&\!\!
     \lambda_u^{(f)}\Big[ C_2 + \Big(1 \!+ \! \frac{1}{\bar u}  \Big)
      \frac{C_1}{N_c}\Big]
      \nn\\
     && 
     - \lambda_t^{(f)}\Big[ \frac{3}{2} C_9 + \Big(1 \!
     +\!\frac{1}{\bar u} \Big)
      \frac{3 C_{10}}{2 N_c} \Big]
     + \Delta b_2^{(f)}
     \,,\nn 
     \\
   b_3^{(f)}  \!\!&=&\!\!
     - \lambda_t^{(f)}\Big[ \frac{3}{2} C_7 
     + \Big(1 \!-\!\frac{1}{u} \Big)
      \frac{3 C_8}{2 N_c} \Big]
     + \Delta b_3^{(f)}
     \,, \nn \\
   b_4^{(f)}  \!\!&=&\!\!
     -\lambda_t^{(f)}\Big[ C_4 \!-\! \frac{C_{10}}{2}\!
     +\!  \Big(1 \!+\! \frac{1}{\bar u}  \Big)
      \Big( \frac{C_3}{N_c} \!-\! \frac{C_9}{2N_c} \Big)\Big]
      + \Delta b_4^{(f)} ,
\nn 
\end{eqnarray}
where $\bar u=1-u$. The $\Delta c_i^{(f)}$ and $\Delta b_i^{(f)}$ denote terms
depending on $\alpha_s$ generated by matching from $H_W$, and will be considered
at ${\cal O}(\alpha_s(m_b))$ below. The displayed terms in $c_4^{(f)}$ and
$b_4^{(f)}$ correspond to $C_{3,4,9,10}$ terms in the first line of
Eq.~(\ref{hatP}), and $\Delta c_4^{(f)}$ and $\Delta b_4^{(f)}$ include the $\alpha_s
C_{1,2,8g}$ terms.


\section{Endpoint Behavior of $\zeta_J^{BM}(z)$} \label{sect:endpt}

In order to model the generalized form factors like $\zeta_J^{BM}(z)$ it is
useful to know their behavior in the endpoints  $z\to 0$ and $z\to 1$. This
behavior along with that of $\phi^M(u)$ determines whether the convolutions
$\int du \, c_i^{(f)}(u)\phi^M(u)$ and $\int du dz\,
b_i^{(f)}(u,z)\zeta_J^{BM}(z) \phi^M(u)$ in the factorization theorem in
Eq.~(\ref{newfact}) converge naively or require zero-bin
subtractions~\cite{Manohar:2006nz}. At tree level $c_i^{(f)}(u)\sim 1$ since it
is independent of $u$, while $b_i^{(f)}(u,z)\sim \bar u^{-1}$ is independent
of $z$ (for the electroweak coefficient $b_3^{(f)}$ replace the $\bar u^{-1}$ by
$u^{-1}$). At one loop the scaling behavior becomes $c_i^{(f)}(u)\sim (u\bar
u)^{-1}$ and $b_i^{(f)}(u,z)\sim (z\, \bar u)^{-1}$ as discussed in
section~\ref{sect:summary} below. Known two-loop corrections do not modify these
one-loop scaling results~\cite{Guido}.

Using a factorization of the generalized form factor we can connect the scaling
of $\zeta_J^{BM}$ to that of $\phi^M(u)$. Separating the scales $\Lambda^2 \ll
m_b\Lambda$ gives the factorization theorem
\begin{align} \label{zetaJfactor}
 & \zeta^{BM}_J(z) = \frac{f_B f_{M}}{m_b}\int_0^1\!\!\!dx \!\int^\infty_0\!\!\!\!  dk^+
  J(z,x,k_+)\phi_B(k_+) \phi^{M_1}(x) ,
\end{align}
where $J$ is a ``jet function'' that can be determined as a power series in
$\alpha_s(\mu_i)$ where $\mu_i\sim \sqrt{m_b\Lambda}$.  At tree level $J$ is given by
\begin{eqnarray}
J(z,x,k_+)=\delta(x-z)C_F\frac{\alpha_s(\mu)\pi}{N_c\, \bar z\,  k_+}\,.
\end{eqnarray}
Using this result in Eq.~(\ref{zetaJfactor}) gives
\begin{eqnarray} \label{zetatree}
\zeta_J^{B M}(z) \Big|_{\rm tree} 
 &=&4 \pi \alpha_s(\mu) \frac{ f_B f_M \beta_B}{3 m_B} \, \frac{\phi^M(z)}{\bar
   z}
\,,
\end{eqnarray}
which demonstrates that that the endpoint scaling $\phi^M(z)\sim (z\bar z)$
implies $\zeta_J^{BM}(z) \sim z$. Beyond tree level the scaling is determined by
corrections to $J(z,x,k_+)$, which is currently known to one-loop
order~\cite{Becher:2004kk,Beneke:2005gs} and still yields $\zeta_J^{BM}(z) \sim
z$. This scaling is expected to persist to all orders. Evidence for this comes
from the argument of Ref.~\cite{Beneke:2005gs} that is based on an assumed
correspondence between soft and collinear endpoint singularities in the form
factor.  A strong argument for why these corresponding contributions must always
arise when endpoint singularities appear was given in
Ref.~\cite{Manohar:2006nz}. Endpoint singularities are simply an artifact of not
properly separating momentum regions in the effective theory, and arise in
situations where a collinear momentum generates a double counting with modes
that account for the region where the momentum is soft (and vice versa). In the
effective field theory this is avoided by including zero-bin subtractions.  Thus
zero-bin subtractions are not expected to arise for $\zeta_J^{BM}(z)$, whereas
the analogous factorization theorem for $\zeta^{BM}$ to Eq.~(\ref{zetaJfactor}),
which exhibits endpoint singularities, requires zero-bin
subtractions~\cite{Manohar:2006nz}.

In Eq.~(\ref{zetaJfactor}) the normalization depends on decay constants and the
inverse moment parameter
\begin{align}
 \beta_B\equiv \int dk^+ \, \phi_B(k^+)/(3k^+)  
       = 1/(3\lambda_B) 
      \sim \Lambda_{\rm QCD}^{-1} \,.
\end{align}
Unfortunately, use of Eq.~(\ref{zetaJfactor}) to determine the normalization,
$\zeta_J^{BM}$, has a large uncertainty due to the unknown parameter $\beta_B$
and the $\alpha_s(\mu_i)$ expansion.  The $\mu_i$ dependence can be reduced by
using one-loop results for the jet function, but this introduces additional
uncertainty from other moments of $\phi_B(k^+)$.

In this paper we avoid using Eq.~(\ref{zetaJfactor}) and instead
reduce the hadronic uncertainties by dealing directly with the
normalization parameters $\zeta^{BM}$ and $\zeta^{BM}_J$ determined
from data.  The analysis in this section constrains the model that we
construct for the shape of $\zeta^{BM}_J(z)$, as discussed in
section~\ref{sect:models} below.


\section{Summary of Leading-Order One-loop Coefficients}  \label{sect:summary}

\noindent In this section we summarize the main results of our
computation of the fourth term on line one of Eq.~(\ref{hatP}), while leaving
the details to follow in the next two sections.  For convenience we will use the
following decomposition of the $\alpha_s$ corrections
\begin{align} \label{cbsplit}
\Delta c_4^{(f)} &= \Delta c_4^{(1c)} + \Delta c_4^{(1u)}
    +\Delta c_4^{(1g)} + {\mathcal O}(\alpha_s^2), \nonumber \\
\Delta b_4^{(f)} &= \Delta b_4^{(1c)} + \Delta b_4^{(1u)} 
     +\Delta b_4^{(1g)}+ {\mathcal O}(\alpha_s^2).
\end{align}
 Here superscripts $(1c)$ and $(1u)$ denote the one loop contribution
due to charm and up quark loops respectively, while $(1g)$ refers to
${\mathcal O}(\alpha_s)$ corrections due to operator $O_{8g}$. We
summarize the results for these terms in the NDR and HV
schemes. For $\Delta c_4^{(f)}$ we have
 \begin{align}
 \label{NDRHVcoefficients}
  \Delta c_4^{(1c)}  &= -\lambda_c^{(f)}\,  \frac{C_F C_1\alpha_s(\mu) }{6N_c\pi} 
    \bigg\{ \frac{2 \rho}{\bar u} h_1^c(u,1,\rho)+h_2^c(u,1,\rho)\nn\\
 &\qquad -\frac{4}{3} + S_c
    \bigg\}, \nonumber \\
    \Delta c_4^{(1u)} &=  -\lambda_u^{(f)} \, \frac{C_F C_1\alpha_s(\mu)
    }{6N_c\pi}
\bigg\{ h_2^u(u,1) -\frac{4}{3}  + S_c \bigg\}, \nonumber \\
\Delta c_4^{(1g)} &= -(\lambda_u^{(f)}+\lambda_c^{(f)}) \frac{C_{8g} \bar m_b}{m_b}
    \: \frac{ C_F\alpha_s(\mu)}{2N_c\pi\, \bar u} \,,
\end{align}
which agrees with the computation in Ref.~\cite{BBNS} and the verification in
Ref.~\cite{chay}. In the NDR scheme the constant $S_c^{\rm NDR}=0$ while in the
HV scheme $S_c^{\rm HV}=1$. One of our main result is the corresponding corrections for
$\Delta b_4^{(f)}$,
\begin{widetext}
\begin{align}  \label{NDRHVcoefficientsB}
 \Delta b_4^{(1c)} &= \lambda_c^{(f)} \, \frac{\alpha_s}{4\pi}\, \Bigg\{
  \frac{h_0^c(u,z,\rho) [3 C_2 (2 z \!-\! 1)\!+\! C_1 (7 z \!+\! 1)]  }
    {9\, \bar u z}
  -\frac{ h_1^c(u,z,\rho) \rho \, C_1  
    \left( 16 
     \bar u \!+\! 16 z \!-\! 27 \right)}
   {27\, \bar u^2 z}+\frac{ h_2^c(u,z,\rho)\, C_1\,(1 \!-\! 8 \bar u)}{27\,  \bar u} \nn \\
&\!\!\!\!\!\!\!\!
    +\frac{h_3^c(u,z,\rho)
   \{ 9 C_2 \bar u z \!+\! C_1 [\bar u z (16 z\!-\! 1)
   \!+\! 2 (16 z^2\!-\!25 z\!+\!27 ) \rho ]\} }
   {54\, \bar u^2 (1\!-\!z) z}
 +\frac{ [27 C_2 (2 z \!-\! 1) \!+\! C_1 \{(64 \bar u \!+\! 55\!+\! S_{b1} \bar
    u \!+\! S_{b2}) z \!-\! 18\}]} 
     {162\, \bar u z} \Bigg\}, \nn \\
 \Delta b_4^{(1u)} &= \lambda_u^{(f)} \, \frac{\alpha_s}{4\pi}\, \Bigg\{
 \frac{ h_2^u(u,z)\, C_1\,(1 \!-\! 8 \bar u)}{27\,  \bar u}+ \frac{h_3^u(u,z) 
   \{ 9 C_2  \!+\! C_1 \, (16 z\!-\! 1)
    \} }
   {54\, \bar u (1\!-\!z) } 
\nn\\
  &
 +\frac{[27 C_2 (2 z \!-\! 1) \!+\! C_1 \{(64 \bar u \!+\! 55 \!+\! S_{b1} \bar
    u \!+\! S_{b2}) z \!-\! 18\}]} 
     {162\, \bar u z} 
  \Bigg\} , \nn \\
  \Delta b_4^{(1g)} &= -(\lambda_u^{(f)}+\lambda_c^{(f)})
       \frac{C_{8g} \bar m_b}{m_b} 
  \: \frac{ C_F\alpha_s(\mu)}{2N_c\pi\, \bar u} \,.
\end{align}
\end{widetext}
In the NDR scheme the constants $S_{b1}^{\rm NDR}=S_{b2}^{\rm NDR}=0$
while in the HV scheme $S_{b1}^{\rm HV}=-48$, and $S_{b2}^{\rm HV}=6$.
The contributions from $O_{8g}$, $\Delta c_4^{(1g)}$ and $\Delta
b_4^{(1g)}$, are generated at tree level and so are scheme independent
at this order. In
Eqs.~(\ref{NDRHVcoefficients},\ref{NDRHVcoefficientsB}),
$\rho=m_c^2/m_b^2$ and 
\begin{align}
 &h_0^c(u,z,\rho)  \nonumber \\
&= \frac{\rho}{ \bar u \bar z}\bigg\{
  \text{Li}_2\bigg[\!\frac{2}{1-g(\rho/\bar u z)}\!\bigg]
  +\text{Li}_2\bigg[\!\frac{2}{1+g(\rho/\bar u z)}\!\bigg] \nn \\
  & - \text{Li}_2\bigg[\!\frac{2}{1+g(\rho/\bar u)} \!\bigg]
  - \text{Li}_2\bigg[\!\frac{2}{1-g(\rho/\bar u)} \bigg]\! \bigg\}  \,,
  \nonumber \\
&h_1^c(u,z,\rho) 
  = G(\bar u z,\rho) - \ln\Big(\frac{\mu^2}{m_b^2\rho}\Big) \,, \nonumber \\
&h_2^c(u,z,\rho) 
  = G(\bar u z,\rho) \,, \\
&h_3^c(u,z,\rho) 
  = G(\bar u,\rho)- G(\bar u z,\rho) \,,\nonumber \\
&h_2^u(u,z) 
  = G_0(\bar u z)  \,,\nonumber \\
&h_3^u(u,z)
   = G_0(\bar u)- G_0(\bar u z) \,, \nn
\end{align}
where we have the usual massive and massless fermion loop functions 
\begin{align}
 G(x,\rho)&= \ln\Big( \frac{\mu^2}{m_b^2 \rho }\Big) 
   \minus 2 \theta(4\rho \!-\! x)\, \bar g(\rho/x) 
   \cot^{-1}\!\big[\bar g(\rho/x) \big]
 \nn\\
  & \plus 2  \minus \theta(x \!-\! 4\rho) g(\rho/x) \,\,
   \bigg\{  \ln\bigg[\frac{1+ g(\rho/x) }
      {1-g(\rho/x)}\bigg] - i\pi \bigg\} \,,
    \nonumber \\
 G_0(x) &= 2 + \ln\Big( \frac{\mu^2}{m_b^2 x}\Big) + i\pi \,,
\end{align}
with $g(x)=\sqrt{1-4x}$ and $\bar g(x) =\sqrt{4x-1}$. The $h_i$ functions are
given in terms of loop integrals in Appendix~\ref{appA}. The factors of
$\alpha_s$ in Eqs.~(\ref{NDRHVcoefficients},\ref{NDRHVcoefficientsB}) should be
evaluated at $\mu\simeq m_b$. One can also look at the endpoint power law
behavior of these matching coefficients, for which we find $\Delta
c_4^{(1c),(1u)}\sim 1$, $\Delta c_4^{(1g)}\sim \Delta b_4^{(1g)}\sim 1/\bar u$,
and $\Delta b_4^{(1c),(1u)}\sim 1/(\bar u z)$.

The SCET Wilson coefficients $c_4^{(f)}$ and $b_4^{(f)}$ should not depend on
the $\gamma_5$-scheme choice for $H_W$.  From the point of view of the
electroweak Hamiltonian, the scheme dependence in
Eqs.~(\ref{NDRHVcoefficients},\ref{NDRHVcoefficientsB}) corresponds to that in
matrix elements, and is compensated by scheme dependence of the electroweak
Hamiltonian's Wilson coefficients. At lowest order, $c_4^{(f)}$ and $b_4^{(f)}$
in Eqs.~(\ref{ci},\ref{bi}) depend on the penguin coefficients $C_3$ and $C_4$.
Since we are calculating $c_4^{(f)}$ and $b_4^{(f)}$ to order $\alpha_s$ we need
to take into account the scheme-dependence of $C_{3,4}$ up to order $\alpha_s$,
which for $\mu\simeq m_b$ is given by
\begin{eqnarray}
\label{NDRHVrel}
C_3^{\rm NDR} &=& C_3^{\rm HV} + \frac{\alpha_s C_1}{36 \pi} +{\mathcal O}(\alpha_s^2) 
   \,,\nn \\
C_4^{\rm NDR} &=& C_4^{\rm HV} - \frac{\alpha_s C_1}{12 \pi}+{\mathcal O}(\alpha_s^2) \,.
\end{eqnarray}
Thus at this order the results in Eqs.~(\ref{ci},\ref{bi}) have to be used in the
same scheme as the $\Delta c_4^{(f)}$ and $\Delta b_4^{(f)}$ corrections. In
$c_4^{(f)}$ and $b_4^{(f)}$ we find that it is only the scheme independent combinations
\begin{align}
  c_4^{(f)} &= - \lambda_t^{(f)} \Big[ C_4 + \frac{C_3}{3} - \frac{2\alpha_s
    C_1  S_c}{27\pi}  \Big] + \ldots \,,
   \\
  b_4^{(f)} &=  - \lambda_t^{(f)} \Big[ C_4 + \frac{1\! +\!\bar u}{3\bar u} C_3
    + \frac{\alpha_s C_1  (S_{b1}\bar u+S_{b2})}{648\pi\bar u} \Big]+\ldots\nn
\end{align}
that occur. This demonstrates that our final results are independent of whether
we use the NDR or HV scheme.


\section{Contributions of the Charm Loop (NDR scheme)} \label{sect:NDR}

\begin{figure}[t!]
  \centerline{
    \mbox{\epsfysize=2.truecm \hbox{\epsfbox{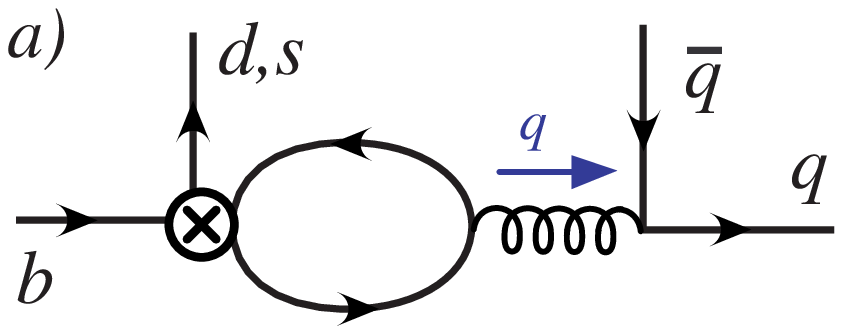}} } \quad
    \mbox{\epsfysize=2.truecm \hbox{\epsfbox{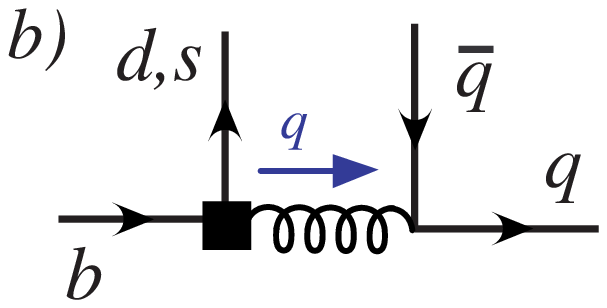}} } } \centerline{
    \mbox{\epsfysize=2.truecm \hbox{\epsfbox{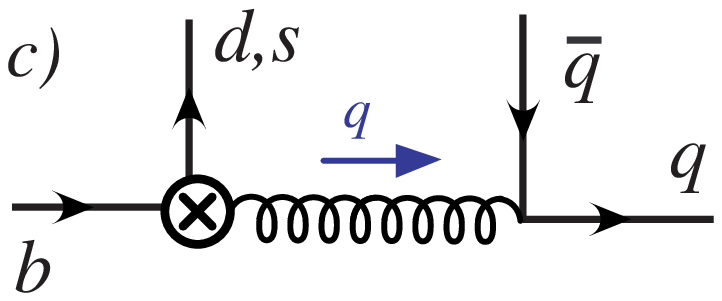}} } } \vskip-0.3cm
  {\caption[1]{ Full-theory graphs for the matching onto the short-distance
      penguin coefficient $c_4$ at ${\cal O}(\alpha_s)$. a) loop graph with
      charm quarks and $C_{1,2}^c$ or up-quarks and $C_{1,2}^u$, b) counterterm
      graph with $O_{DG}$, and c) graph with $C_{8g}$.}
\label{figQCDc} }
\vskip -0.2cm
\end{figure}

\begin{figure}[t!]
   \raisebox{-1cm}{\epsfysize=2.2truecm \hbox{\epsfbox{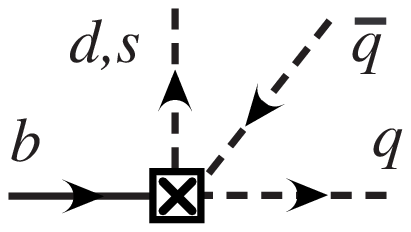}} } 
  \hspace{-0.4cm}
 {\Large \quad $=$  $\frac{2 G_F}{\sqrt{2}}\: c_4^{(d,s)}\: F$ } 
 \vskip-0.3cm {\caption[1]{SCET Feynman rule for the $Q_{4d}^{(0)}$ operator
     where the spinors are $F=[\bar u_{\bn}^{(d)}\bnslash P_L
     u_v^{(b)}][\bar u_n^{(q)} \nslash P_L v_{\bn}^{(q)}]$.}
\label{figSCETc} }
\vskip -0.2cm
\end{figure}

In this section we present the matching calculation in the NDR scheme, where we
have an anticommuting $\gamma_5$ in $d=4-2\epsilon$ dimensions. For $\Delta
c_4^{(f)}$ we calculate the full-theory graphs in Fig.~\ref{figQCDc}a,b,c and
match them onto the SCET graph in Fig.~\ref{figSCETc}.  In order to ensure that
the NDR scheme is consistent, it is important to avoid computing traces from a
closed fermion loop, ${\rm tr}[
\gamma^\mu\gamma^\nu\gamma^\alpha\gamma^\beta\gamma_5]$ (see
Ref.~\cite{Buras:1998ra} for a review). In the basis shown in
Eq.~(\ref{fullops}) the charm fields are not in the same bilinear, so
Fig.~\ref{figQCDc}a does not involve a trace.  Since we are treating $m_c\sim
m_b$, there are no corresponding loop graphs in the effective theory.  Possible
loop corrections stemming from NRQCD loops vanish at leading order in the power
counting, as discussed later in section~\ref{sect:long} on long-distance charm
contributions.

To renormalize Fig.~\ref{figQCDc}a we use the operator
\begin{equation}
O_{DG} =  \bar d [D_\mu,g\,G^{\mu\nu}]\gamma_\nu (1-\gamma^5) b \,,
\end{equation}
which appears in the electro-weak Hamiltonian as
\begin{equation}
\label{Hct}
H_{c.t.}=\frac{G_F}{\sqrt{2}} \sum_{p=u,c} \lambda_p^{(d)} C_{DG}O_{DG}\, .
\end{equation}
In the standard basis for $H_W$ given in Eq.~(\ref{fullops}) the operator
$O_{DG}$ is redundant and has been removed using the gluon equation of motion
(corresponding to an onshell basis of operators). For our computation we keep
$O_{DG}$ with a pure counterterm coefficient, $\delta C_{DG}$, and use it for
renormalization.  This has the advantage that the counterterm graphs maintain
their topological correspondence with the divergent loops (when the divergent
loop is shrunk to a point).  Furthermore, it allows us to obtain the desired
matching results while avoiding the use of $d$-dimensional Fierz relations with
evanescent operators.  At the end of the computation we remove $ O_{DG}$
following Ref.~\cite{Chetyrkin:1997gb}, by writing it in terms of four-quark
operators and an operator that vanishes by the equations of motion, $[D_\mu,
G^{\mu\nu}] = - g T^a \sum_q \bar q \gamma^\nu T^a q$, and transforming the
two-loop anomalous dimension back to that for the standard basis. As shown in
Ref.~\cite{Chetyrkin:1997gb} this gives the usual two-loop anomalous
dimension in the NDR scheme~\cite{fullWilson}. Thus our NDR scheme coefficients
are the standard ones.

The graph in Fig.~\ref{figQCDc}b involves an insertion of the operator $O_{DG}$
with counterterm coefficient $\delta C_{DG}$, where
\begin{equation*}
C^{\text{bare}}_{DG}\, O_{DG}[\psi^0,A^0] = C_{DG}\, O_{DG}^{\text{ren}}+\delta C_{DG}\, O_{DG}^{\text{ren}}.
\end{equation*}
Thus $\delta C_{DG}$ corresponds to a combination of a counterterm for composite
operator renormalization, and wavefunction renormalization, which for our
purposes are not required separately. The choice $\delta C_{DG}= -4 C_1/(3(4
\pi)^2 \epsilon)$ cancels the $1/\epsilon$ divergence in
Fig.~\ref{figQCDc}a.  The same value for $\delta C_{DG}$ will be used to
renormalize the $H_W$ graphs needed for the matching computation for $\Delta
b_4^{(f)}$ below.

At this order in $\alpha_s$ we only have the tree level graph shown in
Fig.~\ref{figSCETc} on the \SCETa side.  Matching Fig.~\ref{figQCDc} and
Fig.~\ref{figSCETc} gives the following contribution to the SCET coefficient
$c_4^{(f)}$ in the notation of Eq.~(\ref{cbsplit}):
\begin{align}
 \label{Deltac4NDR}
 \Delta c_4^{(1c)} \! &= -\lambda_c^{(f)}
 \frac{C_F\alpha_sC_1 }{6\pi N_c} 
    \bigg\{\!\! \frac{2 m_c^2}{q^2} [I_0(q^2)\!-\! I_0]\!+\!I_0(q^2)\! -\! \frac{1}{\epsilon}\! -\!\frac{4}{3}
    \!\bigg \} \nn \\
    \Delta c_4^{(1u)} \! &= 
    -\lambda_u^{(f)}  \frac{C_F\alpha_s C_1 }{6\pi N_c} 
\bigg\{ I_0^{(u)}(q^2)-\frac{1}{\epsilon} -\frac{4}{3} \bigg\}
  , \nn\\
\Delta c_4^{(1g)} \! &= -(\lambda_u^{(f)}+\lambda_c^{(f)}) \frac{C_{8g}\bar m_b}{m_b} \left (\frac{ 2\alpha_s}{9\pi \bar u}\right ),
\end{align}
where definitions for the loop integrals $I_0(q^2)$ etc. are given in the
appendix. The explicit $1/\epsilon$ comes from the counterterm graph and cancels
the divergence in $I_0(q^2)$. In terms of momentum fractions
Eq.~(\ref{Deltac4NDR}) yields the NDR result given above in
Eq.~(\ref{NDRHVcoefficients}).

\begin{figure*}[t!]
  \centerline{
   \mbox{\epsfysize=2.5truecm \hbox{\epsfbox{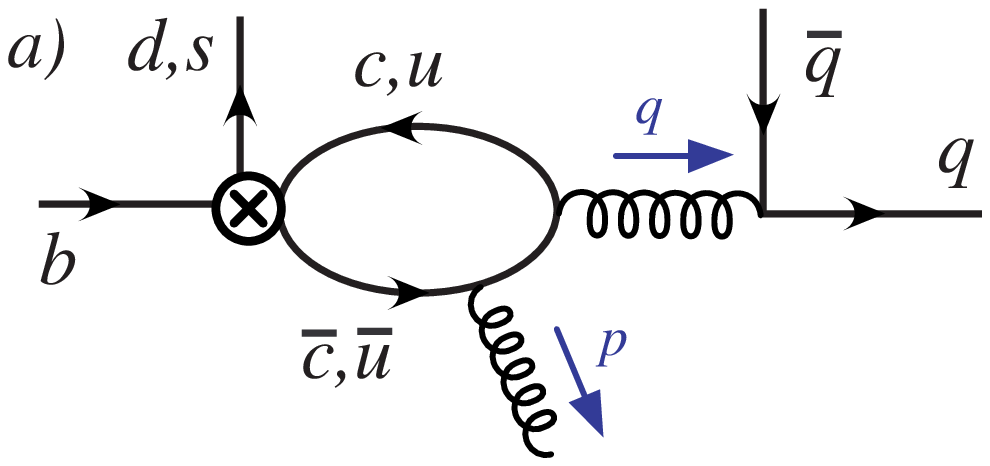}} } \qquad
   \mbox{\epsfysize=2.4truecm \hbox{\epsfbox{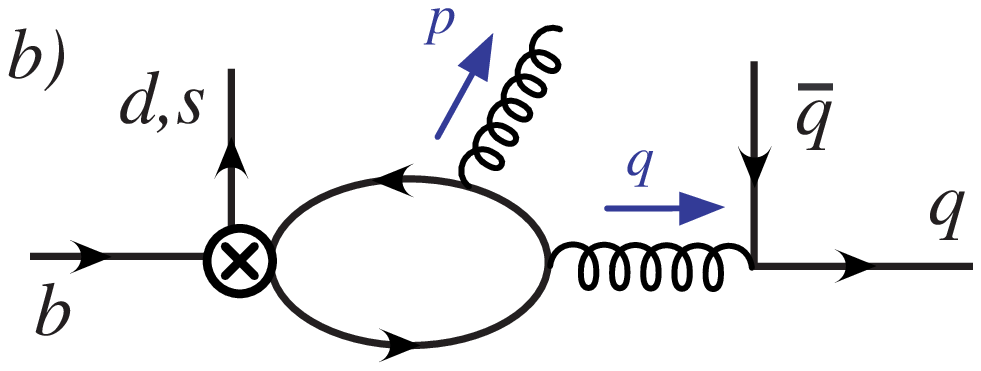}} } \qquad
   \mbox{\epsfysize=2.4truecm \hbox{\epsfbox{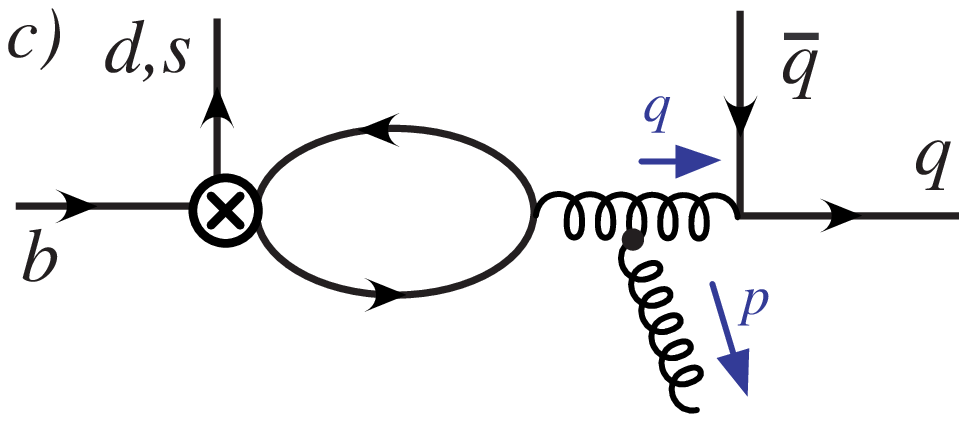}} }
  }
  \centerline{
   \mbox{\epsfysize=2.3truecm \hbox{\epsfbox{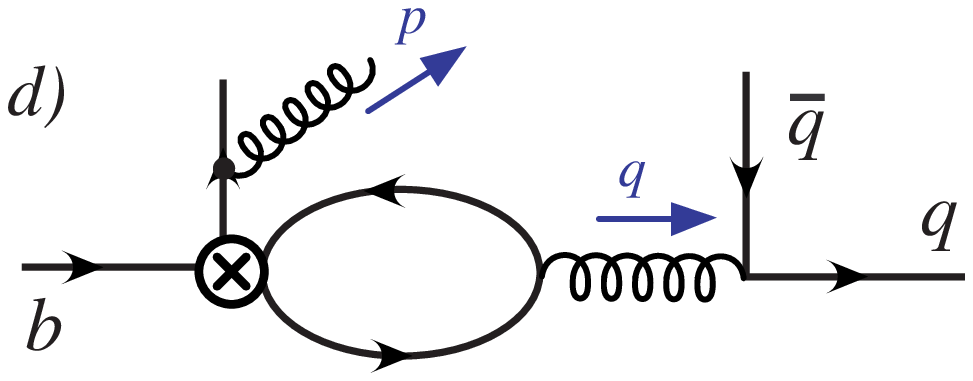}} } \qquad
     \mbox{\epsfysize=2.3truecm \hbox{\epsfbox{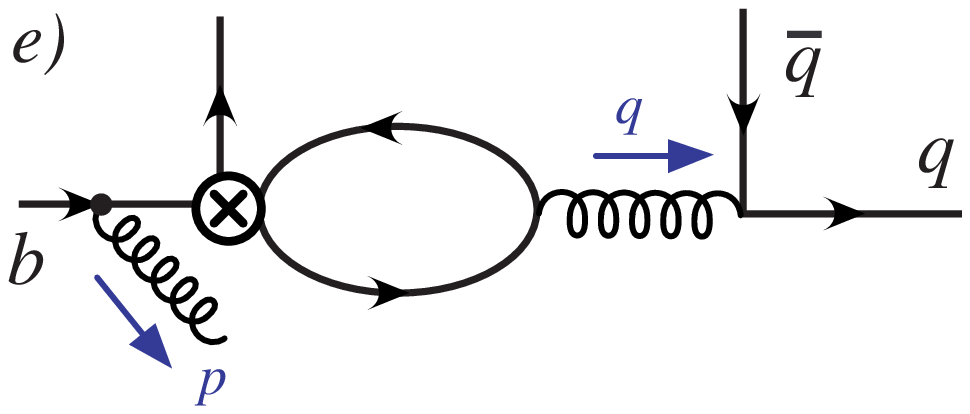}} } \quad
     \mbox{\epsfysize=2.2truecm \hbox{\epsfbox{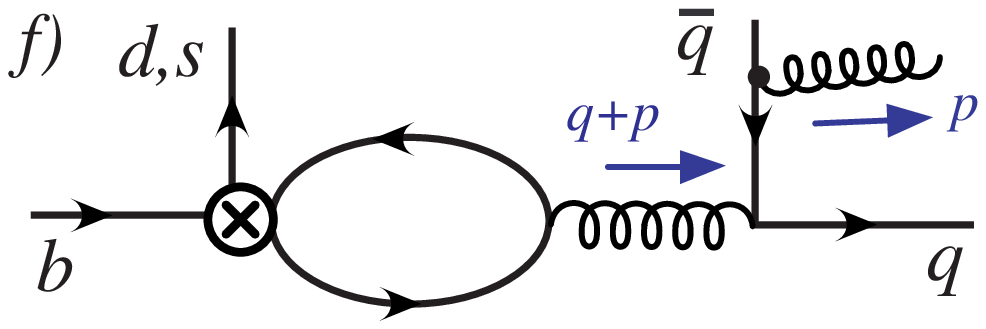}} }
   }
  \centerline{
   \mbox{\epsfysize=2.3truecm \hbox{\epsfbox{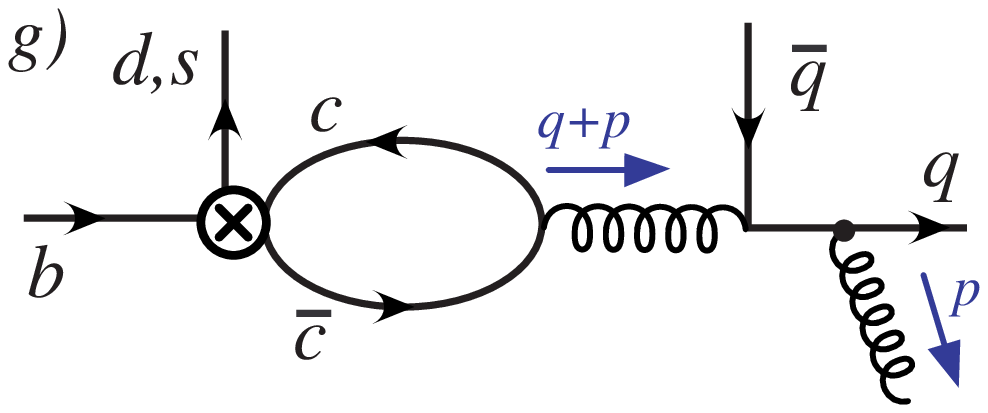}} } \qquad
     \mbox{\epsfysize=2.3truecm \hbox{\epsfbox{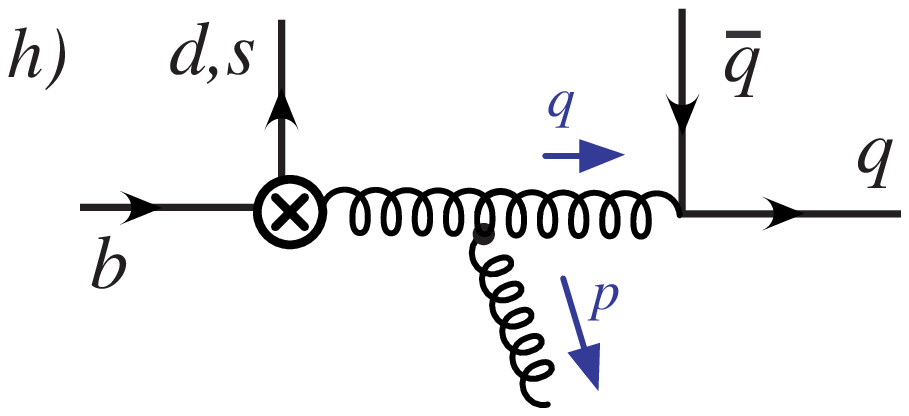}} } \quad
    \mbox{\epsfysize=2.3truecm \hbox{\epsfbox{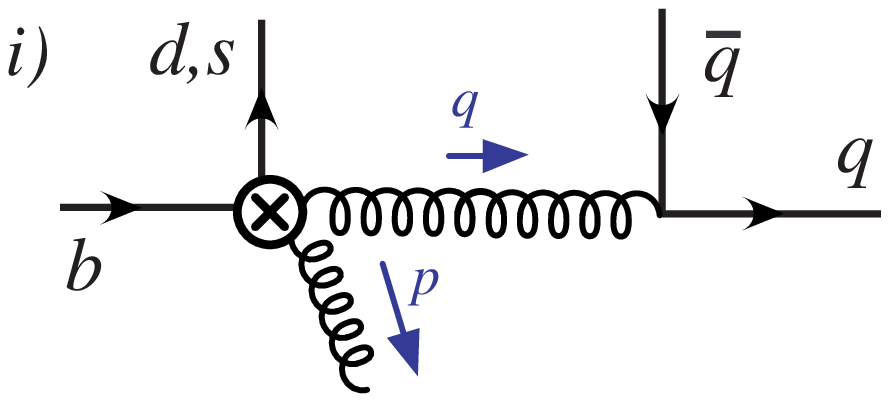}} }
    \hspace{0.3cm}\phantom{x}
    }
    \centerline{
   \mbox{\epsfysize=2.2truecm \hbox{\epsfbox{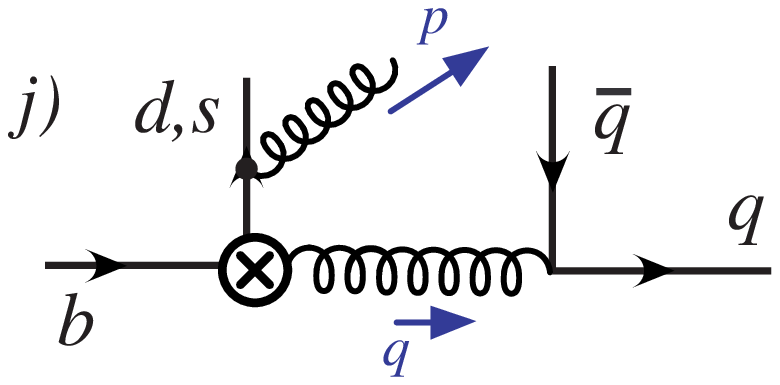}} } \quad
     \mbox{\epsfysize=2.2truecm \hbox{\epsfbox{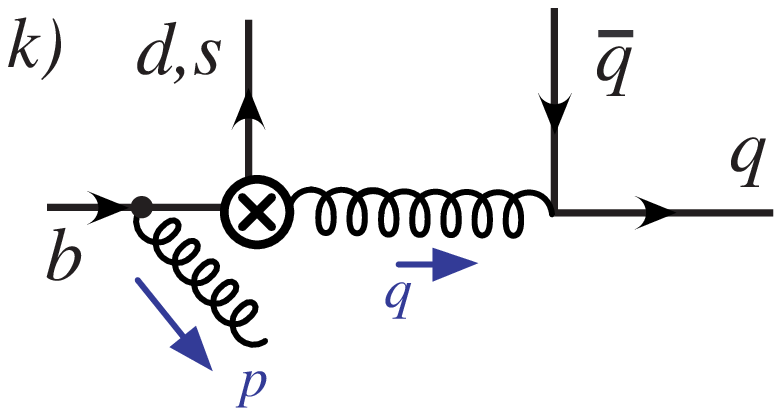}} } \quad
     \mbox{\epsfysize=2.truecm \hbox{\epsfbox{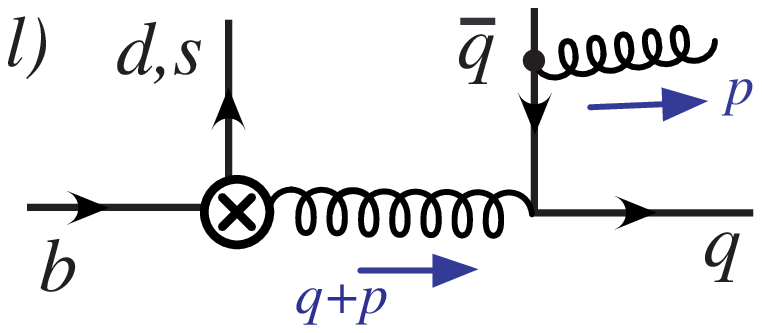}} }
   }
     \centerline{
      \mbox{\epsfysize=2.2truecm \hbox{\epsfbox{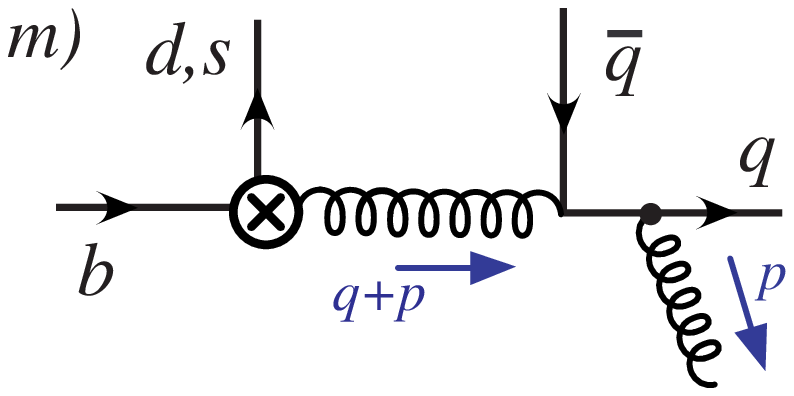}} } \quad
     \mbox{\epsfysize=2.2truecm \hbox{\epsfbox{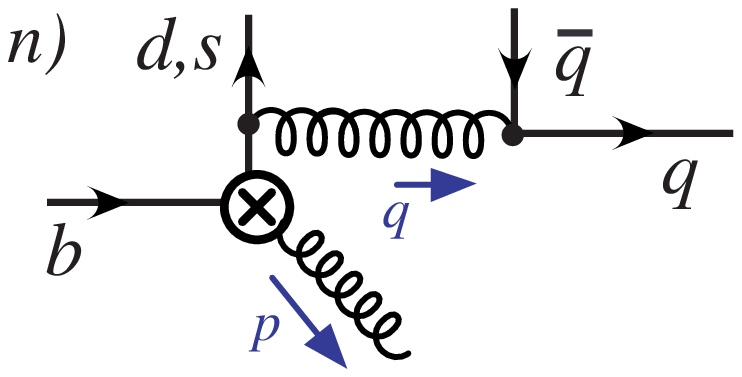}} } \quad
     \mbox{\epsfysize=2.2truecm \hbox{\epsfbox{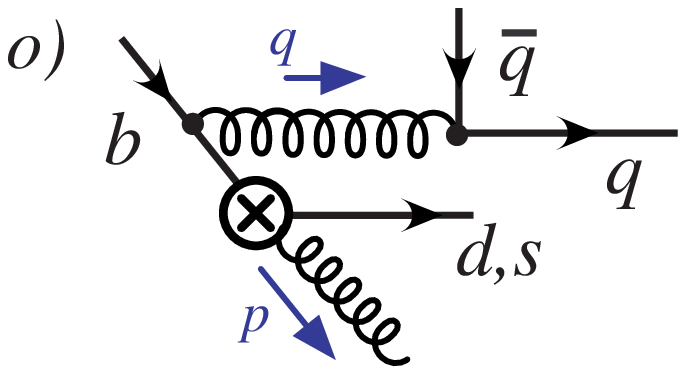}} }
   }
  \vskip-0.3cm
  \caption{Graphs for the matching onto the penguin coefficient $b_4$ at ${\cal
      O}(\alpha_s)$. Not drawn are i) graphs with no gluon attached to the quark
    loop, which vanish in the NDR and HV schemes due to the chirality, and ii)
    graphs with only the gluon of momentum $p$ radiated from the quark loop
    which also vanish.  Here the
    momentum fraction of the gluon is $\bar z$ with $\bn\cdot p=m_b \bar z$, the
    $q$-quark has fraction $z$, with $\bn\cdot q =m_b z$, the $\bar q$-quark has
    momentum fraction $\bar u$, so $n\cdot q =m_b\bar u$, and the $d$ or $s$
    quark has fraction $u$.}
   \label{figQCDb}
\vskip -0.2cm
\end{figure*}

\begin{figure*}[t!]
  \centerline{
   \mbox{\epsfysize=2.5truecm \hbox{\epsfbox{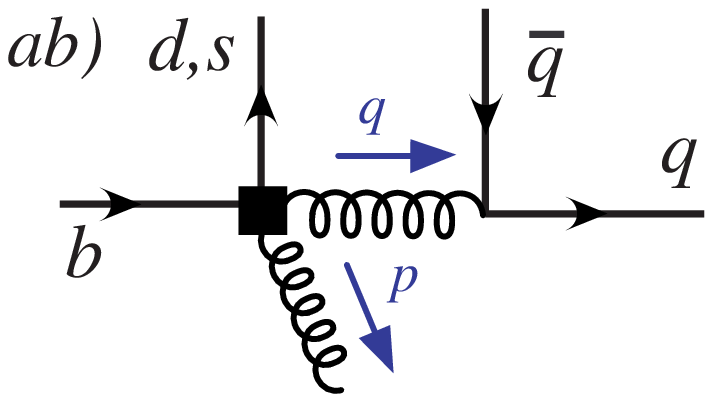}} } \qquad
   \mbox{\epsfysize=2.5truecm \hbox{\epsfbox{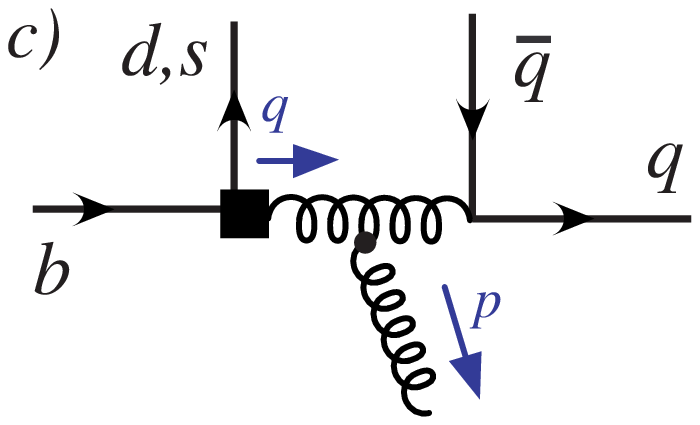}} } \qquad  
   \mbox{\epsfysize=2.5truecm \hbox{\epsfbox{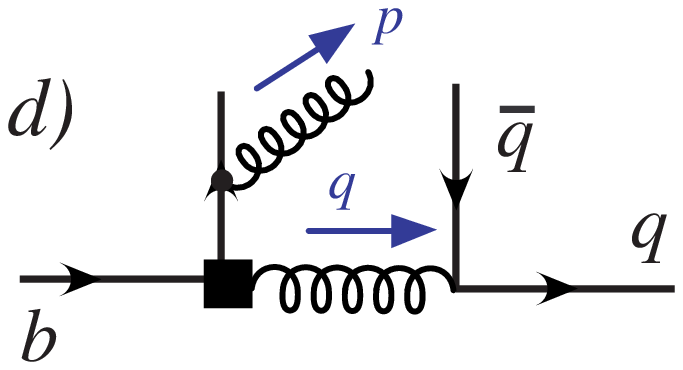}} }  
  }
  \centerline{
    \mbox{\epsfysize=2.3truecm \hbox{\epsfbox{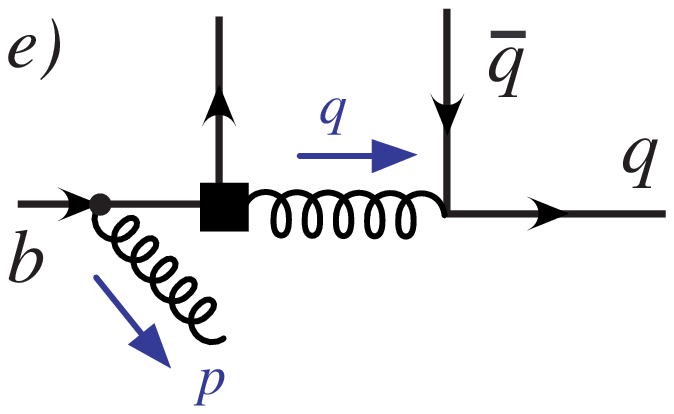}} } \qquad
    \mbox{\epsfysize=2.3truecm \hbox{\epsfbox{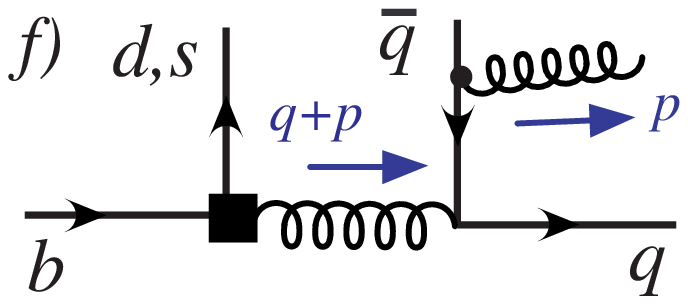}} } \qquad
    \mbox{\epsfysize=2.3truecm \hbox{\epsfbox{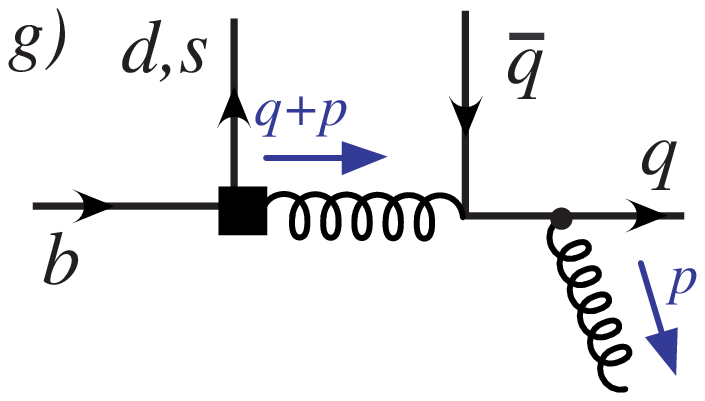}} } } \vskip-0.3cm
  {\caption[1]{Counterterm graphs for Fig.~\ref{figQCDb} involving the operator
      $O_{DG}$. The graph labeled ab) is the counterterm for the sum of graphs
      a) and b) in Fig.~\ref{figQCDb}.}
\label{figct} }
\vskip -0.2cm
\end{figure*}

\begin{figure}[b!]
   \raisebox{-1cm}{\epsfysize=3truecm \hbox{\epsfbox{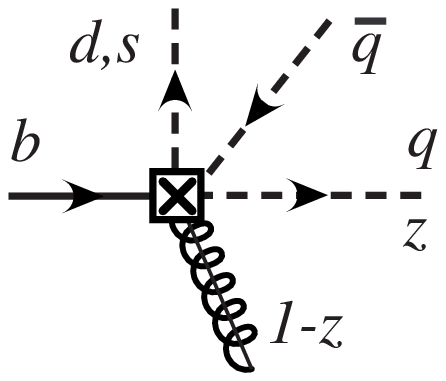}} } 
  \hspace{-.5cm}
  {\Large $=$ 
   $\frac{-4g G_F \varepsilon_{\lambda}^a }{\sqrt{2}m_b}\:  
    b_4^{(d,s)}\: P^{\lambda a} $ }
  \vskip-0.3cm
  {\caption[1]{SCET Feynman rule for the $Q_{4d}^{(1)}$ operator where the
      spinor product is $P^{\lambda a}=[\bar u_{\bn}^{(d)} T^a \gamma_\perp^\mu P_L
      u_v^{(b)}][\bar u_n^{(q)} \nslash P_L v_{\bn}^{(q)}]$.}
\label{figSCETb} }
\vskip -0.2cm
\end{figure}

Next consider the computation of $\Delta b_4^{(f)}$ which comes from matching
the full theory loop graphs in Fig.~\ref{figQCDb} and counterterm diagrams with
$O_{DG}$ shown in Fig.~\ref{figct}, onto the tree level SCET graph in
Fig.~\ref{figSCETb}.  We refer to the graphs in Fig.~(\ref{figQCDb}) as $G_a$,
$G_b$ etc. and those in Fig.~(\ref{figct}) as $\delta G_{ab}$, $\delta G_c$ etc.
%

\begin{widetext}
The results for the graphs with quark-loops, Fig.~\ref{figQCDb}a-f, are
\begin{eqnarray}
\label{NDRgraphs}
G_a &=&  \frac{ig^3}{q^2} \frac{i}{(4 \pi)^2}
\big[\bar{q}_{n}\gamma_{\perp \mu}T^{b}q_{\bar{n}}\big] \left [
  \bar{d}_{\bar{n}}\nslash \left ( \frac{\bar{n}\mcdot f_1}{2}\gamma_\perp^{\lambda}
    \gamma_\perp^\mu +\frac{\bar{n}\mcdot f_2}{2} \gamma_\perp^\mu
    \gamma_\perp^\lambda \right ) \left ( C_1 T^b T^a + \frac{C_2}{2}
    \delta^{ab} \right ) P_{L}b_{v} \right ]  
   , \\
G_b &=&  \frac{-ig^3}{q^2}  \frac{i}{(4 \pi)^2}
\big[\bar{q}_{n}\gamma_{\perp \mu}T^{b}q_{\bar{n}}\big] \left [
  \bar{d}_{\bar{n}}\nslash \left ( \frac{\bar{n}\mcdot
      f_2}{2}\gamma_\perp^{\lambda} \gamma_\perp^\mu +\frac{\bar{n}\mcdot
      f_1}{2} \gamma_\perp^\mu \gamma_\perp^\lambda \right ) \left ( C_1 T^a T^b
    + \frac{C_2}{2} \delta^{ab} \right ) P_{L}b_{v} \right ],
  \nonumber \\
\delta G_{ab} &=& \frac{2iC_1g^3 f^{abc}}{3(4\pi)^2 q^2 \epsilon}
\big[\bar{q}_{n}\gamma_{\perp}^\lambda T^{b}q_{\bar{n}}\big] \big[\bar d_{\bar
  n}(\qslash-\pslash) P_L T^c b_v\big ]
 ,\nonumber \\
G_c+\delta G_c&=&
\frac{-2g^{3}(2C_1)f^{abc}}{3q^2(p+q)^2}\big[\bar{d}_{\bar{n}}\gamma_{\alpha}P_{L}T^{c}b_{v}\big]\!
  \big[\bar{q}_{n}\gamma_{\perp}^{\lambda}T^{b}q_{\bar{n}}\big] \!\frac{i}{(4 \pi)^2} \!\!
  \left \{\!\! \frac{2m_c^2}{(p+q)^2}( I_0((p\!+\!q)^2)\!-\!I_0)\!+\! I_0((p\!+\!q)^2)\!-\!\frac{1}{\epsilon}\! -\!\frac{4}{3} \!\right \} \nonumber \\
  &&\times \left\{ \!q^2 p^\alpha\!+\!p\!\cdot\! q(p^{\alpha}\!\!-\!q^{\alpha}\!) \!\right \} 
 , \nonumber \\
G_d+\delta G_d&=&\frac{ig^3(2C_1)}{6 m_b u} \big[\bar{d}_{\bar{n}}\nslash \gamma_\perp^\lambda
     \gamma_\perp^\mu  T^a T^b P_Lb_v\big ] 
   \big [ \bar q_n \gamma_{\perp\mu}T^bq_{\bar n}\big ] \frac{i}{(4 \pi)^2} 
   \left \{ \frac{2 m_c^2}{q^2} (I_0(q^2)-I_0)+I_0(q^2)-\frac{1}{\epsilon}
  -\frac{4}{3}\right \}
  , \nonumber \\
G_e+\delta G_e&=&\frac{ig^3(2C_1)}{6 m_b} \big [\bar{d}_{\bar{n}}\nslash \gamma_\perp^\mu
      \gamma_\perp^\lambda T^b T^a P_Lb_v\big ] 
   \big [ \bar q_n \gamma_{\perp\mu}T^bq_{\bar n}\big ] \frac{i}{(4 \pi)^2} 
   \left \{ \frac{2 m_c^2}{q^2} (I_0(q^2)-I_0)+I_0(q^2)-\frac{1}{\epsilon}
   -\frac{4}{3} \right \}
   , \nonumber \\
G_f+\delta G_f&=&\frac{ig^3(2C_1)}{6 m_b(1\!-\!u)} \big [\bar{d}_{\bar{n}}\nslash T^b P_L b_v\big ]
    \big [ \bar q_n \gamma_{\perp}^\lambda T^b T^a q_{\bar n}\big ] \frac{i}{(4 \pi)^2} 
    \left \{ \frac{2 m_c^2}{(p+q)^2} (I_0((p\!+\!q)^2)-I_0)+I_0((p\!+\!q)^2)-\frac{1}{\epsilon}
   -\frac{4}{3} \right \}
  , \nonumber \\
G_g+\delta G_g&=&\frac{ig^3(2C_1)}{6 m_b(1\!-\!u)} \big [\bar{d}_{\bar{n}}\nslash T^b P_Lb_v\big ] 
    \big [ \bar q_n \gamma_{\perp}^\lambda T^a T^b q_{\bar n}\big ] \frac{i}{(4 \pi)^2} 
     \left \{ \frac{2 m_c^2}{(p+q)^2} (I_0((p\!+\!q)^2)-I_0)+I_0((p\!+\!q)^2)-\frac{1}{\epsilon}
    -\frac{4}{3} \right \} . \nn
\end{eqnarray}
In $G_a$ and $G_b$ we have objects $f_1^\alpha$ and $f_2^\alpha$ which are
defined by
\begin{align}
\label{f1f2}
f_1^\alpha(p,q,m_c) &=  \Bigg \{  2 m_c^2  \left ( q^\alpha \!-\! p^\alpha
  \!-\! \frac{q^2 p^\alpha}{2 p\mcdot q}   \right )  J_0
+ \left (    \frac{q^2 q^\alpha }{3p\mcdot q}  \!+\!
  \frac{2 m_c^2 q^\alpha }{3p\mcdot q} \!+\! \frac{q^2 p^\alpha}{2p\mcdot q}
  \!-\! \frac{2m_{c}^{2}q^2 p^\alpha }{3(p\mcdot q)^{2}} \!-\! \frac{q^{4}
    p^\alpha}{12(p\mcdot q)^{2}}  \right )  
   \big[ I_0\big((p\!+\!q)^2\big)-I_0(q^2)\big] 
  \nonumber \\ 
&- \frac{2 m^2 p^\alpha}{3 p\mcdot q} \big[I_0\big((p\!+\!q)^2\big)-I_0\big]
   - \frac{(q^\alpha \!-\! p^\alpha )}{3} I_0\big((p\!+\!q)^2\big) 
  +  \left ( \frac{10}{9}(q^\alpha-p^\alpha)-\frac{q^2 p^\alpha}{6 p\mcdot q}  \right ) \Bigg \}
  \,, \nonumber \\ 
f_2^\alpha(p,q,m_c) &= \Bigg \{ - m_c^2  \left (  \frac{q^2 p^\alpha}{
    p\mcdot q} \right )  J_0
+  \left (   - \frac{q^2 q^\alpha }{6p\mcdot q}  +
  \frac{2 m_c^2 q^\alpha }{3p\mcdot q} -\frac{2m_{c}^{2}q^2 p^\alpha }{3(p\mcdot
    q)^{2}} -\frac{q^{4} p^\alpha}{12(p\mcdot q)^{2}}  \right ) 
   \big[I_0\big((p\!+\!q)^2\big)-I_0(q^2)\big]  \nonumber \\ 
& -  \frac{2 m^2 p^\alpha}{3 p\mcdot q} \big[ I_0\big((p\!+\!q)^2\big)-I_0\big]
   - \frac{(q^\alpha - p^\alpha )}{3} I_0\big((p\!+\!q)^2\big)  +  \left ( 
\frac{1}{9}(q^\alpha-p^\alpha)-\frac{q^2 p^\alpha}{6 p\mcdot q}  \right ) \Bigg
\} \,.
\end{align}
\end{widetext}
\noindent Results for the loop integrals $I_0$ and $J_0$ are given in
appendix~\ref{appA}. Note that the only contributions from the $\delta G$-counterterm
graphs are explicit $1/\epsilon$'s, which exactly cancel divergences due to the
loop integrals. We have made some simplifications in the expressions for the
last four graphs, where $u$ is the momentum fraction of the d-quark in $M_1$ and
$z$ is the momentum fraction of the quark in $M_2$. The result $G_g + \delta
G_g$ corresponds to the contribution of the expansion of the external full QCD
$q$-quark field onto the $n$-collinear quark field in SCET. For $\Delta S=1$
decays with $f=s$ one makes the replacement $\bar d_{\bar n} \to \bar s_{\bar n}$
in Eq.~(\ref{NDRgraphs}).

Next we sum the graphs in Eq.~(\ref{NDRgraphs}) and Fierz them to match
onto the SCET operator $Q_{4f}^{(1)}$, and thus obtain $\Delta b_4^{(f)}$. The
results for $G_a+G_b+\delta G_{ab}$, $G_c+\delta G_c$, etc. are finite as
$\epsilon\to 0$, so we can Fierz them in 4-dimensions, and then read off the
prefactor of the spinors $P^{\lambda a}$ shown in Fig.~\ref{figSCETb} to obtain
$\Delta b_4^{(f)}$.  A few useful Fierz formulas are listed in appendix~\ref{appA}.
The definitions for all momentum fractions are summarized in the caption of
Fig.~\ref{figQCDb}.  The result in the NDR scheme is
\begin{widetext}
\begin{align}
\label{charmloopB}
 \Delta b_4^{(1c)} &= \lambda_c^{(f)} \, \frac{\alpha_s}{4\pi}\, \Bigg\{
  \frac{m_c^2 J_0 [3 C_2 (2 z \!-\! 1)\!+\! C_1 (7 z \!+\! 1)]  }
    {9\, \bar u z}
  -\frac{ (I_0(q^2)-I_0) \rho \, C_1  
    \left( 16 
     \bar u \!+\! 16 z \!-\! 27 \right)}
   {27\, \bar u^2 z} \nn \\
& +\frac{ (I_0(q^2)-\frac{1}{\epsilon})\, C_1\,(1 \!-\! 8 \bar u)}{27\,  \bar u}
   +\frac{(I_0((p+q)^2)-I_0(q^2))
   \{ 9 C_2 \bar u z \!+\! C_1 [\bar u z (16 z\!-\! 1)
   \!+\! 2 (16 z^2\!-\!25 z\!+\!27 ) \rho ]\} }
   {54\, \bar u^2 (1\!-\!z) z} \nn \\
 &+\frac{ [27 C_2 (2 z \!-\! 1) \!+\! C_1 \{(64 \bar u \!+\! 55) z \!-\! 18\}]} 
     {162\, \bar u z} \Bigg\} .
\end{align}
The result for the up-quark penguin loops is simply obtained by taking $m_c\to 0$ which gives
\begin{align}
\label{uploopB}
 \Delta b_4^{(1u)} &= \lambda_u^{(d)} \, \frac{\alpha_s}{4\pi}\, \Bigg\{
 \frac{ (I_0^{(u)}(q^2)-\frac{1}{\epsilon})\, C_1\,(1 \!-\! 8 \bar u)}{27\,  \bar u}+ \frac{(I_0^{(u)}((p+q)^2)-I_0^{(u)}(q^2))
   \{ 9 C_2  \!+\! C_1 \, (16 z\!-\! 1)
    \} }
   {54\, \bar u (1\!-\!z) }
  \nn\\
  &+\frac{[27 C_2 (2 z \!-\! 1) \!+\! C_1 \{(64 \bar u \!+\! 55) z \!-\! 18\}]} 
     {162\, \bar u z} 
  \Bigg\} .
\end{align}
These results for $\Delta b_4^{(1c)}$ and $\Delta b_4^{(1u)}$ in NDR are
summarized in a more compact notation in Eq.~(\ref{NDRHVcoefficientsB}).

Next we quote the results for contributions of the graphs with the operator
$O_{8g}$, Fig.~\ref{figQCDb}h-m give
\begin{eqnarray} \label{Gops}
 G_{h}&=&\frac{-ig^{3}C_{8g}\bar m_b}{8\pi^{2}m_{b}^2} \left( \frac{\bar{z}}{\bar{u}z}\right)f^{abc}[\bar{d}_{\bar{n}}P_RT^{c}b_{v}][\bar{q}_{n}\gamma_{\perp}^{\lambda}T^{b}q_{\bar{n}}]  \\
G_{i}&=&\frac{-ig^{3}C_{8g}\bar m_b}{16\pi^{2}m_{b}^2} \left( \frac{1}{\bar{u} z}\right)f^{abc}[\bar{d}_{\bar{n}}(\gamma_{\perp}^{\lambda}\gamma_{\perp}^{\mu}-\gamma_{\perp}^{\mu}\gamma_{\perp}^{\lambda})P_RT^{c}b_{v}][\bar{q}_{n}\gamma_{\perp}^{\mu}T^{b}q_{\bar{n}}] \nonumber \\
G_{j}&=&\frac{-g^{3}C_{8g}\bar m_b}{8\pi^{2}m_{b}^2} \left( \frac{1}{uz}\right)[\bar{d}_{\bar{n}}\gamma_{\perp}^{\lambda}\gamma_{\perp}^{\mu}P_RT^{a}T^{b}b_{v}][\bar{q}_{n}\gamma_{\perp\mu}T^{b}q_{\bar{n}}] \nonumber \\
G_{l}&=&\frac{-g^{3}C_{8g}\bar m_b}{8\pi^{2}m_{b}^2} \left( \frac{1}{\bar{u}}\right)[\bar{d}_{\bar{n}}P_R T^{b}b_{v}][\bar{q}_{n}\gamma_{\perp}^{\lambda}T^{b}T^{a}q_{\bar{n}}] \nonumber \\
G_{m}&=&\frac{-g^{3}C_{8g}\bar m_b}{8\pi^{2}m_{b}^2} \left( \frac{1}{\bar{u}}\right)[\bar{d}_{\bar{n}}P_R T^{b}b_{v}][\bar{q}_{n}\gamma_{\perp}^{\lambda}T^{a}T^{b}q_{\bar{n}}] \nonumber \\
G_o &=& \frac{g^{3}C_{8g}\bar m_b}{8\pi^{2}m_{b}^2} \left( \frac{\bar z}{z(z+\bar u \bar z)}\right)[\bar{d}_{\bar{n}}\gamma_\perp^\lambda \gamma_\perp^\mu P_R T^a T^{b}b_{v}][\bar{q}_{n}\gamma_{\perp\mu}T^{b}q_{\bar{n}}] \nonumber \\
G_{k}&=& G_n = 0 \nonumber .
\end{eqnarray}
\end{widetext}
\noindent  Here $\bar m_b$ is the $\overline{\rm MS}$ mass which always
accompanies $C_{8g}$, while $m_b$ is a short-distance threshold mass (for which
we will use the 1S-mass). For $f=s$ we take $\bar d_{\bar n} \to \bar s_{\bar
  n}$.  Fierzing the results in Eq.~(\ref{Gops}) and matching onto
Fig.~\ref{figSCETb} we obtain the contribution to $\Delta b_4^{(d)}$ due to
$O_{8g}$,
\begin{equation}
 \Delta b_4^{(1g)} = -(\lambda_u^{(f)}+\lambda_c^{(f)}) \frac{C_{8g}\bar m_b}{m_b} \left (\frac{ 2\alpha_s}{9\pi \bar u}\right ),
\end{equation}
which turns out to be identical to the matching result for $\Delta c_4^{(1g)}$.


\section{Contributions of the Charm Loop (HV scheme)} \label{sect:HV}

Next we repeat the calculation of the previous section using the HV scheme. In
the HV scheme, $\gamma_5$ anticommutes with Dirac matrices in $4$-dimensions,
and commutes with the Dirac matrices in the remaining $(-2\epsilon)$-dimensions.
Here we can consistently handle traces with $\gamma^5$ in $d \ne 4$ dimensions.
In the HV scheme the Dirac matrices in the $(V-A)$ interactions in the weak
Hamiltonian are taken in four-dimensions, while all $\gamma$-matrices from the
QCD and QED Lagrangians are in $d$-dimensions (see Ref.~\cite{Buras:1998ra} for
a review of the HV scheme). We will perform the computation in two operator
bases, namely the original one $O_{1,2}^p$ in Eq.~(\ref{fullops}) and a
different basis given by
\begin{eqnarray}
\label{tildebasis}
\tilde O_1^p &=& [\bar d b]_{V-A}[\bar p p]_{V-A},\nn\\
 \tilde O_2^p &=& [\bar d T^a b]_{V-A}[\bar p T^a p]_{V-A},
\end{eqnarray}
with Wilson coefficients $\tilde C_1$ and $\tilde C_2$.  Notice that
in addition to the different Fierz order, the $\tilde O_{1,2}^p$ basis
is also color Fierzed with respect to the $O_{1,2}^p$ basis. In the
$\tilde O_{1,2}^p$ basis we need to evaluate fermion traces like ${\rm
tr}[
\gamma^\mu\gamma^\nu\gamma^\alpha\gamma^\beta\gamma_5]$.  As we explain below,
the answer for the matching computation in either of these bases is the same in
the HV-scheme.  Renormalization with the operator $O_{DG}$ goes the same way as
in the NDR scheme except that we replace $C_1= \tilde C_2/2$.

The result for the charm and up-quark loop in the HV scheme is well known, see
e.g. Ref.~\cite{fullWilson}, and with either basis gives the same result for
$\Delta c_4^{(1c)}$ and $\Delta c_4^{(1u)}$ as the NDR scheme but with $S_c^{\rm
  HV}=+1$ in Eq.~(\ref{NDRHVcoefficients}).

For $\Delta b_4$ we start with the computation in the $\tilde O_{1,2}^p$ basis.
The graphs in Figs.~\ref{figQCDb} and \ref{figct} give
\begin{widetext}
\begin{eqnarray}
&& \hspace{-1cm} G_a+G_b +\delta G_{ab} \nn\\
 &=&  \frac{2i\tilde C_1\, g^3}{(4\pi)^2}\epsilon^{\alpha \lambda \mu \rho} 
  \left[ \frac{(p_{\rho}-q_{\rho})}{q^2} \Big( m_{c}^{2}\, J_{0} + \frac{1}{2}\Big)
    -(q_{\rho}+p_{\rho})\frac{1}{4p.q}(I_{0}((p\!+\!q)^2)-I_{0}(q^2))\right] 
   \big[\bar{d}_{\bar{n}}\gamma_{\alpha}P_{L}b_{v}\big]
    \big[\bar{q}_{n}\gamma_{\perp \mu}T^{a}q_{\bar{n}}\big]
    \nonumber \\
  &+& \frac{i\tilde C_2 g^3}{(4\pi)^2}\Bigg \{\! d^{abc}\epsilon^{\alpha \lambda \mu \rho} \!
    \left[ \frac{(p_{\rho}-q_{\rho})}{q^2} \Big( m_{c}^{2}\, J_{0} +
      \frac{1}{2}\Big)
    -(q_{\rho}+p_{\rho})\frac{1}{4p.q}(I_{0}((p\!+\!q)^2)-I_{0}(q^2))\! \right]\!\!
     \big[\bar{d}_{\bar{n}}\gamma_{\alpha}P_{L}T^cb_{v}\big]\!
     \big[\bar{q}_{n}\gamma_{\perp \mu}T^{b}q_{\bar{n}}\big] \nonumber \\
  && \ \ +
   \frac{ f^{abc}}{q^2} \Bigg [ \left(
     \frac{q^2}{4p.q}(p^{\alpha}+q^{\alpha})+ \frac{2m_{c}^{2}}{3p.q}q^{\alpha}
      -\frac{q^2}{6p.q}q^{\alpha} -\frac{2m_{c}^{2}q^2}{3(p.q)^{2}}p^{\alpha}
      -\frac{q^{4}}{12(p.q)^{2}}p^{\alpha} \right)(I_{0}((p\!+\!q)^2)-I_{0}(q^2)) 
    \nonumber \\
   && \ \ +
   \frac{1}{3}(p^{\alpha}-q^{\alpha})\left ( I_{0}((p\!+\!q)^2) -\frac{1}{\epsilon} \right )-\frac{2m_c^2p^{\alpha}}{3p.q}(I_0((p\!+\!q)^2)-I_0)
   -\left(m_{c}^{2}(p^{\alpha}-q^{\alpha})+\frac{m_{c}^{2}q^2}{p.q}p^{\alpha}\right)J_{0} 
     \nonumber \\
   &&\ \ - p^{\alpha}\frac{q^2}{6p.q}-\frac{5}{18}(p^\alpha-q^\alpha) \Bigg] 
     \big[\bar{d}_{\bar{n}}\gamma_{\alpha}P_{L}T^{c}b_{v}\big]
     \big[\bar{q}_{n}\gamma_{\perp}^{\lambda}T^{b}q_{\bar{n}}\big] \Bigg \}
    ,
\end{eqnarray}
\noindent where we use
$\epsilon^{0123}=+1$.  In Figs.~\ref{figQCDb}c-g only $\tilde O_2^p$
contributes, and we have
\begin{eqnarray}
\label{HVgraphs}
G_c+\delta G_c&=&
\frac{-2 \tilde C_2\, g^{3}f^{abc}}{3q^2(p+q)^2}\big[\bar{d}_{\bar{n}}\gamma_{\alpha}P_{L}T^{c}b_{v}\big]
  \big[\bar{q}_{n}\gamma_{\perp}^{\lambda}T^{b}q_{\bar{n}}\big] \frac{i}{(4 \pi)^2}\! 
  \left \{\! \frac{2m_c^2}{(p+q)^2}( I_0((p\!+\!q)^2)\! -\! I_0) \!+\! I_0((p\!+\!q)^2)\!-\!\frac{1}{\epsilon}\! -\!\frac{1}{3}\! \right \}  \\
  && \times \left\{ \!q^2 p^\alpha\!+\!p\!\cdot\! q(p^{\alpha}\!\!-\!q^{\alpha}\!) \!\right \} 
 , \nonumber \\
G_d+\delta G_d&=&\frac{i\tilde C_2 \, g^3}{6 m_b u} \big[\bar{d}_{\bar{n}}\nslash \gamma_\perp^\lambda
     \gamma_\perp^\mu  T^a T^b P_Lb_v\big ] 
   \big [ \bar q_n \gamma_{\perp\mu}T^bq_{\bar n}\big ] \frac{i}{(4 \pi)^2} 
   \left \{ \frac{2 m_c^2}{q^2} (I_0(q^2)-I_0)+I_0(q^2)-\frac{1}{\epsilon}
  -\frac{1}{3}\right \}
  ,\nonumber \\
G_e+\delta G_e&=&\frac{i \tilde C_2\, g^3}{6 m_b} \big [\bar{d}_{\bar{n}}\nslash \gamma_\perp^\mu
      \gamma_\perp^\lambda T^b T^a P_Lb_v\big ] 
   \big [ \bar q_n \gamma_{\perp\mu}T^bq_{\bar n}\big ] \frac{i}{(4 \pi)^2} 
   \left \{ \frac{2 m_c^2}{q^2} (I_0(q^2)-I_0)+I_0(q^2)-\frac{1}{\epsilon}
   -\frac{1}{3} \right \}
   , \nonumber \\
G_f+\delta G_f&=&\frac{i \tilde C_2\, g^3}{6 m_b(1\!-\!u)} \big [\bar{d}_{\bar{n}}\nslash T^b P_Lb_v\big ]
    \big [ \bar q_n \gamma_{\perp}^\lambda T^b T^a q_{\bar n}\big ] \frac{i}{(4 \pi)^2} 
    \left \{ \frac{2 m_c^2}{(p+q)^2} (I_0((p\!+\!q)^2)-I_0)+I_0((p\!+\!q)^2)-\frac{1}{\epsilon}
   -\frac{1}{3} \right \}
   ,\nonumber \\
G_g+\delta G_g&=&\frac{i \tilde C_2\, g^3}{6 m_b(1\!-\!u)} \big [\bar{d}_{\bar{n}}\nslash T^b P_Lb_v\big ] 
    \big [ \bar q_n \gamma_{\perp}^\lambda T^a T^b q_{\bar n}\big ] \frac{i}{(4 \pi)^2} 
     \left \{ \frac{2 m_c^2}{(p+q)^2} (I_0((p\!+\!q)^2)-I_0)+I_0((p\!+\!q)^2)-\frac{1}{\epsilon}
    -\frac{1}{3} \right \} 
  \nn  .
\end{eqnarray}
\end{widetext}
Results for $I_0$ and $J_0$ can be found in appendix~\ref{appA}. Again we are
free to Fierz these finite results in 4-dimensions. Computing $\Delta c_4^{(f)}$
and $\Delta b_4^{(f)}$ from these expressions gives the results summarized in
Eq.~(\ref{NDRHVcoefficientsB}).

Alternatively one can do the HV scheme calculation in the same basis $O_{1,2}^p$
as the NDR scheme calculation. Although there are no fermion loops in this basis
the HV scheme computation does differ from the NDR scheme. For each graph the results differ
due to an extra ${\cal O}(\epsilon)$ term generated in manipulating the Dirac
matrices.  Therefore it is easy to quote the HV scheme results obtained in this
basis, as replacements to be made in the in NDR result.  For $f_1^\alpha$ in
Eq.~(\ref{f1f2}) we should replace
\begin{equation}
 \left ( 
\frac{10}{9}(q^\alpha-p^\alpha)-\frac{q^2 p^\alpha}{6 p\mcdot q}  \right ) \rightarrow  \left ( 
\frac{7}{9}(q^\alpha-p^\alpha)-\frac{q^2 p^\alpha}{6 p\mcdot q}  \right ),
\end{equation}
and for $f_2^\alpha$ the HV scheme result is obtained by replacing
\begin{equation}
 \left ( 
\frac{1}{9}(q^\alpha-p^\alpha)-\frac{q^2 p^\alpha}{6 p\mcdot q}  \right ) \rightarrow  \left ( 
\frac{-2}{9}(q^\alpha-p^\alpha)-\frac{q^2 p^\alpha}{6 p\mcdot q}  \right ).
\end{equation}
For graphs $G_c+\delta G_c$ to $G_g+\delta G_g$ in Eq.~(\ref{NDRgraphs}) and for
$\Delta c_4^{(1p)}$ in Eq.~(\ref{Deltac4NDR}) we replace
\begin{equation}
\label{c4shift}
-\frac{4}{3}  \rightarrow -\frac{1}{3}.
\end{equation}
Finally the HV scheme result for $\Delta b_4^{(1p)}$ in
Eqns.~(\ref{charmloopB}) and (\ref{uploopB}) is obtained by the
replacement
\begin{align}
\label{b4shift}
& \frac{ [27 C_2 (2 z \!-\! 1) \!+\! C_1 \{(64 \bar u \!+\! 55) z \!-\! 18\}]} 
     {162\, \bar u z} \nn \\
 & \rightarrow \frac{ [27 C_2 (2 z \!-\! 1) \!+\! C_1 \{(16 \bar u \!+\! 61) z \!-\! 18\}]} 
     {162\, \bar u z}.
\end{align}
This is same as the result that we obtained from the HV scheme calculation in
the $\tilde O_{1,2}^p$ basis.  As discussed earlier in
section~\ref{sect:summary}, the scheme dependence in $C_3$ and $C_4$ which
appear at tree level accounts for the shifts in $\Delta c_4^{(f)}$ and $\Delta
b_4^{(f)}$ given by Eqs.~(\ref{c4shift}) and (\ref{b4shift}), thus leaving the
SCET Wilson coefficients independent of choice of NDR or HV scheme. The
calculation in the HV scheme in the $\tilde O_{1,2}^p$ basis differs from that
in the NDR scheme, and provides a non-trivial cross-check on our results.


\section{Chiraly enhanced penguins} \label{sect:chiral}

It is well known that certain power corrections have the potential to be
numerically enhanced in penguin amplitudes. In particular the so-called chiraly
enhanced terms~\cite{BBNS}, which are formally down by a factor of
$\Lambda/m_b$, but are numerically of order $\mu_P/m_b$.\footnote{Although
  chiraly enhanced penguin contributions are large, for tree amplitudes they are
  numerically the same size as other expected power corrections as emphasized in
  Ref.~\cite{Bauer:2005wb}.}  For the pion $\mu_\pi(2\,{\rm GeV})=1.7 ~{\rm
  GeV}$, and this can be understood from the fact that $\mu_\pi \propto
\Lambda_\chi$ rather than $\Lambda_{\rm QCD}$, where $\Lambda_\chi$ is the scale
of chiral symmetry breaking. Thus relative to the other power corrections these
terms have the possibility of being magnified by a numerical factor of
$\mu_\pi/\Lambda\sim 3-4$. A valid factorization theorem for the complete set of
chiraly enhanced corrections has not yet been derived, because previous attempts
encountered endpoint singularities~\cite{BBNS}.  In this section we derive a
factorization theorem for chiraly enhanced tree and penguin amplitudes that does
not suffer from endpoint singularities. Our analysis uses factorization in
\SCETa and the complete result involves only one additional generalized form
factor and one light-cone meson distribution beyond those occurring at leading
order.

To consider chiraly enhanced operators in SCET we can work with a
complete basis of operators suppressed by one power of $\Lambda/m_b$,
and then look for all operators with a $\slash\!\!\!\!  {\cal
P}_\perp$ in the light-quark bilinear as explained in
Ref.~\cite{Arnesen:2006vb}. This provides a unique way to determine
the contributions that are chiraly enhanced, without invoking the
Wandzura-Wilczek (WW) approximation~\cite{Wandzura:1977qf} as was done
in Ref.~\cite{BBNS}.

We therefore construct a complete basis of operators with one $P_\perp^\alpha$, starting with the field
structures:
\begin{align} \label{OABB}
 Q^{(1\chi)}_{A} &=  (\bar\xi_n \Gamma_n h_v)
 (\bar\xi_\bn \Gamma_\bn {\cal P}_\perp^\beta \xi_\bn)  \,,\nn\\
  Q_{B1}^{(2\chi)} &=  
  (\bar\xi_n ig {\cal B}_{n\perp}^\alpha \Gamma_n h_v)(\bar\xi_\bn
\Gamma_\bn  {\cal P}_\perp^\beta \xi_\bn) \, ,\nn\\
  Q_{B2}^{(2\chi)} &=  
  (\bar\xi_n {\cal P}_\perp^{\dagger\beta}  ig {\cal B}_{n\perp}^\alpha \Gamma_n h_v)(\bar\xi_\bn
\Gamma_\bn   \xi_\bn) 
  \,.
\end{align}
Only the color structures shown are required at this order.  Operators
with a $i g{\cal B}_{n\perp}^\alpha$ are needed at the same order as
operators without, because of the additional suppression of the
non-${\cal B}_{n\perp}^\alpha$ operators in the matrix element of the
required time-ordered products. This is the same situation which we
described already at leading order in $\Lambda/m_b$ in
Eq.~(\ref{Tproducts}).
Note that in Eq.~(\ref{OABB}) we do not consider other operators with ${\cal P}_\perp$ or $\partial_\perp$ since they are not chirally enhanced.
To perform
the matching we work with a basis of four-quark operators of definite
chirality, where the possibilities are inherited from the full
electroweak Hamiltonian: $(LH)(LL)$, $(LH)(RR)$, or $(RH)(LR)$. Here
the order corresponds to the quark fields in Eq.~(\ref{OABB}) and we
do not assign a chirality to the heavy quark denoted by $H$. With
definite chirality a complete basis of Dirac structures includes
\begin{align}
  \Gamma_n \in \{ 1,\gamma_\perp^\mu \} \,,\qquad 
  \Gamma_\bn \in \{ \nslash, \nslash \gamma_\perp^\nu \},
\end{align}
where $\Gamma_\bn =\nslash$ contributes only to $(LH)(LL)$ and
$(LH)(RR)$, while $\Gamma_\bn=\nslash \gamma_\perp^\nu$ contributes
only to $(RH)(LR)$.

First let's construct a complete basis of the $Q_{A}^{(1\chi)}$-type operators in
Eq.~(\ref{OABB}). Here $\Gamma_n\otimes \Gamma_\bn$ must have a $\perp$
$\beta$ index, and we find the basis
\begin{eqnarray} \label{Q1chi}
Q_{1(qfq)}^{(1\chi)}&=&\frac{1}{m_b} \big[\bar q_{n\omega_1}^R
b_v\big]\big[ \bar f_{\bn\omega_2}^L \nslash \SppP  q_{\bn\omega_3}^R \big] \,, \nn \\
Q_{2(qfq)}^{(1\chi)} &=& Q_{3(qfq)}^{(1\chi)} \frac{3}{2}e_q \,,\nn\\
 Q_{3(qfq)}^{(1\chi)}&=&\frac{1}{m_b} \big[\bar q_{n\omega_1}^L
\gamma^\perp_\beta b_v\big]\big[ \bar f_{\bn\omega_2}^L \nslash   {\cal
P}_\perp^\beta q_{\bn\omega_3}^L \big] \,, \nn \\
 Q_{4(fuu)}^{(1\chi)}&=&\frac{1}{m_b} \big[\bar f_{n\omega_1}^L
\gamma^\perp_\beta b_v\big]\big[ \bar u_{\bn\omega_2}^R \nslash   {\cal
P}_\perp^\beta u_{\bn\omega_3}^R \big] \,, 
\end{eqnarray}
where we have $f=d,s$. The $(qfq)$ subscripts on the operators indicate the
flavors of the light quarks, and the basis has in addition the operators
$Q_{3(fuu)}^{(1\chi)}$ and $Q_{3(ufu)}^{(1\chi)}$.  Whenever a flavor label $q$
appears we implicitly sum over $q=u,d,s$. The operators $Q_{1,2}^{(1\chi)}$ give
contributions to $PP$, $PV$, and $V_0V_0$ final states, whereas
$Q_{3,4}^{(1\chi)}$ only contribute for transversely polarized vector mesons.
If operators that produce $\bn$-isosinglet mesons are included, we have in
addition $Q_{3(fqq)}^{(1\chi)}$ and $Q^{(2\chi)}_{4(fqq)}$. Since the
$Q^{(1\chi)}_{1,2,4}$ operators have right-handed quarks, only $O_{5,6,7,8}$ in
the electroweak Hamiltonian can contribute to their matching at tree level,
while other operators start contributing at one-loop.

Next we construct a complete basis for the $Q_{B1}^{(2\chi)}$ and
$Q_{B2}^{(2\chi)}$-type operators.  For chirality $(LH)(LL)$ we must
have $\Gamma_n=1$ and $\Gamma_\bn=\nslash$ and we have two choices for
contracting the $\perp$ indices, $g_\perp^{\alpha\beta}$ or
$i\epsilon_\perp^{\alpha\beta}$. To avoid having the epsilon symbol in
our basis we trade $i\epsilon_\perp^{\alpha\beta}$ for a pair of
$\gamma_\perp$'s. Here the possible flavor structures are $(qfq)$,
$(ufu)$, $(fuu)$, and $(fqq)$ from matching the operators in the
original $H_W$. For chirality $(LH)(RR)$ the same Dirac basis applies,
with flavor choices $(fuu)$ and $(fqq)$. The latter flavor structure
only produces $\bn$-collinear isosinglet mesons. Finally for
$(RH)(LR)$ we must have $\Gamma_n=\gamma_\perp^\mu$ and
$\Gamma_\bn=\nslash\gamma_\perp^\nu$ and there are only two
inequivalent ways of contracting the $\perp$ indices
$(\alpha\beta\mu\nu)$. This follows since contractions with an
$i\epsilon_\perp$ do not lead to independent structures because of the
fixed chirality, and the identity $\bnslash \gamma_\perp^\mu
P_L\otimes \nslash \gamma_\mu^\perp P_R=0$ which allows an additional
contraction to be eliminated. For $(RH)(LR)$ only the flavor structure
$(qfq)$ contributes. All together these results lead us to define the
basis
\begin{align} \label{Q2chi}
 Q_{1(qfq)}^{(2\chi)}&=\frac{-1}{m_b} 
 \Big[\bar q_{n\omega_1}^L \frac{1}{\bn\mcdot \cP}\, 
  {\cal P}_\perp \mcdot ig{\cal B}_{n\perp}  b_v\Big]
 \big[ \bar f^L_{\bn\omega_2} \nslash    q^L_{\bn\omega_3}\big] 
  \,,\nn   \\
 Q_{2(fuu)}^{(2\chi)}&=\frac{-1}{m_b} 
 \Big[ \bar f_{n\omega_1}^L \frac{1}{\bn\mcdot \cP}\,
  {\cal P}_\perp \mcdot ig{\cal B}_{n\perp}  b_v\Big]
 \big[ \bar u^R_{\bn\omega_2} \nslash    u^R_{\bn\omega_3}\big] \,,  \nn \\
 Q_{3(qfq)}^{(2\chi)}&=\frac{-1}{m_b^2} 
 \big[\bar q_{n\omega_1}^R
  ig\: \slash\!\!\!\! {\cal B}_{n\perp}   b_v\big]
 \big[ \bar f_{\bn\omega_2}^L \nslash  \SppP 
  q_{\bn\omega_3}^R \big]\,,
\nn \\
  Q_{4(qfq)}^{(2\chi)} &=  \frac{3}{2} e_q\: Q_{3(qfq)}^{(2\chi)} 
 \,, \nn \\[5pt]
 Q_{5(qfq)}^{(2\chi)}&=\frac{-1}{m_b} 
 \Big[\bar q_{n\omega_1}^L \frac{1}{\bn\mcdot \cP}\,  ig{\cal B}_{n\perp}^\alpha b_v\Big]
 \big[ \bar f_{\bn\omega_2}^L \nslash {\cal P}^\perp_\alpha q_{\bn\omega_3}^L
 \big]
 \,, \nn \\
 Q_{6(fuu)}^{(2\chi)}&=\frac{-1}{m_b} 
 \Big[\bar f_{n\omega_1}^L  \frac{1}{\bn\mcdot \cP}\, ig{\cal B}_{n\perp}^\alpha b_v\Big]
 \big[ \bar u_{\bn\omega_2}^R \nslash {\cal P}^\perp_\alpha u_{\bn\omega_3}^R
 \big]
  \,, \nn \\
 Q_{7(qfq)}^{(2\chi)}&=\frac{-1}{m_b^2} 
 \big[\bar q_{n\omega_1}^R {\cal P}_\perp^{\dagger\alpha} 
  ig\: \slash\!\!\!\! {\cal B}_{n\perp}   b_v\big]
 \big[ \bar f_{\bn\omega_2}^L \nslash    \gamma^\perp_\alpha   q_{\bn\omega_3}^R
 \big]\,,
\nn \\
 Q_{8(qfq)}^{(2\chi)} &=  \frac{3}{2} e_q\: Q_{7(qfq)}^{(2\chi)} \,,
\end{align}
plus operators with the same Dirac structure but different flavors,
$Q^{(2\chi)}_{i(ufu)}$ and $Q^{(2\chi)}_{i(fuu)}$ for $i=1$ and $5$. The
operators in Eq.~(\ref{Q2chi}) also incorporate electroweak penguins, since we
can write $e_q \, q\bar q = u\bar u - 1/3\, q\bar q $. Operators
$Q_{1-4}^{(2\chi)}$, contribute for $B\to PP$, $B\to PV$, and $B\to V_0V_0$
decays, whereas the operators $Q_{5-8}^{(2\chi)}$ only contribute for decays
with transverse vectors in the final state, $B\to V_TV_T$. If $\bn$-isosinglet
operators are included we have in addition the operators $Q_{i(fqq)}^{(2\chi)}$
where $i=1,2,5,6$.

For the basis in Eq.~(\ref{Q2chi}) we have only written operators that
contribute to $B$ decays. The remaining operators which only contribute for weak
$B^*$ decays are
\begin{align} \label{Bstar}
 Q_{1(qfq)}^{(2\chi)*}&=\frac{-1}{m_b^2} 
 \big[\bar q_{n\omega_1}^L \cP_\alpha^{\dagger} 
  ig\: \slash\!\!\!\! {\cal  B}_{n\perp} \gamma_\perp^\alpha  b_v\big]
 \big[ \bar f^L_{\bn\omega_2} \nslash    q^L_{\bn\omega_3}\big] \,,
\nn \\
 Q_{2(fuu)}^{(2\chi)*}&=\frac{-1}{m_b^2} 
 \big[\bar f_{n\omega_1}^L \cP^{\dagger}_\alpha 
 ig\: \slash\!\!\!\! {\cal B}_{n\perp}  \gamma_\perp^\alpha b_v\big]
 \big[ \bar u^R_{\bn\omega_2} \nslash    u^R_{\bn\omega_3}\big] \,,
\nn \\
 Q_{3(qfq)}^{(2\chi)*}&=\frac{-1}{m_b^2} 
 \big[\bar q_{n\omega_1}^R  \gamma_\perp^\beta ig{\cal B}_{n\perp}^\alpha  b_v\big]
 \big[ \bar f_{\bn\omega_2}^L \nslash   \gamma^\perp_\alpha {\cal
P}_\beta^\perp q_{\bn\omega_3}^R \big] \nn\\
 & + \frac{1}{m_b^2} 
 \big[\bar q_{n\omega_1}^R
  ig\: \slash\!\!\!\! {\cal B}_{n\perp}   b_v\big]
 \big[ \bar f_{\bn\omega_2}^L \nslash  \SppP 
  q_{\bn\omega_3}^R \big]
 \,, \nn\\
 Q_{5(qfq)}^{(2\chi)*}&=\frac{-1}{m_b^2} 
 \big[\bar q_{n\omega_1}^L ig\: \slash\!\!\!\! {\cal B}_{n\perp}
\gamma^\perp_\alpha  b_v\big]
 \big[ \bar f_{\bn\omega_2}^L \nslash  {\cal P}_\perp^\alpha
 q_{\bn\omega_3}^L \big]\,, 
\nn \\[5pt]
 Q_{6(fuu)}^{(2\chi)*}&=\frac{-1}{m_b^2} 
 \big[\bar f_{n\omega_1}^L ig\: \slash\!\!\!\! {\cal B}_{n\perp}
\gamma^\perp_\alpha  b_v\big]
 \big[ \bar u_{\bn\omega_2}^R \nslash  {\cal P}_\perp^\alpha
 u_{\bn\omega_3}^R \big]\,, 
\nn \\[5pt]
 Q_{7(qfq)}^{(2\chi)*}&=\frac{-1}{m_b^2} 
 \big[\bar q_{n\omega_1}^R  \SppP^\dagger ig{\cal B}_{n\perp}^\alpha b_v\big]
 \big[ \bar f_{\bn\omega_2}^L \nslash   \gamma^\perp_\alpha q_{\bn\omega_3}^R
 \big] \nn\\
 & + \frac{1}{m_b^2} 
 \big[\bar q_{n\omega_1}^R {\cal P}_\perp^{\dagger\alpha} 
  ig\: \slash\!\!\!\! {\cal B}_{n\perp}   b_v\big]
 \big[ \bar f_{\bn\omega_2}^L \nslash    \gamma^\perp_\alpha   q_{\bn\omega_3}^R
 \big]
 \,, 
\end{align}
and $ Q_{4,8(qfq)}^{(2\chi)*} = \frac{3}{2} e_q\: Q_{3,7(qfq)}^{(2\chi)*}$. 
Taken together the results in Eqs.~(\ref{Q2chi}) and (\ref{Bstar}) form a
complete basis for decays to non-isosinglet final states. We demonstrate
that the $Q_{i(F)}^{(2\chi)*}$ only contribute for $B^*$ decays in appendix~\ref{appB}.

The Hamiltonian for the full basis of $(1\chi)$ and $(2\chi)$ type-operators
contributing to $B$-decays is
\begin{align}
H &=\frac{ 4 G_F}{\sqrt 2}
\sum_{i,F} \bigg[ \!\int\! [d\omega_{1,2,3}] \, c_{i(F)}^{\chi} (\omega_j)\,
Q_{i(F)}^{(1\chi)}(\omega_j) \nn\\ &\qquad + \!\int\! [
d\omega_{1-4}] \, b_{i(F)}^{\chi} (\omega_j) \,
Q_{i(F)}^{(2\chi)}(\omega_j) \bigg] 
\,,
\end{align}
where the indices run over the operator number $i$ and possibilities for the
flavors $F$ for the $Q_{i(F)}$'s shown in Eqs.~(\ref{Q1chi}) and (\ref{Q2chi}), and
$c_{i(F)}^{\chi}$ and $b_{i(F)}^{\chi}$ are short-distance Wilson
coefficients.

Next we match from $H_W$ onto the operators in Eqs.~(\ref{Q1chi})
and (\ref{Q2chi}) to determine the Wilson coefficients
$c_{i(F)}^{\chi}$ and $b_{i(F)}^{\chi}$ at lowest order in the
$\alpha_s(m_b)$ expansion. At lowest order $c_{i(F)}^{\chi}$ are
simply given by the matrix elements of the $O_i$'s expanded to
next-to-leading order in the $\lambda$ power counting with 
\begin{eqnarray} \label{fieldexpn}
 q= \Big(1+\frac{1}{\bn\mcdot \cP}\SppP \frac{\bnslash}{2} \Big) q_n.  
\end{eqnarray} 
For the $(LH)(LL)$ and $(LH)(RR)$ chirality only the expansion of the
$\bn$-bilinear contributes, and for the non-isosinglet operators we find
\begin{align}
  c_{3(ufu)}^{\chi} &= -\frac{1}{\bar u}\bigg[ \lambda_u^{(f)} \Big(
C_1\plus
\frac{C_2}{N_c} \Big) \minus \lambda_t^{(f)} \frac32 \Big( C_{10}\plus
\frac{C_9}{ N_c}\Big) \bigg]\nn\\
 &\qquad 
+\Delta c_{3(ufu)}^{\chi}
, \nn \\
  c_{3(fuu)}^{\chi} &= -\frac{1}{\bar u}\Big[ \lambda_u^{(f)} \Big(C_2\plus
\frac{C_1}{N_c} \Big) - \lambda_t^{(f)} \frac32 \Big( C_{9}+
\frac{C_{10}}{N_c}\Big) \bigg]\nn\\
 &\qquad 
 +\Delta c_{3(fuu)}^{\chi}
 \, , \nn\\
  c_{3(qfq)}^{\chi} &= \lambda_t^{(f)} \frac{1}{\bar u}\Big[C_4\plus
\frac{C_3}{N_c} -\frac{C_{10}}{2}-\frac{C_{9}}{2 N_c}\Big] 
+\Delta c_{3(qfq)}^{\chi} \, , \nn \\
 c_{4(fuu)}^{\chi} &= 
   -\lambda_t^{(f)} \frac{3}{2 u}\left ( C_7+\frac{C_8}{N_c} \right )
   + \Delta c_{4(fuu)}^{\chi} \, .
\end{align}
\begin{figure}[t!]
  \centerline{
    \mbox{\epsfysize=2.3truecm \hbox{\epsfbox{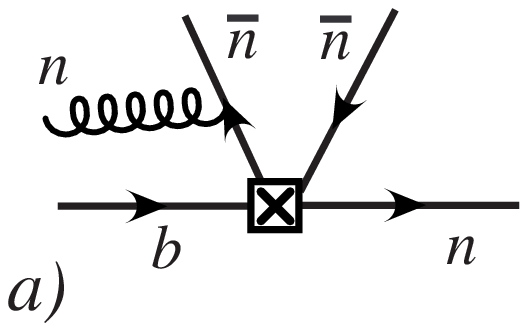}} } \quad
    \mbox{\epsfysize=2.3truecm \hbox{\epsfbox{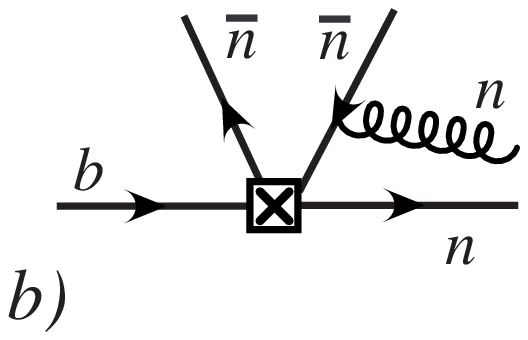}} } }
   \centerline{
    \mbox{\epsfysize=2.52truecm \hbox{\epsfbox{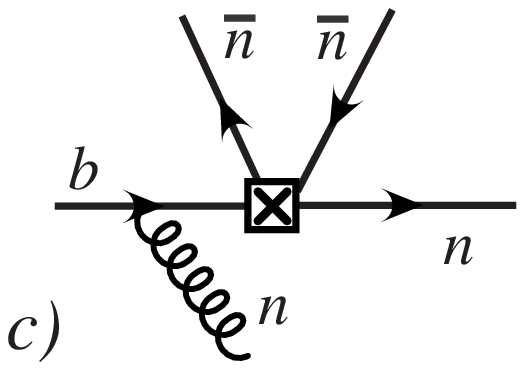}} }\quad\
    \mbox{\epsfysize=2.5truecm \hbox{\epsfbox{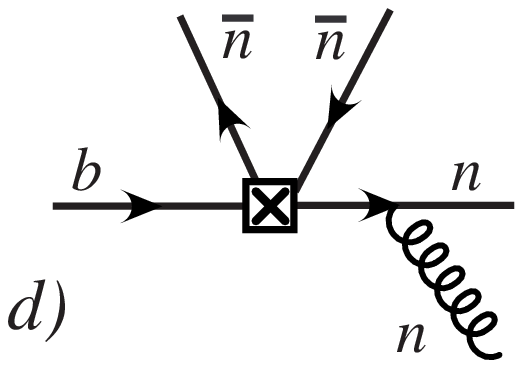}} } } \vskip-0.3cm
  {\caption[1]{ Full theory graphs for the matching onto the short-distance
      penguin coefficients for the $Q_{B1}$ and $Q_{B2}$ type operators.}
\label{figQCDchi} }
\vskip -0.2cm
\end{figure}
As usual the $\Delta c_{i(F)}^{\chi}$ terms denote perturbative corrections.
Numerically they will not always be suppressed due to the competition between
$C_{3,4}$ and $\alpha_s(m_b) C_1$.  For the operators $Q_{1,2(F)}^{(1\chi)}$
only $O_{5-8}$ from $H_W$ contribute at tree-level since the operator involves
right handed quarks. We find
\begin{align}
\label{chiralcoefficientsc}
c_{1(qfq)}^{\chi}&= \lambda_t^{(f)} \Big(C_6 \plus \frac{C_5}{N_c} \Big)
\frac{1}{u\bar u} + \Delta c_{1(qfq)}^{\chi}
 \,, \nn\\
c_{2(qfq)}^{\chi} &=   \lambda_t^{(f)} \Big(C_8 \plus \frac{C_7}{N_c} \Big) 
  \frac{1}{u\bar u} + \Delta c_{2(qfq)}^{\chi} \,.
\end{align}
We find that the loop and magnetic penguin graphs in Fig.~\ref{figQCDc} can only
contribute to the matching when a factor of $\perp$-momentum is generated by
expanding the spinors. They give the following ${\mathcal O}(\alpha_s)$
corrections to the matching
\begin{align} \label{alphaschi}
\Delta c_{1(qfq)}^{\chi} &= \frac{C_F\alpha_s(\mu)}{6N_c\pi}\,  
    \bigg\{  \frac{\lambda_u^{(f)} C_1}{u\bar u} \Big[
 h_2^u(u,1) -\frac{4}{3}  + S_c \Big]
     \\
 &\hspace{-0.4cm}
  +  \frac{\lambda_c^{(f)} C_1}{u\bar u} \Big[ \frac{2 \rho}{\bar u}
    h_1^c(u,1,\rho)+h_2^c(u,1,\rho)-\frac{4}{3} + S_c \Big]
  \nn \\
&\hspace{-0.4cm}
 + \, (\lambda_u^{(f)}+\lambda_c^{(f)}) \frac{3 C_{8g} \bar m_b}{m_b\, u \bar u}   
    \bigg\} \, ,
  \nn\\
\Delta c_{3(qfq)}^{\chi} &=\frac{C_F\alpha_s(\mu)}{6N_c\pi}\, 
    \bigg\{\lambda_u^{(f)} C_1 \frac{1}{\bar u}
   \Big[ h_2^u(u,1) -\frac{4}{3}  + S_c \Big] 
 \nn\\
 & \hspace{-0.4cm}
 +  \lambda_c^{(f)} C_1\,  \frac{1}{\bar u} \Big[\frac{2 \rho}{\bar u}
 h_1^c(u,1,\rho) \plus h_2^c(u,1,\rho)
  \minus \frac{4}{3} \plus S_c \Big]
   \bigg\} \, ,\nn
\end{align}
where the other $\Delta c_{i(F)}$ coefficients are zero at this level, and
$S_c=0$ for the NDR scheme while $S_c=1$ for the HV scheme. The scheme
dependence in these results cancels against that in the tree level $C_{3,4}$
terms in $c_{3(qfq)}^{\chi}$ just as for the LO $c_4^{(f)}$ Wilson
coefficient, and also in an identical manner against the scheme dependence in
the tree-level $C_{5,6}$ terms in $c_{1(qfq)}^{\chi}$. 

Next consider the matching calculation which determines $b_{i(F)}^{\chi}$. At
tree-level this involves computing the graphs in Fig.~\ref{figQCDchi}, and
involves non-zero contributions from expanding the propagators in graphs a), b),
and c), and from expanding the spinors with Eq.~(\ref{fieldexpn}) for graphs a),
b), and d). We find
\begin{align}
\label{chiralcoefficientsb1}
 b_{1(qfq)}^{\chi} &= 2  \lambda_t^{(f)} \bigg[ 
  \frac{(1 \plus u z)}{u z} 
    \Big(\frac{C_3}{N_c} \minus \frac{C_9}{2N_c}\Big)
   \plus
  C_4 \minus \frac{C_{10}}{2} 
  \bigg] 
  \nn \\ 
  &\quad 
  + \Delta  b_{1(qfq)}^{\chi}
  \, , \nn\\
 b_{2(fuu)}^{\chi} &= 3 \lambda_t^{(f)} \bigg[ C_7\plus
 \frac{C_8}{N_c}  -  \frac{1}{\bar u z}\frac{C_8}{N_c}
  \bigg]  +  \Delta b_{2(fuu)}^{\chi}
   \, , \nn \\ 
 b_{3(qfq)}^{\chi} &= \lambda_t^{(f)}  \frac{1}{u\bar u }
   \Big(C_6\plus \frac{C_5}{N_c} \Big)
   +\Delta b_{3(qfq)}^{\chi}
   \, ,  \nn \\
 b_{4(qfq)}^{\chi} &=  \lambda_t^{(f)} \frac{1}{u\bar u }
    \Big( C_8 \plus \frac{C_7}{N_c} \Big)
  + \Delta b_{4(qfq)}^{\chi}  
    \,, \nn\\
 b_{5(qfq)}^{\chi} &= -\lambda_t^{(f)} \frac{2}{u\bar u} \: \Big(\frac{C_3}{N_c}
    \minus \frac{C_9}{2N_c}\Big) 
    + \Delta  b_{5(qfq)}^{\chi}
    \,, \nn \\
    b_{6(fuu)}^{\chi} &= -\lambda_t^{(f)} \frac{3C_8}{N_c} \frac{1}{ u\bar u } 
+ \Delta b_{6(fuu)}^{\chi}
    \, , \nn \\
 b_{7(qfq)}^{\chi} &=  \lambda_t^{(f)} \frac{1}{u\bar u z}
    \frac{C_5}{N_c}
 + \Delta b_{7(qfq)}^{\chi}
    \,,
   \nn \\
 b_{8(qfq)}^{\chi} &=  \lambda_t^{(f)} \frac{1}{u\bar u z}
    \frac{C_7}{N_c} 
  + \Delta b_{8(qfq)}^{\chi}  
    \,,
\end{align} 
and 
\begin{align}
\label{chiralcoefficientsb2}
 b_{1(ufu)}^{\chi} &=   
  \frac{2(1\plus uz)}{u z} 
    \Big(\minus  \frac{C_2}{N_c} \lambda_u^{(f)}  
    \plus \frac{3C_9}{2N_c} \lambda_t^{(f)}  \Big)
   \nn \\
 &\qquad  \minus \Big( 2 C_1 \lambda_u^{(f)} 
  \minus  3 C_{10} \lambda_t^{(f)} \Big) 
  + \Delta  b_{1(ufu)}^{\chi}
  \, ,  \nn\\
 b_{5(ufu)}^{\chi} &=  \frac{2}{u\bar u} \: \Big( \frac{C_2}{N_c} \lambda_u^{(f)}
    \minus  \frac{3C_9}{2N_c} \lambda_t^{(f)} \Big) 
    +\Delta  b_{5(ufu)}^{\chi} 
    \,, \nn \\
 b_{1(fuu)}^{\chi} &=
  \frac{2(1\plus uz)}{u z} 
    \Big(\minus\lambda_u^{(f)}  \frac{C_1}{N_c}
    \plus  \lambda_t^{(f)} \frac{3C_{10}}{2N_c}\Big) \nn\\
  &\qquad - \Big(2C_2 \lambda_u^{(f)} 
   \minus 3C_{9} \lambda_t^{(f)} \Big)
   +\Delta  b_{1(fuu)}^{\chi} 
   \, , \nn\\
 b_{5(fuu)}^{\chi} &=  \frac{2}{u\bar u} \: \Big(\frac{C_1}{N_c}\lambda_u^{(f)}
    \minus  \frac{3C_{10}}{2N_c} \lambda_t^{(f)} \Big) +\Delta
    b_{5(fuu)}^{\chi}
   \,. 
\end{align}
Results for the Wilson coefficients for cases with isosinglet $\bn$-mesons can
be determined in an analogous way, but in this case operators with gluons also
become necessary. We leave this for future work.

Since we are only using factorization of effects at $\mu\sim m_b$ the matrix
elements of the operators $Q^{(1\chi)}_{i(F)}$ and $Q^{(2\chi)}_{i(F)}$ give
hadronic parameters. For the $\bn$-collinear field we need the matrix elements
in Eq.~(\ref{zeta}), 
\begin{table*}[t!]
\begin{tabular}{|c|c|c|c|c|}
\hline\hline
$M_1 M_2$ & $R_{1}$ & $R_{2}$ & $R^{\chi}_{1}$ & $R^{\chi}_{2}$
\\ \hline\hline
$\pi^-\pi^+ $,  $\rho^-\pi^+$
  & $\,c^{\chi}_{1(qfq)} \plus \, c^{\chi}_{2(qfq)}$
&  $0$ 
  &  $ b^{\chi}_{1(qfq)}\plus b^{\chi}_{1(ufu)}$
&  $0$ 
\\[3pt]
 $\pi^-\rho^+$
  & $\minus c^{\chi}_{1(qfq)} \minus \, c^{\chi}_{2(qfq)}$
&  $0$ 
  &  $ b^{\chi}_{1(qfq)}\plus b^{\chi}_{1(ufu)}$
&  $0$ 
\\[3pt]
  $\pi^-\pi^0 $  
 & $\frac{1}{\sqrt2} \big[ c^{\chi}_{1(qfq)}\plus c^{\chi}_{2(qfq)} \big]$
 & $\frac{-1}{\sqrt2}\big[ c^{\chi}_{1(qfq)}\minus\frac12 c^{\chi}_{2(qfq)} \big]$
 & $\frac{1}{\sqrt2} \big[b^{\chi}_{1(ufu)} \plus  b^{\chi}_{1(qfq)}  \big]$
 & $\frac{1}{\sqrt2} \big[ b^{\chi}_{1(fuu)}
    \minus b^{\chi}_{2(fuu)} \minus b^{\chi}_{1(qfq)} \big] $
   \\[3pt]
   $ \rho^-\pi^0$  
 & $\frac{1}{\sqrt2} \big[ c^{\chi}_{1(qfq)}\plus c^{\chi}_{2(qfq)} \big]$
 & $\frac{1}{\sqrt2}\big[ c^{\chi}_{1(qfq)}\minus\frac12 c^{\chi}_{2(qfq)} \big]$
 & $\frac{1}{\sqrt2} \big[b^{\chi}_{1(ufu)} \plus  b^{\chi}_{1(qfq)}  \big]$
 & $\frac{1}{\sqrt2} \big[ b^{\chi}_{1(fuu)}
    \minus b^{\chi}_{2(fuu)} \minus b^{\chi}_{1(qfq)} \big] $
   \\[3pt]
   $\pi^-\rho^0$  
  & $\frac{-1}{\sqrt2} \big[ c^{\chi}_{1(qfq)}\plus c^{\chi}_{2(qfq)} \big]$
 & $\frac{-1}{\sqrt2}\big[c^{\chi}_{1(qfq)}\minus\frac{1}{2} c^{\chi}_{2(qfq)}\big]$
 & $\frac{1}{\sqrt2} \big[b^{\chi}_{1(ufu)} \plus  b^{\chi}_{1(qfq)}  \big]$
 & $\frac{1}{\sqrt2} \big[ b^{\chi}_{1(fuu)}
    \plus b^{\chi}_{2(fuu)} \minus b^{\chi}_{1(qfq)} \big] $
 \\[3pt]
$\pi^0 \pi^0$ 
& $ \minus \frac{1}{2} c^{\chi}_{1(qfq)}\plus \frac{1}{4} c^{\chi}_{2(qfq)} $
&   $ \frac{-1}{2}\,c^{\chi}_{1(qfq)}\plus \frac{1}{4}\, c^{\chi}_{2(qfq)} $
& $\frac{1}{2} \big[ b^{\chi}_{1(fuu)} \minus b^{\chi}_{2(fuu)}
    \minus b^{\chi}_{1(qfq)} \big]$
& $\frac{1}{2} \big[ b^{\chi}_{1(fuu)} \minus b^{\chi}_{2(fuu)}
     \minus b^{\chi}_{1(qfq)} \big] $
  \\[3pt]
  $ \rho^0 \pi^0$
 & $  \frac{-1}{2} c^{\chi}_{1(qfq)}\plus \frac{1}{4} c^{\chi}_{2(qfq)} $
&   $ \frac{1}{2}\,c^{\chi}_{1(qfq)}\minus \frac{1}{4}\, c^{\chi}_{2(qfq)} $
& $\frac{1}{2} \big[ b^{\chi}_{1(fuu)} \plus b^{\chi}_{2(fuu)}
    \minus b^{\chi}_{1(qfq)} \big]$
& $\frac{1}{2} \big[ b^{\chi}_{1(fuu)} \minus b^{\chi}_{2(fuu)}
     \minus b^{\chi}_{1(qfq)} \big] $
 \\[3pt]
  $K^{(*)0} K^{-}$\!, $ K^{(*)0} \bar K^{0}$
  &  $- c^{\chi}_{1(qfq)}\plus \frac{1}{2} c^{\chi}_{2(qfq)}$
  & $0$
  & $- b^{\chi}_{1(qfq)}$
   & $0$
  \\[3pt]
  $K^{0} K^{*-}$\!, $ K^{0} \bar K^{*0}$
  &  $ c^{\chi}_{1(qfq)}\minus \frac{1}{2} c^{\chi}_{2(qfq)}$
  & $0$
  & $- b^{\chi}_{1(qfq)}$
   & $0$
  \\[3pt]
$K^{(*)-} K^{(*)+}$ &
  {\bf ---} & {\bf ---}
  & {\bf ---}
  & {\bf ---}
  \\
\hline
$\pi^+ K^{(*)-} $
   &  $0$
   &  $\,c^{\chi}_{1(qfq)}\plus \, c^{\chi}_{2(qfq)}$
   & $0$
   & $ b^{\chi}_{1(ufu)}\plus b^{\chi}_{1(qfq)}$
\\[3pt]
$\rho^+ K^{-}$
   &  $0$
   &  $- c^{\chi}_{1(qfq)}\minus \, c^{\chi}_{2(qfq)}$
   & $0$
   & $ b^{\chi}_{1(ufu)}\plus b^{\chi}_{1(qfq)}$
\\[3pt]
$\pi^0 K^{(*)-}$
   & $0 $
   & $\frac{1}{\sqrt2}\big[ c^{\chi}_{1(qfq)} \plus  c^{\chi}_{2(qfq)}\big] $
   & $\frac{1}{\sqrt2} \big[ b^{\chi}_{1(fuu)}\minus b^{\chi}_{2(fuu)} \big]$
   & $\frac{1}{\sqrt2} \big[  b^{\chi}_{1(ufu)} \plus b^{\chi}_{1(qfq)} \big]$
   \\[3pt]
$\rho^0 K^-$
  & $0 $
   & $\frac{-1}{\sqrt2}\big[ c^{\chi}_{1(qfq)} \plus c^{\chi}_{2(qfq)}\big] $
   & $\frac{1}{\sqrt2} \big[ b^{\chi}_{1(fuu)}\plus b^{\chi}_{2(fuu)} \big]$
   & $\frac{1}{\sqrt2} \big[ b^{\chi}_{1(ufu)} \plus b^{\chi}_{1(qfq)} \big]$
   \\[3pt]
$\pi^- \bar K^{(*)0}$
   & $ 0 $ & $\minus c^{\chi}_{1(qfq)} \plus \frac{1}{2} c^{\chi}_{2(qfq)} $
  & $0$
   & $\minus b^{\chi}_{1(qfq)}$
  \\[3pt]
 $\rho^- \bar{K}^{0}$
   & $ 0 $ & $c^{\chi}_{1(qfq)} \minus \frac{1}{2} c^{\chi}_{2(qfq)} $
  & $0$
   & $\minus b^{\chi}_{1(qfq)}$
  \\[3pt]
$\pi^0 \bar K ^{(*)0}$
   & $ 0 $
   & $\frac{-1}{\sqrt2} \big[c^{\chi}_{1(qfq)}\minus
      \frac{1}{2}  c^{\chi}_{2(qfq)} \big] $
   & $\frac{1}{\sqrt2} \big[ b^{\chi}_{1(fuu)}\minus
   b^{\chi}_{2(fuu)}\big] $
   &$\minus \frac{1}{\sqrt2} b^{\chi}_{1(qfq)}$
   \\[3pt]
$\rho^0 \bar K ^{0}$
  & $ 0 $
   & $\frac{1}{\sqrt2} \big[ c^{\chi}_{1(qfq)}\minus
        \frac{1}{2}  c^{\chi}_{2(qfq)} \big]$
   & $\frac{1}{\sqrt2} \big[ b^{\chi}_{1(fuu)}\plus
   b^{\chi}_{2(fuu)}\big] $
   & $\minus \frac{1}{\sqrt2} b^{\chi}_{1(qfq)}$
 \\
  \hline\hline
\end{tabular}
\caption{Hard functions for the chiraly enhanced amplitudes in
Eq.~(\ref{chienhanced}) for $\bar B^0$ and $B^-$ decays to $PP$ and $PV$
channels.  We have not listed results for $R_{1,2\,}^J$, but they have the same
Clebsch-Gordan coefficients as $R_{1,2 \, }$ and so can be simply obtained by
the replacements $c^{\chi}_{1(qfq)} \rightarrow b^{\chi}_{3(qfq)} $ and 
$c^{\chi}_{2(qfq)} \rightarrow b^{\chi}_{4(qfq)}$ in the columns above.
}
\label{table2a}
\end{table*}
and the corresponding results with right-handed light quarks
and Clebsch-Gordan coefficients
\begin{align}
    \big\langle M_n \big| T_1\big[ \bar q_{n\omega_1}^R \bnslash b_v \big]
   \big| B\big\rangle 
  & =  {\cal C}_{q_R}^{BM}  
   \bar \delta_{\omega_1} \: {m_B} \: \zeta^{BM} , \\
  \big\langle M_n \big| T_2\big[ \bar q_{n\omega_1}^R \,
  ig\,\slash\!\!\!\!{\cal B}^\perp_{n\omega_4}  b_v \big] \big| B\big\rangle 
 &= - {\cal C}_{q_R}^{BM} 
   \bar \delta_{\omega_1\omega_4} \frac{m_B}{2} \zeta_J^{BM}(z)\,. \nn
\end{align}
We also need a new form factor $\zeta_\chi^{BM}(z)$ and the chiral-enhanced function
$\phi_{pp}^P(x)$ defined by
\begin{align} \label{phipp}
& \big\langle M \big\vert T_2 \Big [ \bar q_{n\omega_1}^L 
   \frac{1}{\bn\mcdot \cP}\, {\cal P}_\perp \mcdot ig{\cal B}_{n\omega_4}^\perp  b_v \Big ] \big
   \vert B \big\rangle 
  \\
 & \hspace{1 cm}
  = - {\cal C}_{q_L}^{BM} \bar \delta_{\omega_1\omega_4}\, \frac{\mu_M}{12}\,
   \zeta_\chi^{BM}(z) \, , \nn  \\
&\langle P(p) |\, \bar q^L_{n\omega_2}\, \bnslash\,\,
  {\slash\!\!\!\! \cP_\perp} q^{\prime R}_{n\omega_3}\, | 0 \rangle
 =  - \frac{i}{6}\, {\cal C}_{q_Lq'}^P\,
  \bar \delta_{\omega_2\omega_3}\, f_P \mu_P\, \phi_{pp}^P(x) 
    ,\nn
\end{align}
where $\bar \delta_{\omega\omega'} = \delta(\omega-\omega'-m_B)$ and
the momentum fractions $z=\omega_1/\bn\mcdot p_M$ and $x=\omega_2/\bn
\cdot p_P$.  where ${\cal C}_{q_L}^{BM}$ and ${\cal C}_{q_Lq'}^M$ are
the same Clebsch-Gordan coefficients that appeared already at leading
order in Eq.~(\ref{Tz}). For the chiral-enhanced distribution function
$\phi_{pp}^M$ we used the definition in Ref.~\cite{Arnesen:2006vb},
and take the other twist-3 meson distribution to be the three-body
$\phi^{3M}(x,\bar x)$ which does not generate chiraly enhanced
contributions.  In a more traditional basis there is a redundancy at
this order in $1/m_b$ (see for example~\cite{Hardmeier:2003ig}), and
$\phi_{pp}^P(x) =3x[\phi_p^P(x)+\phi_\sigma^{P\prime}(x)/6+ 2
f_{3P}/(f_P \mu_P) \int dy'/y'
\phi_{3P}(y-y',y)]$. In the Wandzura-Wilczek approximation one would
set $\phi_{pp}(x) = 6 x (1-x)$.

Taking the matrix element of the above operators leads to a
factorization theorem for the chiraly enhanced amplitude for
non-isosinglet charmless $B$-decays to $PP$ and $PV$ channels
\begin{align}\label{chienhanced}
A^\chi(\bar B\to M_1 M_2) &=
\frac{G_F m_B^2}{\sqrt2}  \bigg\{
   \\
&\hspace{-2.3cm}
  - \frac{\mu_{M_1}  f_{M_1}}{3m_B}\, \zeta^{BM_2}  \int_0^1\!\!\!\! du\: R_{1}(u)\,
   \phi_{pp}^{M_1}(u) 
   + (1\leftrightarrow 2)
  \nn \\
 &\hspace{-2.3cm} -  \frac{\mu_{M_1}  f_{M_1}}{3m_B}\!\! \int_0^1\!\!\!\! du\,dz\:
   R_{1}^{J}(u,z) \zeta_J^{BM_2}(z) \phi_{pp}^{M_1}(u) 
 \!+\! (1\leftrightarrow 2) \nn \\
 &\hspace{-2.3cm} -   \frac{\mu_{M_2}  f_{M_1}}{6 m_B}\!\! \int_0^1\!\!\!\! du\,dz\:
   R_{1}^{\chi}(u,z) \zeta_{\chi}^{BM_2}(z) \phi^{M_1}(u) 
 \!+\! (1\leftrightarrow 2) \bigg\}. \nn
\end{align}
This amplitude only includes the chiraly-enhanced power corrections
where factors of $\mu_M$ are generated by pseudo-scalars, and so for
vectors we define $\mu_V=0$. (Note that we include the symmetry factor
of $1/2$ in the branching ratio prefactor for $B\to \pi^0\pi^0$ rather
than the amplitude.) In terms of Clebsch-Gordan coefficients for the
different final states, the hard functions $R_i$, $R_i^J$,and
$R_i^\chi$ for the chiraly enhanced amplitudes are
\begin{align}
\label{chihard}
R_{1}(u) &=  {\cal C}_{q_R}^{BM_{2}} {\cal C}_{f_Lq}^{M_{1}}
  \Big[ c^{\chi}_{1(qfq)}+\frac{3}{2}e_q \,c^{\chi}_{2(qfq)} \Big]
\,, \\
R_{1}^J(u,z) &=  {\cal C}_{q_R}^{BM_{2}} {\cal C}_{f_Lq}^{M_{1}}
  \Big[ b^{\chi}_{3(qfq)}+\frac{3}{2}e_q \,b^{\chi}_{4(qfq)} \Big]
\,, \nn \\
R_{1}^{\chi}(u,z) &= 
  {\cal C}_{q_L}^{BM_{2}} {\cal C}_{f_Lq}^{M_{1}} \, b^{\chi}_{1(qfq)} 
 + {\cal C}_{u_L}^{BM_{2}} {\cal C}_{f_Lu}^{M_{1}}\,  b^{\chi}_{1(ufu)} \nn \\
& +{\cal C}_{f_L}^{BM_{2}} {\cal C}_{u_L u}^{M_{1}}\, b^{\chi}_{1(fuu)}
  +{\cal C}_{f_L}^{BM_{2}} {\cal C}_{u_R u}^{M_{1}}\, b^{\chi}_{2(fuu)} \, .\nn 
\end{align} 
Summation over $q=u,d,s$ is implicit. Results for these hard functions in
different channels are listed in Table~\ref{table2a}.
Equation~(\ref{chienhanced}) with Eq.~(\ref{chihard}) corresponds to the
contributions given in the second line of Eq.~(\ref{hatP}) (when we extract the
coefficients of the $\lambda_c^{(f)}$ terms).  

From the matching results we find that the endpoint behavior of the
Wilson coefficients is $c_{i(F)}^{\chi}\sim 1/(u\bar u)$ and
$b_{i(F)}^{\chi}\sim 1/(z u\bar u)$.  Since we know the endpoint
behavior $\phi_{pp}(u)\sim u\bar u$ and $\zeta_J(z)\sim z$, it remains
to determine the behavior of $\zeta_\chi(z)$.  The operator defining
$\zeta_\chi(z)$ has an extra ${\cal P}_\perp/ \bn\cdot {\cal P}$
relative to the operator defining the distribution $\zeta_J(z)$. Now
from the collinear power counting ${\cal P}_\perp \ll \bn\cdot {\cal
P}$, so consistency of the power counting in \SCETa implies that the
scaling of $\zeta_\chi(z)$ as $z\to 0$ and $z\to 1$ can be no worse
than $\zeta_J(z)$. Thus we take $\zeta_\chi(z)\sim z$.  This
demonstrates that all the terms in the factorization theorem for
chiraly enhanced penguin and tree contributions given in
Eq.~(\ref{chienhanced}) converge, just like the leading order
factorization theorem in Eq.~(\ref{newfact}).  In appendix~\ref{appB}
we argue that the same conclusion about the $z$-convolution is
obtained if one considers the direct computation of $\zeta_\chi(z)$ in
\SCETb.

As already noted, the operators in Eqs.~(\ref{Q1chi}) and (\ref{Q2chi}) also
generate contributions with two transverse vectors in the final state.  To take
the matrix element of these terms requires
\begin{align}
\big\langle V \big\vert \bar q^{\prime L}_{\bn\omega_1} \nslash  \gamma_\perp^\alpha q^R_{\bn\omega_2}
  \big \vert 0 \big\rangle 
&= {\cal C}^{V}_{q_L'q}  \bar\delta_{\omega_1\omega_2} f_T^V
  \phi_{\perp}^{V}(u) \epsilon_{\perp}^\alpha , 
  \nn\\
 \big\langle V \big\vert \bar q^{\prime L,R}_{\bn\omega_1} \nslash  \cP_\perp^\alpha q^{L,R}_{\bn\omega_2}
  \big \vert 0 \big\rangle 
 &= {\cal C}^{V}_{q_{L,R}'q}  \bar\delta_{\omega_1\omega_2} f_{pp}^V
\phi_{pp\perp}^V(u) \epsilon_{\perp}^\alpha   \,,
\end{align}
where $u=\omega_1/m_b$ and three form factors
\begin{align}
\big\langle V \big\vert \bar q_{n\omega_1}^L  \gamma_\perp^\alpha b_v
  \big \vert B \big\rangle 
&= {\cal C}^{BV}_{q_L}  \bar\delta_{\omega_1} m_B\,
  \zeta_{\perp}^{BV} \epsilon_{L\perp}^\alpha , 
  \\
\big\langle V \big\vert \bar q_{n\omega_1}^L  ig{\cal
  B}_{n\omega_4}^{\perp\alpha} b_v
  \big \vert B \big\rangle 
&= -{\cal C}\:  \bar\delta_{\omega_1\omega_4} m_B\,
  \zeta_{J\perp}^{BV}(z) \epsilon_\perp^\alpha , 
  \nn \\
\big\langle V \big\vert \bar q_{n\omega_1}^R {\cal P}_\perp^{\dagger\alpha}
   ig\: \slash\!\!\!\! {\cal B}_{n\omega_4}^\perp   b_v \big \vert B \big\rangle 
&= -{\cal C}^\prime\: \bar\delta_{\omega_1\omega_4} m_B\,
  \zeta_{K\perp}^{BV}(z) \epsilon_\perp^\alpha , \nn
\end{align}
where $z=\omega_1/m_b$. Thus, our complete basis of operators with $\cP^\alpha$
terms generates a contribution to the amplitude to produce two transverse vector
mesons that involve two types of light-cone meson distributions, and three types
of form-factors.  These analogs of the chiraly enhanced terms were displayed as
the contributions on the second line of Eq.~(\ref{hatPT}). Our analysis
demonstrates that only these terms will be generated from the ${\cal P}_\perp$
operators considered in this section, however a full analysis of these terms will
not be given here.  Hence we have not bothered to specify the Clebsch-Gordan
coefficients ${\cal C}$ and ${\cal C}^\prime$ relative to our other conventions.


\section{Long-Distance Charm} \label{sect:long}

In order to properly determine the short-distance coefficients by matching we
must make sure that we subtract any effective theory diagrams. Earlier we stated
that there were no SCET loop graphs to subtract. In this section we further
justify this claim and discuss long-distance charm contributions.  We take
$m_c\sim m_b$ and so do not have collinear charm quarks.  Furthermore, graphs
with collinear or soft up quarks are power suppressed. The only remaining term
to consider are soft non-relativistic charm that propagate in the EFT.  While a
factorization theorem for this type of long distance charm effect has not yet
been derived, we may nevertheless match systematically by including in the
effective theory the proper SCET-NRQCD hybrid operators as discussed in
\cite{Bauer:2005wb}.

We begin by showing that non-zero contributions from the hybrid operators
requires a non-zero residual momentum and hence do not affect the matching
computations in earlier sections. We may write the momenta of the charm quarks as
\begin{eqnarray}
  p_1^\mu&=&\frac{q^\mu}{2}+L^\mu_{~i} r^i \,, \nn \\
  p_2^\mu &= & \frac{q^\mu}{2}-L^\mu_{~i} r^i \,,
\end{eqnarray}
where $q^\mu$ is the total momentum of the charm quark pair and $r^i$ is the relative 3-momentum in the $c\bar c$ rest frame.  $L^\nu_{~\mu}(q)$ is the Lorentz boost from the center-of-mass frame to the B rest frame and has
components
\begin{eqnarray}
L^0_{~0} &=& 1+\frac{\vec q^{\,2}}{4m_c^2}\,,
 \qquad L^0_{~i} = \frac{q_i}{2m_c} \,,
 \nn \\
L^j_{~i}&=&\delta^j_i+\left( \frac{E_q}{2m_c}-1\right)\frac{q^j q_i}{\vec q^2} \,.
\end{eqnarray}
When matching onto NRQCD at lowest order in $\alpha_s$ we generate a generic set
of the four quark operators.  As with the operators in Eqs.~(\ref{Q0},\ref{Q1})
there will be a set of operators with and without gluon external lines. Thus
there will be two generic forms of the operators
\begin{eqnarray}
O^{a}_{prod}&=&( \eta^\dagger L(\Gamma_{NR})\chi) 
  ( \bar f_{\bn,\omega_1} \Gamma_{hl}  b_v)\, ,  \nn \\
O^{b}_{prod}&=&( \eta^\dagger L( \Gamma_{NR}) \chi) 
   ( \bar f_{\bn,\omega_1}B^\perp_n \Gamma_{hl} b_v)\, ,  
\end{eqnarray}
$\Gamma_{NR},\Gamma_{hl}$ are the possible bilinear Dirac structures for NRQCD
and a heavy-light bilinear in SCET. The gluon field $B_\perp$ has a four vector
index that is either contracted with $\Gamma_{hl}$ or $L( \Gamma_{NR}) $.  For
$O^a$ the only possible structure is
\begin{equation}
 L(\Gamma_{NR})\otimes \Gamma_{hl}= \sigma^i L_\mu^i\otimes \gamma^\mu_\perp P_L
\end{equation}
For the $O^{b}$ operators the possible structures are
\begin{equation}
 L(\Gamma_{NR})\otimes \Gamma_{hl}=\{1\otimes~ \slash\!\!\!\!{\cal B}_n^\perp
 P_L,\sigma^i L_0^i \otimes~\slash\!\!\!\!{\cal B}_n^\perp  P_L \}
\end{equation}
In addition we have four quark operators that are generated by integrating out
one hard gluon exchange. They have the general form
\begin{eqnarray}
O^{a}_{ann}&=&\frac{\alpha_s(2m_c)}{4m_c^2}( \eta^\dagger L(\Gamma_{NR})\chi) 
  ( \bar q_{\bn,\omega_1} \Gamma_{n\bn}  q_{n,\omega_3})\, ,  \nn \\
O^{b}_{ann}&=&\frac{\alpha_s(2m_c)}{4m_c^2}( \eta^\dagger L( \Gamma_{NR}) \chi) 
   ( \bar q_{\bn,\omega_1}B^\perp_{n} \Gamma_{n\bn} q_{n,\omega_3})\, .  \nn\\
\end{eqnarray}
For the $O^a_{ann}$ operators the possible structures are
\begin{equation}
 L(\Gamma_{NR})\otimes \Gamma_{n\bn}=\{1\otimes 1 ,\sigma^i L_0^i \otimes 1 ,
   \sigma^i L_\mu^i\otimes \gamma^\mu_\perp \}
\end{equation}
while for the $O^b_{ann}$ operators the possible structures are
\begin{equation}
 L(\Gamma_{NR})\otimes \Gamma_{n\bn}=\{1\otimes~ \slash\!\!\!\!{\cal
B}_n^\perp  ,\sigma^i L_0^i \otimes~\slash\!\!\!\!{\cal B}_n^\perp
,\sigma^i L_\mu^i\otimes {\cal B}^\mu_{n \perp} \}.
\end{equation}
In general both $1\otimes 1$ and $T\otimes T$ color structures are allowed in
the operators $O^{a,b}_{prod}$ and $O^{a,b}_{ann}$.

$A_{c\bar c}$ then follows from the time ordered products of the form
\begin{align} \label{Tproducts2}
  T_1^{c\bar c}  &\equiv  \nn\\
 &\hspace{-0.3cm}
 \mbox{\large $\int$} d^4zd^4y\, d^4y'\,T 
   \big[ O^{a}_{prod}(0)O^a_{ann}(z) \: i{\cal L}^{(1)}_{\xi_n q}(y) \: i{\cal
    L}_{\xi_n\xi_n}^{(1)}\!(y') \big] \!\nn\\
  &\hspace{-0.3cm}
  + \mbox{\large $\int$}d^4z d^4y\, d^4y'\,T 
   \big[O^{a}_{prod}(0)O^a_{ann}(z) \: i{\cal L}^{(1)}_{\xi_n q}(y) \: 
   i{\cal L}_{cg}^{(1)}(y')\} \big]\nn \\
  &\hspace{-0.3cm}
   +  \mbox{\large $\int$} d^4z d^4y\, 
    T \big[ O^{a}_{prod}(0)O^a_{ann}(z),i{\cal L}^{(1,2)}_{\xi_n q}(y) \big], \ \nn\\
  T_2^{c\bar c}&\equiv \mbox{\large $\int$} d^4zd^4y \:
    T \big[O^{b}_{prod}(0)O^a_{ann}(z),i{\cal L}^{(1)}_{\xi_n q}(y) \big] \nn \\
  &+   T \big[O^{a}_{prod}(0)O^b_{ann}(z),i{\cal L}^{(1)}_{\xi_n q}(y) \big] .
\end{align} 
These operators could be factored into soft and collinear components, however
the details of this factorization will not be carried out here. The
factorization for semi-inclusive decays was discussed in
Ref.~\cite{Chay:2006ve}.

Now let us review some aspects of the power counting for these terms (refering
to the appendix of Ref.~\cite{Bauer:2005wb} for further details).  First note
that implicit in these operators is a label conserving delta function.  Recall
that the NRQCD fields have two large labels \cite{Luke:1999kz} $\chi_{m,mv}$
which have been suppressed in these operators.  For instance, repristinating the
momentum conserving delta function and momentum labels for $O^{a}_{prod}$, we
write
\begin{eqnarray}
O^{a}_{prod}&=&( \eta^\dagger_{mv} T^A (\sigma ^i L^i_0) \chi_{-mv}) 
 ( \bar q_{n,\omega_1} \bnslash P_L  T^A  b_v)\nn \\
&& \delta(m_b-\omega_1-n_\mu L^\mu_0 (2m_c)).  
\end{eqnarray}
Note that these delta functions do not imply that we are only including a
single point in the phase space, since the residual momenta of the HQET and SCET
field in the operator may flow through the charm loop. Furthermore since the
residual momentum scales as $\Lambda\sim mv^2$, the fluctuations in the external
momenta effectively smear over the non-relativistic region.

\begin{figure}[t!]
 \centerline{ \epsfysize=2.4truecm \hbox{\epsfbox{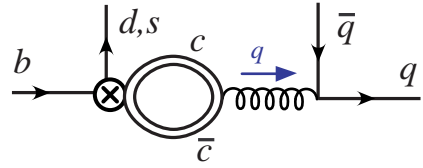}} }
  \vskip-0.3cm
  {\caption[1]{Charm loop contribution from the non-relativistic region.}
\label{NRcharm} }
\vskip -0.2cm
\end{figure}

The delta function constraint simplifies the matching since the
contribution of these hybrid operators to the matching vanishes at the lowest
order in $v$. To see this, we may work slightly away from threshold, by giving
the heavy quark (without loss of generality) some small residual momentum $k\sim
\Lambda$ , such that the invariant mass of the charm quark pair is 
\begin{eqnarray} 
\label{eqnq2}
 q^2=(m_b v +k -p)^2=4m_c^2 -\frac{4m_c^2}{m_b} \, \bn\mcdot k \,,
\end{eqnarray} 
where $\bn\cdot k\sim \Lambda$ and $p$ is the momentum of the $d,s$ quark. In deriving Eq.~(\ref{eqnq2}) we used the heavy quark equation of motion, $v\cdot k=0$, and expanded about the threshold, $m_b-n\cdot p\simeq 4 m_c^2 / m_b$.
An explicit calculation of the one loop diagram shown in Fig.~\ref{NRcharm}, shows that
this contribution is proportional to 
\begin{align}
 \sqrt{q^2-4m_c^2} \sim \sqrt{\Lambda m_b} \,.
\end{align}
Given our scaling, $mv^2\sim \Lambda$ we find that this contribution is order
$v$ as anticipated by power counting arguments.  Hence, only if we were
interested in matching explicitly onto $v$-suppressed corrections would be need
to include these hybrid operator diagrams.
Since the suppression by $v$ leaves these terms larger than other $\Lambda/m_b$
power corrections it is quite feasible that long-distance charm contributions
are  numerically relevant.

Finally, we explain why in general we expect these long distance charm
contributions to be complex.  As was shown in Ref.~\cite{Mantry:2003uz} a
sufficient condition for the generation of a complex phase is the presence of
soft Wilson lines in both the $n$ and $\bn$ directions. In most observables
these Wilson lines cancel, however, it is clear that this will not be the case
for the long-distance charm contribution. The underlying reason for the lack of
such a cancellation is the fact that the charm quark propagate over long
distances. Thus when we rescale the light quark fields
\begin{eqnarray}
q_n(x) \rightarrow Y_n(x) q_n(x)\,,  \qquad   q_\bn(y) \rightarrow Y_\bn(y) q_\bn(y)\,,
\end{eqnarray}
the argument of the Wilson lines will be at different positions. Furthermore in
general the charmed quarks will not decouple from the $B$ meson, thus the soft
matrix element will be of the form
\begin{eqnarray}
 && \big\langle 0 \big| [\eta^\dagger\chi] (0) [\chi^\dagger   \eta](y)
  [ \bar q Y_n](z^-) [Y^\dagger_n b_v](0)\nn \\
  &&\quad  Y_n(y) Y^\dagger_n(z^-)Y_\bn(0)Y^\dagger_\bn(y) \big| B \big\rangle,
\end{eqnarray}
where the spin contractions and the color contractions of the fields and Wilson
lines have been suppressed.  Note that since the charm quark pair propagates
over long distances the point $y^\mu$ is displaced away from zero along the
light cone as well as transverse directions. Since this matrix element for
long-distance charm knows about the two final state hadrons,
it can have a nonperturbative complex phase.


\section{Penguin Annihilation} \label{sect:ann}

In this section we review the penguin annihilation contributions occurring on the
third lines of Eqs.~(\ref{hatP}) and (\ref{hatPT}). For our purposes they are
defined as the $\lambda_c^{(f)}$ terms in the amplitude obtained when the
spectator quark is annihilated by the weak operator. These contributions start
at ${\cal O}(\Lambda/m_b)$ relative to the leading power penguin terms. They
include the well known terms $f_B \phi^M\phi^{M'}$ from spectator annihilation
with a subsequent pair creation~\cite{Keum,BBNS2}, as well as terms of the same
parametric size $f_B \phi_B^+ \phi^{3M} \phi^{M'}$ where the spectator emits an
energetic collinear gluon prior to its annihilation~\cite{Arnesen:2006dc}. The
former require zerobin subtractions to obtain finite
convolutions~\cite{Arnesen:2006vb}, while the latter do not.  With these
subtractions the leading penguin annihilation contribution to the amplitude is
real. The scheme dependence of the zerobin procedure is compensated by terms
involving the exchange of a soft quark in the annihilation process,
$A_{Tann}^{(1)}$, which come from time-ordered products in \SCETa. These
time-ordered product terms start at order $\alpha_s^2/m_b$ and have a
nonperturbative strong rescattering phase.  At ${\cal O}(\Lambda^2/m_b^2)$ one
also has chiral-enhanced penguin annihilation terms $f_B \mu_M
\phi_{pp}^M\phi^{M'}$~\cite{BBNS2} which can also be factorized with zerobin
subtractions as in Ref.~\cite{Arnesen:2006dc}.

The annihilation terms $f_B \phi^M\phi^{M'}$ and $f_B \mu_M
\phi_{pp}^M\phi^{M'}$ do not involve a hard-collinear propagator and so appear
to be insensitive to the intermediate scale $\mu_i\simeq \sqrt{m_b\Lambda}$.
However the zero-bin subtraction procedure is needed to distinguish soft and
collinear regions, and so they are not defined independent of $A_{Tann}^{(1)}$
at ${\cal O}(\alpha_s^2(\mu_i))$. Since this procedure has an $\alpha_s(\mu_i)$
expansion we consider all penguin annihilation contributions with an expansion
at the intermediate scale, unlike our analysis of amplitudes in earlier
sections. This increases the theoretical uncertainty, and will be accounted for
in our error analysis.
 
From Refs.~\cite{Arnesen:2006vb} and~\cite{Arnesen:2006dc} the penguin
annihilation amplitudes at ${\cal O}(\alpha_s)$ for $B\to M_1M_2$ are
\begin{align}
 \hat A_{Lann}^{(1)} &=  - \frac{\hat f_B f_{M_1} f_{M_2}}{m_B^2} 
   \big\langle H_c^{M_1M_2} \big\rangle
   \,, \\
  \hat A_{hcann}^{(1)} &= - \frac{\hat f_B \beta_B}{m_B m_b}\Big[
     f_{3M_1} f_{M_2} \big\langle H_{hc1}^{M_1M_2}\big\rangle
   \nn\\
    &\qquad
     +\eta_{M_1} f_{M_1} f_{3M_2} \big\langle H_{hc1}^{M_1M_2}\big\rangle
  \Big] \,,
    \nn\\
   \hat A_{Lann}^{(2\chi)} &=  - \frac{\hat f_B f_{M_1} f_{M_2}}{m_B^2}  \Big[
   \frac{\mu_{M_1}}{m_b} \big\langle H_{\chi 1}^{M_1M_2}\big\rangle
  \plus \frac{\mu_{M_2}}{m_b} \big\langle H_{\chi 2}^{M_1M_2}\big\rangle
  \Big]   
   . \nn
\end{align}
Here we have the inverse moment of the $B$ light-cone distribution
\begin{align}
 \beta_B &= \int_0^\infty\! \frac{dk}{3k}\: \phi_B^+(k) = \frac{ \lambda_{B}^{-1} }{3}
    \,, 
\end{align}
and the five factors $\langle H_c^{M_1M_2} \rangle$, $\langle
H_{hc1}^{M_1M_2}\rangle$, $\langle H_{hc1}^{M_1M_2}\rangle$, $\langle H_{\chi
  1}^{M_1M_2}\rangle$, and $\langle H_{\chi 2}^{M_1M_2}\rangle$ are linear
combinations of the moment parameters
\begin{align} \label{betadef}
 \beta_{ic}^{M_1M_2} &= \ o\hspace{-0.32cm} \int_0^1\!\! dx dy\:
    (a_{ic}\plus a_{i+4})(x,y)\, \phi^{M_1}(y) \phi^{M_2}(x)
    \,,\nn\\
 \beta_{hc1,3}^{M_1 M_2} &= \int_0^1\!\! dxdyd\bar y\: 
   \frac{a_{1,3}^{hc}(x,y,\bar y)}{1\minus y\minus \bar y} \phi^{3M_1}(y,\bar y)\phi^{M_2}(x)
   \,,\nn \\
 \beta_{hc2,4}^{M_1 M_2} &= \int_0^1\!\! dxd\bar xd y\: 
   \frac{a_{2,4}^{hc}(x,\bar x, y)}{1\minus x\minus \bar x}
   \phi^{M_1}(y)\phi^{3M_2}(x,\bar x)
   \,,\nn \\
 \beta_{\chi 1,5}^{M_1M_2} &=
    \frac{1}{6}\ o\hspace{-0.32cm} \int_0^1\!\! dx dy\:
     a_{1,5}^{\chi} (x,y)\: \phi_{pp}^{M_1}(y) \phi^{M_2}(x)
    \,,\nn \\
  \beta_{\chi 2,6}^{M_1M_2} &=
    \frac{1}{6}\ o\hspace{-0.32cm} \int_0^1\!\! dx dy\:
      a_{2,6}^{\chi} (x,y)\: \phi^{M_1}(y) \phi_{pp}^{M_2}(x) 
   \,,
\end{align}
respectively. The appropriate linear combinations for each channel are given by
the entries in Tables~II, III, IV, and V of Ref.~\cite{Arnesen:2006vb} for $\hat
A_{Lann}^{(1)}$ and $\hat A_{Lann}^{(2\chi)}$, and in Table~I of
Ref.~\cite{Arnesen:2006dc} for $\hat A_{hcann}^{(1)}$. The distribution
functions $\phi^{M}$ and $\phi_{pp}^M$ appearing in Eq.~(\ref{betadef}) were
defined above in Eqs.~(\ref{zeta}) and (\ref{phipp}), while the three-body
distribution is defined by the matrix element
\begin{align}
  \big\langle M \big| \bar q^{\prime L}_{n\omega_1} \bnslash\,
  i g\, \slash\!\!\!\!{\mathcal B}^\perp_{n\omega_3}  q^R_{n\omega_2}
   \big| 0 \big\rangle 
&  =  \frac{i {\cal C}_{q'_Lq}^M}{\omega_3} 
   \bar\delta_{\omega_1\omega_3}^{\omega_2}\, 
   f_{3P}\, \phi_{3P}(y,\bar y) , 
\end{align}
where
$\bar\delta_{\omega_1\omega_3}^{\omega_2}=\delta(\omega_1+\omega_3-\omega_2-m_b)$,
$y=\omega_1/m_b$, and $\bar y=-\omega_2/m_b$. The circle on some of the
integrations in Eq.~(\ref{betadef}) indicates the terms which require zero-bin
subtractions. These subtractions modify the hadronic distributions by inducing
dependence on a rapidity parameter, which increases the uncertainty from these
terms. For our numerical analysis we adopt the models used in
Refs.~\cite{Arnesen:2006vb} and~\cite{Arnesen:2006dc} to determine the
$\beta$'s in section~\ref{sect:inputs} below.


\section{Analysis strategy and models for the shape of
$\zeta_J^{BM}(z)$ and $\zeta_{\chi}^{BM}(z)$ }
\label{sect:models}

To make predictions for the $P^{M_1M_2}$ penguins at leading order we need
values for the twist-2 meson distribution $\phi_M(u)$ and the form factors
$\zeta^{BM}$, $\zeta^{BM}_J(z)$. To compute the $\Lambda/m_b$ suppressed
chiral-enhanced amplitudes we need in addition the twist-3 distribution
$\phi_{pp}^M(u)$ and the form factor $\zeta_\chi^{BM}(z)$.

A common model for $\phi^M$ and $\phi_{pp}^M$ capturing the essential features
is given by the first few terms in the Gegenbauer series
\begin{align}
\label{phiM}
\phi^M(x) &= 6 x(\!1\!-\!x\!) \big\{1 + a_{1}^{M}(\!6x\!-\!3)
 + 6 a_{2}^{M}(\!1\!-\!5 x \!+\!5 x^{2})\!\} \nn \\
 &\quad + 15 a_4^M (1\minus 14 x\plus 56 x^2\minus 84 x^3\plus 42 x^4) \big\},
  \nn\\
\phi_{pp}^M(x) &=  6 x(\!1\!-\!x\!) \big\{a_{0pp}^M + a_{1pp}^{M}(\!6x\!-\!3)
  \nn\\
 &\quad  + 6 a_{2pp}^{M}(\!1\!-\!5 x\! +\!5 x^{2}) \big\} \,,
\end{align}
where $x$ corresponds to the momentum fraction for the quark field (dressed by a
Wilson line).  Our ranges for the model parameters $a_i^M$ and $a_{ipp}^M$ are
summarized in the next section. Note that we include both $a_2^M$ and $a_4^M$ in
$\phi^M(x)$. This allows us to account for data on $\langle x^{-1}\rangle_\pi$
which constrains $a_2^\pi+a_4^\pi$, while varying $a_2^\pi-a_4^\pi$ to obtain a
range of models. From charge conjugation and isospin,
$\phi^{\pi,\rho}(x)=\phi^{\pi,\rho}(1-x)$. Thus we will set
$a_1^\pi=a_1^\rho=0$. With SU(3) flavor symmetry $a_1^K=0$, so smaller values
are adopted for this parameter than are used for $a_2^K$ (we keep a non-zero
$a_1^M$ for SU(3) violation, but do not include a non-zero $a_3^M$ or any other
odd-moment parameter).  These restrictions from charge conjugation also apply to
the distribution $\phi_{pp}^M$. To see this we follow the same argument given in
Ref.~\cite{Bauer:2002nz} but for the matrix element defining the chiral-enhanced
distribution:
\begin{align} \label{PSunderC}
 & \big\langle \pi^0 \big| C^\dagger C \bar{\xi}_{n} W_n  C^\dagger C \bnslash\gamma_5
  \cPslash_\perp 
  \delta(\omega-\bnP_+ ) W_n^\dagger \xi_{n} C^\dagger C \big| 0 \big\rangle \nn\\
 & 
  =  (+1) \big\langle \pi^0 \big| ({\cal C}W_n^\dagger\xi_{n})^T 
  \bnslash\gamma_5 \cPslash_\perp \delta(\omega-\bnP_+) (\bar\xi_{n} W_n{\cal C})^T 
  \big| 0 \big\rangle\nonumber\\
 & 
  = + \big\langle \pi^0 \big| \bar{\xi}_{n,p'} W_n 
  \bnslash\gamma_5 \cPslash_\perp \delta(\omega+ \bnP_+ ) W_n^\dagger \xi_{n,p} \big| 0
  \big\rangle\,.
\end{align}
Here $\bar \cP_+=\bar\cP^\dagger +\bar\cP$, $C$ is the charge-conjugation
operator and ${\cal C}$ is the charge-conjugation matrix. To obtain the last
line we note that the sign from $\cP_\perp^\alpha \bar\xi_n = - \bar\xi_n
\cP_\perp^{\alpha\dagger}$ cancels the sign from ${\cal C}^\dagger
(\gamma_\alpha^\perp)^T {\cal C}=\gamma_\mu^\perp$. Hence the matrix element is
even under $\omega=(1-2x)\bn\cdot p_\pi\to -\omega$, which implies
$\phi_{pp}^{\pi^0}(x)=\phi_{pp}^{\pi^0}(1-x)$. Hence we have
$a_{1pp}^\pi=a_{1pp}^\rho=0$ and small values of $a_{1pp}^K$. Note that in Wandzura-Wilczek approximation $a_{0pp}^M=1$ and $a_{1,2 \, pp}^M=0$.

Potentially larger uncertainty comes from the values for $\zeta^{BM}$ and
$\zeta^{BM}_J(z)$.  At lowest order the $b_i(u,z)$ coefficients in
Eq.~(\ref{newfact}) are independent of $z$ and thus only $\zeta^{BM}$ and the
zeroth $z$-moment, $\zeta_J^{BM}$, from Eq.~(\ref{zJnorm}) are required.  This
yields two form factor parameters to be fit to tree amplitude data. For
convolutions with the $\alpha_s$ $\Delta b_i(u,z)$ terms we need more
information about the $z$-dependence. However it is still very useful to fit the
norm $\zeta_J^{BM}$ to the nonleptonic tree amplitude data. In particular, by
only modeling the shape of $\zeta_J(z)$ we reduce the model uncertainty
considerably. The fit to nonleptonic decay data currently provides the most
accurate way of determining the normalization of $\zeta_J^{BM}(z)$.  Thus, our
strategy allows us to make predictions for the penguin amplitudes based on input
about the parameters from the tree amplitudes, while avoiding expanding in
$\alpha_s(\mu_i)$.

We adopt a polynomial model for the $z$-dependence by using the
parameterization
\begin{align}\label{zJBM}
 &\zeta_J^{B M}(z) \\
&=  \, z \{ A_{0}^{BM} + A_{1}^{BM} (6z-3) + A_{2}^{BM}(1-5z+5z^{2}) \} .\nn
\end{align}
One of these parameters is then eliminated by constraining
$\zeta_{J}^{B M}=\int dz \, \zeta_{J}^{B M} (z)$ to its central value
obtained from experiment. We then eliminate $A_{0}^{B\pi}$ in favor of
$\zeta_J^{B\pi}$ to obtain
\begin{align} \label{zJ1}
\zeta_J^{B M}(x) &=  2x\, \zeta_J^{B M}  - A_{1}^{B M}(4x -6 x^2 )
  \nn\\
 &\quad + \frac{5}{6} A_2^{B M} (x - 6 x^2 + 6 x^3) \,.
\end{align}
Note that the remaining $A_i^{BM}$ terms in Eq.~(\ref{zJ1}) must
integrate to zero. As we will see in the next section, this
considerably reduces the uncertainty generated by these form factor
parameters. For $M=\pi$ we will simply set $A_{1}^{B\pi}=0$, thus
leaving $A_{2}^{B\pi}$ as the remaining parameter.\footnote{The choice
$A_{1}^{B\pi}=0$ can be justified by isospin and tree-level \SCETb
factorization, but we instead view this choice as part of the model. The
value of $A_1^{BK}$ then parameterizes SU(3) violation.}

The polynomial form in Eq.~(\ref{zJBM}) could be justified by the \SCETb
factorization theorem in Eq.~(\ref{zetaJfactor}), where it is inherited from
that in Eq.~(\ref{phiM}) at lowest order.  However we do not view our model in
this context, and thus do not fix the coefficients $A_i^M$ to values determined
by $a_i^M$. Instead we consider Eq.~(\ref{zJ1}) as a model specified in \SCETa
without reference to \SCETb-factorization, and take $A_i^M$ as parameters to be
varied in a suitably large range.  This ensures that our model for
$\zeta^{BM}_J$ covers a wider range of $z$-dependence than the restrictive
approximation in Eq.~(\ref{zetatree}) would.

Similar to $\zeta^{BM}_J(z)$ we write a model for $\zeta^{BM}_{\chi}(z)$ as
\begin{eqnarray}
\zeta_{\chi}^{B\pi}(x) &=&  
  2x \zeta_{\chi}^{B\pi} - A_{\chi 1}^{B\pi}(4x -6 x^2 ) \nn\\
 && + \frac{5}{6} A_{\chi 2}^{B\pi} (x - 6 x^2 + 6 x^3) \,, 
\end{eqnarray}  
We have taken $\zeta_{\chi}^{BM}(0)=0$ due to the constraint on this function
derived in section~\ref{sect:chiral}.  

For $M=\rho$, the simple polynomial model of Eq.~(\ref{phiM}) does not support the
value of $\langle x^{-1} \rangle_\rho = 2.2^{+0.6}_{-0.2}$ obtained from data in
section~\ref{sect:zeta}, unless we include  higher order polynomial terms
in Gegenbauer expansion. Values of $\langle x^{-1} \rangle_M$ close to $2.0$
require $\phi^M(x)$ to peak around $x=1/2$ with smaller widths. Therefore we
choose the following model for $\phi^\rho(x)$, which has all the desired properties
\begin{equation}
\label{phirho}
\phi^\rho(x) = N(a_\rho)\,x(1-x)\,\text{sech}\left ( \frac{x-1/2}{a_\rho} \right ) \, .
\end{equation}
Here $a^\rho$ is a parameter whose value is motivated by the inverse moment
$\langle x^{-1} \rangle_\rho$ determined from data in section~\ref{sect:zeta},
and $N(a^\rho)$ is chosen to normalize $\phi^\rho(x)$ to $1$. For
$\zeta_J^{B\rho}$ we will use a polynomial model like Eq.~(\ref{zJ1})
\begin{equation}
\label{zJ3}
\zeta_J^{B\rho}(x) =  2x\, \zeta_J^{B\rho} + \frac{5}{6} A_2^{B\rho} (x - 6 x^2 + 6 x^3) \,,
\end{equation}
where for simplicity we take $A_1^{B\rho}=0$.  Alternatively we could
have based our model for $\zeta_J^{B\rho}(x)$ on Eqs.~(\ref{phirho})
and~(\ref{zetatree}) where it would inherit features of the ${\rm
sech}$ function, however we find that using this alternative
functional form does not significantly change our error analysis.
Therefore we adhere to the simple polynomial model of Eq.~(\ref{zJ3}).
Numerical estimates for the model parameters introduced in this
section are presented in the next section.


\section{Input Parameters} \label{sect:inputs}

Several well determined parameters that are needed for our analysis
include~\cite{Yao:2006px} $m_b^{1S}=4.7\,{\rm GeV}$, $\bar m_b(4.7\,{\rm
  GeV})=4.1\,{\rm GeV}$, $m_c^{1S}=1.4\,{\rm GeV}$, $\alpha_s(m_b)=0.22$,
$\mu_\pi(m_b)=2.5\,{\rm GeV}$, and $\mu_K(m_b)=2.8\,{\rm GeV}$. Defining $\hat
f_M=f_M/(1\,{\rm GeV})$ we take $\hat f_\pi=0.131$, $\hat f_K=0.160$, $\hat
f_\rho = 0.209$, and from recent lattice data~\cite{Gray:2005ad} $\hat
f_B=0.22$. We also require the Wilson coefficients of the weak effective
Hamiltonian, which are known at NLL order~\cite{fullWilson}. In the NDR scheme
taking $\alpha_s(m_Z)=0.118$, $m_t=170.9\,{\rm GeV}$, and $m_b=4.7\,{\rm GeV}$
gives $C_{7\gamma}(m_b)=-.316$, $C_{8g}(m_b) =-0.149$ and the NLL results
%
\begin{align} \label{CiNLL}
 &
C_{1-10}(m_b) = \{
  1.080\,, 
  -.179\,,
  .012\,,
 -.033\,, 
  .0096\,, 
 -.040 \,, 
  \nn\\
 & \ \ 
  4.2 \!\times\! 10^{-4} \,,
  4.2 \!\times\! 10^{-4} \,,
  -9.7 \!\times\! 10^{-3} \,,
  1.9 \!\times\! 10^{-3} \} \,.
\end{align}
In varying $\mu$ to estimate uncertainties we will also need
\begin{align}
 &
C_{1-10}(2m_b) = \{
  1.04\,, 
  -.104\,,
  .0080\,,
 -.023\,, 
  .0074\,, 
 -.026 \,, 
  \nn\\
 & \ \ 
  4.2 \!\times\! 10^{-4} \,,
  2.8 \!\times\! 10^{-4} \,,
  -9.3 \!\times\! 10^{-3} \,,
  1.3 \!\times\! 10^{-3} \} \,,\nn\\
 &\quad  C_{7\gamma}(2m_b) = -0.281\,,\quad
 C_{8G}(2m_b) = -0.135
  \,,\nn\\[5pt]
 &
C_{1-10}(m_b/2) = \{
  1.13\,, 
  -.279\,,
  .019\,,
 -.047\,, 
  .012\,, 
 -.061 \,, 
  \nn\\
 & \ \ 
  5.7 \!\times\! 10^{-4} \,,
  7.1 \!\times\! 10^{-4} \,,
  -10.0 \!\times\! 10^{-3} \,,
  2.8 \!\times\! 10^{-3} \} \,, \nn\\
 &\quad  C_{7\gamma}(m_b/2) = -0.358\,,\quad
 C_{8G}(m_b/2) = -0.166 \,.
\end{align}
With 2-loop running the $\overline {\rm MS}$ mass $\overline{m_b}(2m_b)=3.7\,{\rm
GeV}$, $\overline{m_b}(m_b/2)=4.7\,{\rm GeV}$, and the
chiral-enhancement parameters $\mu_\pi(2m_b)=2.8\,{\rm GeV}$,
$\mu_\pi(m_b/2)=2.2\,{\rm GeV}$, $\mu_K(2m_b)=3.1\,{\rm GeV}$, and
$\mu_K(m_b/2)=2.5\,{\rm GeV}$.

The $\gamma^*\gamma\to \pi^0$ data constrains the inverse pion moment,
and based on the analysis in
Ref.~\cite{Bakulev:2003cs} gives
\begin{align}
 a_2^\pi + a_4^\pi = -0.03 \pm 0.14 \,.
\end{align} 
For the other linear combination we take $a_2^\pi - a_4^\pi=0.2\pm 0.3$. In our
error analysis we do a Gaussian scan over these ranges in order to properly take
into account the correlation in the individual errors of $a_2^\pi$ and
$a_4^\pi$, which is large. Based on recent lattice data for moments of the $\pi$
and $K$ distributions \cite{Braun:2006dg} we take $a_2^{K}=0.2\pm 0.2$ and set
$a_4^K=0$. Here the lattice error on $a_2^K$ was doubled to give an estimate for
higher moments. For $M=\pi$ isospin and charge conjugation imply
$a_1^\pi=a_{1pp}^\pi=0$, while for $M=K$ we use \cite{Braun:2006dg} $a_1^K=
-0.05 \pm 0.02$. For simplicity we take $a_{0pp}^{\pi , K}=1$. We also take $a_{2pp}^{\pi,K}=0.1 \pm 0.3$ and
$a_{1pp}^K=0.0\pm 0.1$. For our model of $\phi^\rho(x)$ in Eq.~(\ref{phirho}) we
use $a^\rho=0.1^{+0.3}_{-0.1}$.

In section~\ref{sect:zeta} we obtained values for the nonleptonic 
 form factors $\zeta^{B\pi}$, $\zeta_J^{B\pi}$, $\zeta^{B\rho}$,
$\zeta_J^{B\rho}$, from a fit to nonleptonic data for the tree
amplitudes. Because the uncertainty in these parameters are highly
correlated we scan over their values by doing a Gaussian scan over the
range specified by the experimental errors in Eqs.~(\ref{flzJ}) and
(\ref{flzJrho}) and for the form factors and $\langle x^{-1}\rangle_M
\zeta_J^{BM}$ with $M=\pi,\rho$. Since data is being used for these
normalization parameters this does not introduce model uncertainty.
The choice of the remaining parameters $A_i$ introduces model
dependence to $\zeta_J(z)$.  We take $A_{1}^{\pi,\rho}=0 $,
$A_{2}^{\pi}=(0.25 \pm 0.30)$ and $A_2^{\rho}=-(0.05 \pm 0.05)$.  We
also will use $\zeta_{\chi}^{B\pi} = 0.0 \pm 0.2$ and set $A_{\chi
1}^{\pi}=0$, and $A_{\chi 2}^{\pi}=(0.0 \pm 0.5)$. Note that to
predict the $\hat P^{K\pi}$ amplitudes we do not need values of
$\zeta^{BK}$ and $\zeta_J^{BK}$, since only pion form factors appear
in the $B\to K\pi$ amplitudes. The kaon form factors are needed for
$B\to K\bar K$.

Finally we will need values for the model parameters appearing in the
annihilation amplitudes in section~\ref{sect:inputs}. The three-body
decay constants are taken as \( f_{3\pi} \simeq 4.5 \times 10^{-3}\,
{\rm GeV}^2\), \(f_{3K} \simeq 4.5 \times 10^{-3}\, {\rm GeV}^2\), and
\(f_{3\rho} \simeq 0.13 \, {\rm GeV}^2\) from recent QCD sum rule 
results~\cite{Ball:2006wn,Ball:2007rt}, where $f_{3\rho}=m_\rho
f_\rho^T$.  For the $B$-meson inverse moment appearing in the the
three-body annihilation amplitude, $\hat A_{hcann}^{(1)}$, we take
$\beta_B = (2.5\pm 1.0) {\rm GeV}^{-1}$, where the central value is
consistent with our value for $\zeta_J^{B\pi}$ using
Eq.~(\ref{zetatree}) and (\ref{phiM}), and the error takes into
account the uncertainty from the $\alpha_s(\mu_i)$ expansion.  For the
remaining ingredients we simply quote results for the necessary
moments at $\mu=m_b$
\begin{widetext}
\begin{align} \label{betas}
\beta_{1c}^{\pi\pi} &= (-3.0\pm 1.6)\times 10^{-2}
   \,, 
 & \beta_{3c}^{\pi\pi} &= 0.63 \mpm 0.32
   \,,
 & \beta_{4c}^{\pi\pi} &=  -0.15 \mpm 0.09
   \,,  \\
\beta_{hc 1}^{\pi\pi} &= -1.32 \pm 0.42 
   \,, 
 & \beta_{hc 2}^{\pi\pi} &= 0.13 \pm 0.12 
   \,,
 & \beta_{hc 3}^{\pi\pi} &= (-2.4\pm 2.2)\times 10^{-3} 
   \,,
 & \beta_{hc 4}^{\pi\pi} &= (-4.9\pm 1.6)\times 10^{-2} 
   \,,\nn\\
 \beta_{\chi 1}^{\pi\pi} &=  0.0 \pm 5.1
   \,, 
 & \beta_{\chi 2}^{\pi\pi} &= 0.0 \pm 4.7 
   \,,
 & \beta_{\chi 5}^{\pi\pi} &= 0.0 \pm 0.067
   \,,
 & \beta_{\chi 6}^{\pi\pi} &= 0.0 \pm 0.084
   \,,\nn\\
\beta_{4c}^{\pi K} &= -0.159 \pm 0.087
   \,, \nn\\
\beta_{hc 1}^{\pi K} &= -1.37 \pm 0.44 
   \,, 
 & \beta_{hc 2}^{\pi K} &= 0.13 \pm 0.12 
   \,,
 & \beta_{hc 3}^{\pi K} &= (-2.3\pm 2.3) \times 10^{-3}  
   \,,
 & \beta_{hc 4}^{\pi K} &= (-4.9\pm 1.5) \times 10^{-2}  
   \,,\nn\\
 \beta_{\chi 1}^{\pi K} &= 0.0 \pm 6.4 
   \,, 
 & \beta_{\chi 2}^{\pi K} &= 0.0 \pm 5.8
   \,,
 & \beta_{\chi 5}^{\pi K} &= 0.0 \pm 0.085
   \,,
 & \beta_{\chi 6}^{\pi K} &= 0.0 \pm 0.10 
   \,,\nn\\
\beta_{1c}^{\rho\rho} &= (5.1 {}^{+4.2}_{-1.4})\times 10^{-3}
   \,, 
 & \beta_{3c}^{\rho\rho} &= -0.11 {}^{+.09}_{-0.03}
   \,,
 & \beta_{4c}^{\rho\rho} &=  (2.5 {}^{+2.1}_{-0.7})\times 10^{-2}
   \,, \nn\\
\beta_{hc 1}^{\rho\rho} &= (-3.9 {}^{+3.4}_{-3.0})\times 10^{-2}
   \,, 
 & \beta_{hc 2}^{\rho\rho} &= (-1.7 {}^{+2.9}_{-1.6})\times 10^{-3}
   \,,
 & \beta_{hc 3}^{\rho\rho} &= (-1.6 \pm 1.2)\times 10^{-4}
   \,,
 & \beta_{hc 4}^{\rho\rho} &= (1.5 {}^{+1.3}_{-1.1})\times 10^{-3}
   \,. \nn
\end{align}
\end{widetext}
The values are computed as in Refs.~\cite{Arnesen:2006vb}
and~\cite{Arnesen:2006dc} with inputs for $C_i$, $\mu_\pi$, and $\mu_K$
consistent with those given above. For the case of $\beta_{hc i}^{\rho\rho}$ we
used $\phi_{3\rho}(x,\bar x)= 360 x\bar x (1-\bar x-x)^2 (x-\bar x) w_3^\rho$,
where $w_{3}^\rho =-0.20\pm 0.15$ is taken from QCD sum-rules~\cite{Ball-Zwicky}
with an inflated error to account for higher Gegenbauer terms (the relation
between our notation and theirs is $\phi_{3\rho}=-\Phi_{3;\rho}^\perp/2$). Note
that our central value of $0.0$ for the $\beta_{\chi i}$ terms in
Eq.~(\ref{betas}) indicates that we do not have information on the sign of these
terms. Results for the $\beta$'s at $\mu=m_b/2$ and $\mu=2m_b$ are quoted in
appendix~\ref{appC}.


\section{Numerical Analysis} \label{sect:analysis}

In this section we make predictions for the penguin amplitudes $\hat P_{M_1
  M_2}$ in the standard model, focusing on $\pi^+\pi^-$, $K^+\pi^-$, and
$\rho^+\rho^-$ final states. Our sign convention for the penguin amplitudes
was given in Eqs.~(\ref{A}) and (\ref{Psign}).  To facilitate comparing the size
of various contributions we introduce the notation
\begin{align} \label{PP}
\hat P_{M M'} &=  \big (\hat P_{M M'}^\zeta+\hat P_{M M'}^{\zeta_J} \big )\plus
 \big (\hat P_{M M'}^{\chi\zeta}+\hat P_{M M'}^{\chi\zeta_J} + 
  \hat P_{M M'}^{\chi\zeta_{\chi}} \big ) 
  \nn\\
& + \hat P_{c\bar c} 
 + \big(\hat P^{Lann}_{MM'} + \hat P^{Gann}_{MM'} + \hat P^{Lann\chi}_{MM'} \big)  \,.
\end{align}
The two terms in the first parentheses correspond to the leading power terms in
line 1 of Eq.~(\ref{hatP}), the second parentheses to the chiraly enhanced terms
in line 2, and $\hat P_{c\bar c}$ corresponds to the long-distance charm penguin
in line 3. In the last parentheses the first two are LO annihilation terms from
local annihilation and hard-collinear annihilation respectively, while the term
$\hat P^{Lann\chi}_{MM'}$ stands for chiral-enhanced annihilation.

The leading power terms can be written as moments over the distribution
functions
\begin{eqnarray}
\label{Pamplitude}
\hat P^{\zeta}_{\pi \pi} & = & - \hat f_\pi \big\langle
   c_{1}^{c}+c_{4}^c \big\rangle_\pi \zeta^{B \pi} \,, \nn \\ 
\hat P^{\zeta_J}_{\pi \pi} &=& - \hat f_\pi \big\langle 
   (b_{1}^{c}+b_{4}^c )\, \zeta_J^{B \pi} \big\rangle_\pi
   \,, \nonumber \\
\hat P^{\zeta}_{K \pi} & = & - \hat f_K \big\langle 
   c_{1}^{c}+c_{4}^c \big \rangle_K \zeta^{B \pi} 
   \,,\nn \\ 
  \hat P^{\zeta_J}_{K \pi} & = & - \hat f_K  \big\langle 
   (b_{1}^{c}+b_{4}^c ) \zeta_J^{B \pi} \big\rangle_K  \,,
   \nonumber\\
\hat P^{\zeta}_{\rho\rho} & = & - \hat f_\rho \big\langle 
   c_{1}^{c}+c_{4}^c \big \rangle_\rho \zeta^{B \rho} 
   \,,\nn \\ 
  \hat P^{\zeta_J}_{\rho} & = & - \hat f_\rho  \big\langle 
   (b_{1}^{c}+b_{4}^c ) \zeta_J^{B \rho} \big\rangle_\rho  \,,
\end{eqnarray}
where $\hat f_\pi=f_\pi/(1\,{\rm GeV})$ and $\hat f_K=f_K/(1\,{\rm GeV})$. In an
analogous fashion we can define moments for the chiraly enhanced penguin
amplitudes. For $B\to \pi\pi$ and $B\to K\pi$ we obtain from
Eq.~(\ref{chienhanced}) and Table~\ref{table2a}
\begin{align}
\hat P^{\chi\zeta}_{\pi \pi} & =  \frac{\hat f_\pi \mu_\pi}{3 m_B}\,
 \zeta^{B \pi} \Big\langle  c_{1(qfq)}^{\chi c} \plus c_{2(qfq)}^{\chi c}
  \Big\rangle^{pp}_\pi 
  \,, \nn \\ 
\hat P^{\chi\zeta_J}_{\pi \pi} & =  \frac{\hat f_\pi \mu_\pi}{3 m_B}\!
\Big \langle \Big[ b_{3(qfq)}^{\chi c} \plus b_{4(qfq)}^{\chi c}\Big] \,\zeta_J^{B \pi}
\Big \rangle^{pp}_\pi
  \,, \nn \\ 
\hat P^{\chi\zeta_{\chi}}_{\pi \pi} & =  \frac{\hat f_\pi \mu_\pi}{6 m_B}\!
\Big \langle \Big[ b_{1(qfq)}^{\chi c} \plus b_{1(ufu)}^{\chi c}
\Big] \, \zeta_{\chi}^{B \pi} \Big\rangle_\pi
  \,, \nn \\ 
\hat P^{\chi\zeta}_{K \pi} & =  \frac{\hat f_K \mu_K}{3 m_B}\,
\zeta^{B \pi}\Big \langle c_{1(qfq)}^{\chi c} +
 c_{2(qfq)}^{\chi c} \Big \rangle^{pp}_K \,, \nn \\ 
\hat P^{\chi\zeta_J}_{K \pi} & =  \frac{\hat f_K \mu_K}{3 m_B}
\Big \langle \Big[ b_{3(qfq)}^{\chi c} +
b_{4(qfq)}^{\chi c} \Big ] \zeta_J^{B \pi} \Big
\rangle^{pp}_K
  \,, \nn \\ 
\hat P^{\chi\zeta_{\chi}}_{K \pi} & =  \frac{\hat f_K \mu_\pi}{6 m_B}\!
\Big \langle \Big[ b_{1(qfq)}^{\chi c} \plus b_{1(ufu)}^{\chi c}
\Big] \,\zeta_{\chi}^{B \pi}\!\! \Big
\rangle_K
  \,.
%
%
%
\label{Pchiamplitude}
\end{align}
In Eq.~(\ref{Pamplitude}) we have decomposed the leading Wilson coefficients 
into terms proportional to the two CKM structures,
\begin{align}
c_i^{(f)} & =  \lambda_u^{(f)} c_{i}^u +\lambda_t^{(f)} c_{i}^t 
   = \lambda_u^{(f)} \tilde c_{i}^u  + \lambda_c^{(f)} c_{i}^c , \nonumber \\
b_i^{(f)} & =  \lambda_u^{(f)} b_{i}^u +\lambda_t^{(f)} b_{i}^t 
    = \lambda_u^{(f)} \tilde b_i^u + \lambda_c^{(f)} b_{i}^c ,
\end{align}
where some coefficients (such as $c_1^c$ and $b_1^c$) are purely from
electroweak penguins~\cite{Bauer:2004ck}.  Similarly we split the Wilson
coefficients $c_{i(F)}^{\chi}$ and $b_{i(F)}^{\chi}$ for the chiraly enhanced
amplitudes in Eqs.~(\ref{chiralcoefficientsc}-\ref{chiralcoefficientsb2}) as
\begin{eqnarray}
c_{i(F)}^{\chi} &=& \lambda_u^{(f)}\, c_{i(F)}^{\chi u} + \lambda_c^{(f)}\, c_{i(F)}^{\chi c} 
  \, ,\nn \\
b_{i(F)}^{\chi} &=& \lambda_u^{(f)}\, b_{i(F)}^{\chi u} + \lambda_c^{(f)}\, b_{i(F)}^{\chi c} 
  \, .
\end{eqnarray}
The moments appearing in Eqs.~(\ref{Pamplitude}) and (\ref{Pchiamplitude}) are 
\begin{align}
\label{average}
\langle c_i \rangle_M & =
   \int_0^1\!\!\! du \:  c_i(u) \phi^M(u) \,,  \\
\langle b_i \zeta_J^{B M_2} \rangle_{M_1} & = 
     \int_0^1\!\!\! du\! \int_0^1\!\!\! dz \: b_i(u,z) \phi^{M_1}(u) \zeta_J^{BM_2}(z)\,, 
   \nonumber \\
\langle c_i \rangle_M^{pp} & =  
    \int_0^1\!\!\! du \:  c_i(u) \phi^M_{pp}(u) \,,  \nn \\
\langle b_i \zeta_J^{B M_2} \rangle_{M_1}^{pp} & =  
    \int_0^1\!\!\! du\! \int_0^1\!\!\! dz \: b_i(u,z)
\phi_{pp}^{M_1}(u) \zeta_J^{BM_2}(z) \,,
   \nonumber \\
\langle b_i \zeta_{\chi}^{B M_2} \rangle_{M_1} & =
     \int_0^1\!\!\! du\! \int_0^1\!\!\! dz \: b_i(u,z) \phi^{M_1}(u) \zeta_{\chi}^{BM_2}(z).
   \nonumber 
\end{align}
Generically power counting alone gives $\hat P^\zeta\sim \hat
P^{\zeta_J}$, where the exact size is modified by numerical
coefficients. For the chiraly enhanced moments the power counting is
$\hat P_{M_1 M_2}^{\chi\zeta} \sim \hat P_{M_1 M_2}^{\chi\zeta_J} \sim
\hat P_{M_1 M_2}^{\chi\zeta_{\chi}}$ since $\zeta
\sim \zeta_J \sim \zeta_{\chi}$.

The penguin annihilation amplitudes can also be written in terms of moments of
distributions.  Using the notation in Refs.~\cite{Arnesen:2006vb}
and~\cite{Arnesen:2006dc} the necessary amplitudes are
\begin{align}
 \hat P_{\pi\pi}^{\rm Lann} &= 
   -\frac{\hat f_B f_\pi^2}{m_B^2}\, \Big( 
   \beta_{1c}^{\pi\pi} + 2 \beta_{3c}^{\pi\pi} + \beta_{4c}^{\pi\pi} \Big)
   \,,\\
%
 \hat P_{\pi\pi}^{\rm Gann} &= 
    - \frac{\hat f_B \beta_B f_{3\pi} f_\pi}{m_B m_b}
   \Big( \beta_{hc1}^{\pi\pi}  \plus \beta_{hc2}^{\pi\pi} 
  \minus \frac12 \beta_{hc3}^{\pi\pi} \minus \frac12 \beta_{hc4}^{\pi\pi} \Big)
   \,, \nn\\
  \hat P_{\pi\pi}^{\rm Lann\chi} &= 
   - \frac{\hat f_B f_\pi^2 \mu_\pi }{m_B^2 m_b}
  \, \Big( \beta_{\chi 1}^{\pi\pi} \minus \beta_{\chi 2}^{\pi\pi}
  - \frac12 \beta_{\chi 5}^{\pi\pi} + \frac12 \beta_{\chi 6}^{\pi\pi} \Big)
  \,,\nn \\[5pt]
\hat P_{K\pi}^{\rm Lann} &= 
   - \frac{\hat f_B f_\pi f_K}{m_B^2}\, \beta_{4c}^{\pi K}
   \,,\nn \\
%
 \hat P_{K\pi}^{\rm Gann} &= 
  -\frac{\hat f_B \beta_B }{m_B m_b}\bigg[ f_{3\pi}f_K  \Big(\beta_{hc1}^{\pi K} 
  \minus \frac12 \beta_{hc3}^{\pi K}\Big) \nn\\
 &\quad 
   + f_{3K} f_\pi \Big( \beta_{hc2}^{\pi K} \minus \frac12 \beta_{hc4}^{\pi K} \Big) 
  \bigg]
 \,, \nn\\
\hat P_{K\pi}^{\rm Lann\chi} &= 
   - \frac{\hat f_B f_\pi f_K}{m_B^2} \bigg[
   \frac{\mu_\pi}{m_b} \Big(\beta_{\chi 1}^{\pi K}\minus \frac12 \beta_{\chi 5}^{\pi K}\Big)
  \nn\\
 &\quad 
  + \frac{\mu_K}{m_b} \Big(-\beta_{\chi 2}^{\pi K}\plus \frac12 \beta_{\chi 6}^{\pi K}\Big)
  \bigg]
   \,,\nn \\[5pt]
\hat P_{\rho\rho}^{\rm Lann} &= 
    - \frac{\hat f_B f_\rho^2}{m_B^2}\, \Big( 
   \beta_{1c}^{\rho\rho} + 2 \beta_{3c}^{\rho\rho} + \beta_{4c}^{\rho\rho} \Big)
   \,, \nn \\
%
 \hat P_{\rho\rho}^{\rm Gann} &= 
    -\frac{\hat f_B \beta_B f_{3\rho} f_\rho}{m_B m_b}
   \Big( \beta_{hc1}^{\rho\rho}  \minus \beta_{hc2}^{\rho\rho} 
  \minus \frac12 \beta_{hc3}^{\rho\rho} \plus \frac12 \beta_{hc4}^{\rho\rho} \Big)
   \,, \nn
\end{align}
where the $\beta$-moment parameters were defined above in Eq.~(\ref{betadef})
and numerical values were given in Eq.~(\ref{betas}).

To evaluate the remaining penguin amplitudes in Eqs.~(\ref{Pamplitude}) and
(\ref{Pchiamplitude}) we use the form of the distributions from
section~\ref{sect:models}. It is useful to write Eq.~(\ref{average}) as
integrals over short-distance coefficients, $i_\alpha$ and $j_{\alpha\beta}$,
multiplying model parameters $a_\alpha$ and $A_\beta$:
\begin{align} \label{ijdecompose}
\langle c_{4}^c \rangle_M &= i^{(4c)}_0 + \sum_{\alpha\ne 0}
i^{(4c)}_\alpha a_\alpha^M , \\
\langle c_{1}^{c} \rangle_M &=
 i_0^{(1c)} + \sum_{\alpha\ne 0} i_\alpha^{(1c)} a_\alpha^M \,, \nn \\
\langle b_{4}^c \zeta_J^{B M_2} \rangle_{M_1} &= j^{(4c)}_{00} \,
   \zeta_J^{B M_2}   +\!\!\! \sum _{(\alpha,\beta)\ne(0,0)}  \!\!\! j^{(4c)}_{\beta\alpha}
   \, A^{B M_2}_\beta \, a^{M_1}_\alpha \,, \nn \\
\langle b_{1}^{c} \zeta_J^{B M_2} \rangle_{M_1} &=
  j^{(1c)}_{00}\zeta_J^{B M_2}+ \!\!\!\!\!\! \sum_{(\alpha,\beta)\ne
   (0,0)} \!\!\! j^{(1c)}_{\beta\alpha} \,A^{B M_2}_\beta \, a^{M_1}_\alpha
  \nn \,,
\end{align}
where $\alpha=0,1,2,4$ with $a^M_\alpha=(1,a_1^M,a_2^M,a_4^M)$, and
$\beta=0,1,2$ with $A^{BM}_\beta=(\zeta_J^{BM},A_1^{BM},A_2^{BM})$. This step
is useful because the short-distance coefficients, $i_\alpha,~j_{\alpha\beta}$
are integrals which can be evaluated numerically independent of the choice of
the model parameters. This makes it easier to propagate errors from parameter
uncertainties into the final amplitude predictions. It also makes it possible to
study the short-distance uncertainties (such as the $\mu$-dependence) directly
in terms of $i_\alpha$ and $j_{\beta\alpha}$.  In Eq.~(\ref{ijdecompose}) we
have separated out the dominant term from the sum.  Since our values of
$\zeta_J^{B\pi}$ and $\zeta^{B\pi}$ are extracted from independent experimental
data, these dominant terms in the penguin amplitudes become model independent.
For the chiral enhanced amplitudes the analog of Eq.~(\ref{ijdecompose}) is
\begin{align} \label{ijdecompose2}
\langle c_{1(qfq)}^{\chi c} \rangle_M^{pp} &= k^{(1c)}_0 + \sum_{\gamma\ne 0}
  k^{(1c)}_\gamma p_\gamma^{M} ,
 \\
\langle c_{2(qfq)}^{\chi c}\rangle_M^{pp} &= k^{(2c)}_0 + \sum_{\gamma\ne 0}
  k^{(2c)}_\gamma p_\gamma^M , 
  \nn \\
\langle b_{3(qfq)}^{\chi c } \zeta_J^{B M_2} \rangle_{M_1}^{pp} &=
 \ell^{(3c)}_{00} \,
   \zeta_J^{B M_2}   +\!\!\!\!\!\! \sum _{(\gamma,\beta)\ne(0,0)}  \!\!\!
  \ell^{(3c)}_{\gamma\beta}
   \, A^{B M_2}_\gamma \, p^{M_1}_\beta \,, 
\nn \\
\langle b_{4(qfq)}^{\chi c} \zeta_J^{B M_2} \rangle_{M_1}^{pp} &=
  \ell^{(4c)}_{00} \,
   \zeta_J^{B M_2}   +\!\!\!\!\!\! \sum _{(\gamma,\beta)\ne(0,0)}  \!\!\!
  \ell^{(4c)}_{\gamma\beta}
   \, A^{B M_2}_\gamma \, p^{M_1}_\beta \,, 
\nn \\
\langle b_{1(qfq)}^{\chi c} \zeta_\chi^{B M_2} \rangle_{M_1} &= 
   \ell^{(1c)}_{00} \,
   \zeta_\chi^{B M_2}   +\!\!\!\!\!\! \sum _{(\alpha,\beta)\ne(0,0)}  \!\!\!
   \ell^{(1c)}_{\beta\alpha}
   \, A^{B M_2}_{\chi\beta} \, a^{M_1}_\alpha \,, 
\nn \\
\langle b_{1(ufu)}^{\chi c} \zeta_\chi^{B M_2} \rangle_{M_1} &= 
   \ell^{(2c)}_{00} \,
   \zeta_\chi^{B M_2}   +\!\!\!\!\!\! \sum _{(\alpha,\beta)\ne(0,0)}  \!\!\!
   \ell^{(2c)}_{\beta\alpha}
   \, A^{B M_2}_{\chi \beta } \, a^{M_1}_\alpha \,, \nn
\end{align}
where $\beta,\gamma=0,1,2$, $p_\gamma^M=\{1,a_{1pp}^M,a_{2pp}^M\}$ and $A_{\chi\beta}^{BM}=
\{\zeta_\chi^{BM}, A_{\chi 1}^{BM},A_{\chi 2}^{BM}\}$.  In terms of the $i$,
$j$, $k$, $\ell$ coefficients, Eqs.~(\ref{Pamplitude}) and (\ref{Pchiamplitude})
are given by
\begin{align}
\hat P^{\zeta}_{\pi \pi} & =  
  - \hat f_\pi \big[ i^{(4c)}_\alpha\, +i_\alpha^{(1c)}\big]
   \zeta^{B \pi}  a^\pi_\alpha 
   \,, \nn \\
\hat P^{\zeta_J}_{\pi \pi} & =  
  - \hat f_\pi \big[ j^{(4c)}_{\beta\alpha} 
   +  j^{(1c)}_{\beta\alpha}\big] A^{B\pi}_\beta \, a^\pi_\alpha
   \,,\nonumber \\
\hat P^{\chi \zeta}_{\pi\pi} &= 
    \frac{\hat f_\pi\mu_\pi}{3m_B} \big[ k_\gamma^{(1c)}+k_\gamma^{(2c)}\big]
   \zeta^{B \pi}  p_\gamma^\pi
  \,,\nn\\
\hat P^{\chi \zeta_J}_{\pi\pi} &= 
   \frac{\hat f_\pi\mu_\pi}{3m_B} \big[ \ell_{\gamma\beta}^{(3c)}
   +\ell_{\gamma\beta}^{(4c)}\big] A_\gamma^{B \pi}  p_\beta^\pi
  \,,\nn\\
\hat P^{\chi \zeta_\chi}_{\pi\pi} &= 
   \frac{\hat f_\pi \mu_\pi}{3m_B} \big[ \ell_{\beta\alpha}^{(1c)}
   +\ell_{\beta\alpha}^{(2c)}\big] A_{\chi \beta}^{B \pi}  a_\alpha^\pi
  \,,
\end{align}
and
\begin{align}
\hat P^{\zeta}_{K \pi} & =  
  -\hat f_K \big[ i^{(4c)}_\alpha\, +i_\alpha^{(1c)}\big] \, a^K_\alpha \,
   \zeta^{B \pi} \,,\nn \\
\hat P^{\zeta_J}_{K \pi} & = 
  - \hat f_K \, \big[ j^{(4c)}_{\beta\alpha} 
   +  j^{(1c)}_{\beta\alpha}\big]  
   \, A^{B\pi}_\beta  \, a^K_\alpha
    \,, \nn \\
\hat P^{\chi \zeta}_{K\pi} &= 
   \frac{\hat f_K \mu_K}{3m_B} \big[ k_\gamma^{(1c)}+k_\gamma^{(2c)}\big]
   \zeta^{B \pi}  p_\gamma^\pi
  \,,\nn\\
\hat P^{\chi \zeta_J}_{K\pi} &= 
    \frac{\hat f_K\mu_K}{3m_B} \big[ \ell_{\gamma\beta}^{(3c)}+\ell_{\gamma\beta}^{(4c)}\big]
   A_\gamma^{B \pi}  p_\beta^K
  \,,\nn\\
\hat P^{\chi \zeta_\chi}_{K\pi} &= 
   \frac{\hat f_K \mu_\pi}{3m_B} \big[ \ell_{\beta\alpha}^{(1c)}
   +\ell_{\beta\alpha}^{(2c)}\big] A_{\chi \beta}^{B \pi}  a_\alpha^K
  \,,
%
\end{align}
where a sum over $\alpha=0,1,2,4$ and $\beta,\gamma=0,1,2$ is understood.

Evaluating the short-distance integrals at zeroth order in $\alpha_s$
with the $C_i$'s in Eq.~(\ref{CiNLL})  the $i$'s and $j$'s are
\begin{align} \label{ijLO1}
i^{(4c)}\! \times\! 10^{3}&= (- 28.4,~0,~0,~0)  \,, \nn \\
i^{(1c)} \! \times\! 10^{3} &= (- 1.96,~0,~0,~0)\,,  \nn \\
j^{(4c)}\!\times\! 10^3 &=  \left [
{\begin{array}{cccc}
\minus 11.1 &  17.3 &  17.3 &  17.3 \\    
0 & 0 & 0 & 0  \\
0 & 0 & 0 & 0  
\end{array}} \right ] \,, \nn \\
j^{(1c)} \!\times\! 10^{3} &= \left [
{\begin{array}{cccc}
\minus 16.4 & \minus 14.5 & \minus 14.5 & \minus 14.5 \\    
0 & 0 & 0 & 0 \\
0 & 0 & 0 & 0
\end{array}} \right ]\, .
\end{align}
Note that the short-distance ``$i$'' coefficients for $\zeta^{BM}$ are comparable
in size to the short-distance ``$j$'' coefficients for $\zeta_J^{BM}(z)$.  For the
chiraly enhanced integrals we find
\begin{align} \label{ijLO2}
k^{(1c)}\! \times\! 10^{3}&= (218,~0,~218)  \,, \nn \\
k^{(2c)} \! \times\! 10^{3} &= (-3.38,~0,~-3.38)
 \,,  \nn \\
\ell^{(1c)}\!\times\! 10^3 &=  \left [
{\begin{array}{cccc}
-12.3 & 69.1 & \minus 69.1 & \minus 69.1 \\    
34.5 & -34.5 & 34.5 & 34.5 \\
0 & 0 & 0 & 0
\end{array}} \right ] \,, \nn \\
\ell^{(2c)} \!\times\! 10^{3} &= \left [
{\begin{array}{cccc}
61.9 &  -57.9 & 57.9 & 57.9 \\    
-29.0 & 29.0 & -29.0 & -29.0 \\
0 & 0 & 0 & 0 
\end{array}} \right ]\,, \nn\\
\ell^{(3c)}\!\times\! 10^3 &=  \left [
{\begin{array}{ccc}
218 & 0 &  218 \\    
0 & 0 & 0 \\
0 & 0 & 0 
\end{array}} \right ] \,, \nn \\
\ell^{(4c)} \!\times\! 10^{3} &= \left [
{\begin{array}{ccc}
-3.38 & 0 & -3.38 \\    
0 & 0 & 0 \\
0 & 0 & 0 
\end{array}} \right ]\, .
\end{align}
Relative to the size of $i^{(4c)}$ and $j^{(1c,4c)}$ the enhanced size of the
$k^{(1c)}$ and the $\ell^{(ic)}$ short-distance coefficients is quite striking.
Comparing the matching coefficients in Eqs.~(\ref{ci},\ref{bi}) and
(\ref{chiralcoefficientsc},\ref{chiralcoefficientsb1}) we see that the
combinations of coefficients from $H_W$ are similar in size ($C_{3,4}$ versus
$C_{5,6}$). However, the $k^{(1c)}$ and $\ell^{(ic)}$ moments are enhanced by a
factor of $\simeq 6$ due to the inverse moment fraction factor $1/u\bar u$. This
numerical factor provides additional enhancement beyond the numerical
enhancement in $\mu_M/m_b$, and is the essential reason why the chiraly enhanced
penguin amplitudes are numerically important.

Next we evaluate the short-distance integrals $i^{(4c)}$, $j^{(4c)}$,
and $k^{(1c)}$ up to order $\alpha_s$, by including the one-loop
results for $b_4^{(f)}$ and $c_4^{(f)}$ given earlier in
section~\ref{sect:summary}. Where known we also evaluate the chiral
enhanced short-distance integrals up to ${\cal O}(\alpha_s)$ (from 
Eq.(\ref{alphaschi})).
\begin{widetext}
At the scale $\mu=m_b$ we find
\begin{align} \label{ijNLO}
i^{(4c)}\! \times\! 10^{3}&  = (- 36.2\minus i 9.58, 4.49\plus i8.12,
   10.5\plus i2.06,  4.42\minus i 2.36) , \nn \\
j^{(4c)}\!\times\! 10^3 &= \!\! \left [
{\begin{array}{cccc}
 \minus 14.0\minus i 4.52 &  27.6\plus i 4.47 &  28.1\minus i 0.15 
& 25.8\minus i 0.55 \\    
 \minus 1.05\minus i 3.63 & \minus 2.30\plus i 2.10 &  1.32 \plus i 1.95 
& \minus 0.20 \minus i 1.64 \\
 0.15\plus i 0.04 & \minus 0.10\minus i 0.08 &   0.08\minus i 0.14
& \minus 0.07 \plus i 0.14
\end{array}} \right ] \,, 
\nn\\
k^{(1c)}\! \times\! 10^{3}&= (281\plus i 54.7, 8.6\minus i 69.1,
   ~240 \plus  i 41.1) \,.
\end{align}
We will also analyze how stable our results are to variations in
$\mu$. For the LO results in Eqs.~(\ref{ijLO1}) and (\ref{ijLO2}) a change in
$\mu$ simply reflects changes in the $C_i(\mu)$ and so will not be shown. At NLO
in the perturbative expansion we find for $\mu=m_b/2=2.35\,{\rm GeV}$
\begin{align}
i^{(4c)}\! \times\! 10^{3}&= (- 38.9\minus i 12.3,  7.46\plus i 10.5,
   15.2\plus i2.65, 7.37 \minus i 3.05),  \nn \\
j^{(4c)}\!\times\! 10^3 &= \!\! \left [
{\begin{array}{cccc}
\minus 12.6\minus i 6.67 & 34.6\plus i 6.12 &  36.3\plus i 0.45
 & 33.1\minus i 1.23 \\    
 \minus 0.92\minus i 4.01 & \minus 2.51\plus i 2.30 &  1.40 \plus i 2.14 
 & \minus 0.13 \minus i 1.82 \\
 0.20\plus i 0.09 &  0.17\plus i 0.09 &  0.08 \minus i 0.21
 & \minus 0.08 \plus i 0.20
\end{array}} \right ], \nn\\
k^{(1c)}\! \times\! 10^{3}&= (359\plus i 70.6,  11.1\minus i 89.2,
   306 \plus i53.0) \,, 
\end{align}
while for $\mu=2 m_b=9.4\,{\rm GeV}$ we find
\begin{align}
i^{(4c)}\! \times\! 10^{3}&= (- 32.5\minus i 7.57, 2.81\plus i 6.42,
  7.58\plus i 1.63,  2.75 - i1.87),  \nn \\
j^{(4c)}\!\times\! 10^3 &= \!\! \left [
{\begin{array}{cccc}
 \minus 13.2\minus i 3.10 &  22.7\plus i 3.31 &  22.6 \minus i 0.45 
 & 21.0 \minus i 0.16 \\    
 \minus 1.05\minus i 3.19 & \minus 2.05\plus i 1.85 &  1.20 \plus i 1.72 
 & \minus 0.22\minus i1.44 \\
 0.11\plus i 0.01 &  0.06\plus i 0.08 &   0.07\minus i 0.09
 & \minus 0.07 \plus i 0.10
\end{array}} \right ] ,\nn\\
k^{(1c)}\! \times\! 10^{3}&= (228 \plus i 43.3, 6.8\minus i 54.7,
   195\plus i32.5) \,. 
\end{align}
\end{widetext}
From Eq.~(\ref{ijNLO}) we observe that these $\alpha_s$ corrections induce
imaginary contributions which are often appreciable since the $\alpha_s C_{1,2}$
terms can compete with $C_{3-6}$. For example, the imaginary part of
$j_{00}^{(4c)}$ determined from our result for the one-loop matching given in
Eq.~(\ref{NDRHVcoefficientsB}), is $\sim 30\%$ of the real part.

Because we have neglected terms $\alpha_s C_{3-6}$ we must also
neglect the $\mu$ dependence of $\zeta$, $\zeta_J$, and the $\phi$'s
for consistency. These terms induce a $\alpha_s\ln(\mu)$ that
multiplies the tree-level penguin coefficients involving $C_{3-6}$ and
are hence compensated by $\alpha_s \ln(\mu) C_{3-6}$ corrections to
the short-distance coefficients.  The dominant coefficients have
$\alpha=\beta=0$.  At zeroth order in $\alpha_s$ the central values
for the coefficients $i^{(4c)}_0$ and $j^{(4c)}_{00}$ vary by $\pm
30$-$50\%$ when we take $\mu=m_b/2$ and $\mu=2m_b$.  We find this change is
reduced to $\le 10\%$ at NLO.  At LO the chiraly enhanced $k^{(1c)}_0$
varies by $\pm 35$-$55\%$, and this is reduced by about a factor of two, to
$\pm 20$-$25\%$ at NLO.  The imaginary parts first appear at ${\cal
O}(\alpha_s(\mu))$, and exhibit a $\pm 20$-$30\%$ range for
$i^{(4c)}_0$ and $k^{(1c)}_0$, and $\pm 30$-$50\%$ dependence for
$j^{(4c)}_{00}$. The LO coefficients $\ell^{(1c,3c,4c)}_{00}$ also
have a sizeable $\mu$-dependence ($20$-$50\%$) and it will be
important to compute their $\alpha_s$ corrections in the future. Below
we will take this residual scale uncertainty as a way of estimating
the size of missing higher order perturbative corrections on our final
result.

%
\definecolor{Light}{gray}{0.86}
\begin{table*}[t!]
\begin{tabular}{|r|c|c|c|c|c||c|c|c|c|}
\hline
  \multicolumn{5}{|c}{Penguin Amplitudes, $\hat P^i_{M_1 M_2}\!\times\! 10^4$}  
& &  \multicolumn{4}{c|}{Penguin Annihilation, $\hat P^i_{M_1 M_2}\!\times\! 10^4$}
 \\[5pt]
& \colorbox{Light} {$\hat P_{\pi\pi}^{\zeta} $}
&  \colorbox{Light}{$\hat P_{\pi\pi}^{\zeta_J}$}
&  \colorbox{Light}{$\hat P_{\pi\pi}^{\chi\zeta}$}
&  \colorbox{Light}{$\hat P_{\pi\pi}^{\chi\zeta_J}$}
&  \colorbox{Light}{$\hat P_{\pi\pi}^{\chi\zeta_\chi}$}
&
&  \colorbox{Light}{$\hat P_{\pi\pi}^{\rm Lann}$}
&  \colorbox{Light}{$\hat P_{\pi\pi}^{\rm Gann}$}
&  \colorbox{Light}{$\hat P_{\pi\pi}^{\rm Lann\chi}$}
  \\
\hline
$ C_{3-10}$
& 3.58 $\mpm$ 1.02 & 3.34 $\mpm$ 0.88 & $4.41 \mpm 1.78$ 
 & 4.51 $\mpm$ 1.71 & $0.00 \mpm 1.03$
&
& {\bf --} & {\bf --}  & {\bf --}
\\
$\alpha_{s} C_{1,2,8g}$
& $(0.86 \mpm 0.25) $ & $(0.32 \mpm 0.26) $
& $(1.21 \mpm 0.37) $ & {\bf --} & {\bf --}
& $\alpha_s C_{1-10}$ & $-1.46 \mpm 0.88$ & $0.15 \mpm 0.08$ & $0.00 \mpm 5.00$
\\
&   $ + i (1.08 \mpm 0.32)$ & $+ i (0.53 \mpm 0.13)$
  & $+ i (1.10 \mpm 0.40)$ & &
  & & & &
\\
\hline
$\hat P^i_{K\pi}\!\times\! 10^4$:
& \colorbox{Light}{$\hat P_{K\pi}^{\zeta}$}
&  \colorbox{Light}{$\hat P_{K\pi}^{\zeta_J}$}
&  \colorbox{Light}{$\hat P_{K\pi}^{\chi\zeta}$}
&  \colorbox{Light}{$\hat P_{K\pi}^{\chi\zeta_J}$}
&  \colorbox{Light}{$\hat P_{K\pi}^{\chi\zeta_\chi}$}
&
&  \colorbox{Light}{$\hat P_{K\pi}^{\rm Lann}$}
&  \colorbox{Light}{$\hat P_{K\pi}^{\rm Gann}$}
&  \colorbox{Light}{$\hat P_{K\pi}^{\rm Lann\chi}$}
\\
\hline
$ C_{3-10}$
& 4.37 $\mpm$ 1.25 & 4.00 $\mpm$ 1.02
 & $6.02 \mpm 2.42$ & 6.15 $\mpm$ 2.33 & $0.00 \mpm 1.18$
& 
& {\bf --} & {\bf --}  & {\bf --}
  \\
$\alpha_{s} C_{1,2,8g}$
& $(0.86 \mpm 0.40) $ & $(0.11 \mpm 0.35) $
& $(1.66 \mpm 0.50)$ & {\bf --} & {\bf --}
& $\alpha_s C_{1-10}$ & $0.26 \mpm 0.14$ & $0.20 \mpm 0.11$ & $0.00 \mpm 8.03$
\\
& $+ i (1.38 \mpm 0.40)$ &  $+ i (0.70 \mpm 0.18)$
  & $ + i (1.50 \mpm 0.57)$  & &
  & & & &
\\
\hline
$\hat P^i_{\rho\rho}\!\times\! 10^4$:
& \colorbox{Light} {$\hat P_{\rho\rho}^{\zeta}$}
&  \colorbox{Light}{$\hat P_{\rho\rho}^{\zeta_J}$}
& & &
%
&
&  \colorbox{Light}{$\hat P_{\rho\rho}^{\rm Lann}$}
&  \colorbox{Light}{$\hat P_{\rho\rho}^{\rm Gann}$}
&
\\
\hline
$ C_{3-10}$
& 14.8 $\mpm$ 3.5 & 1.64 $\mpm$ 2.99
& & &
%
& & & &
\\
$\alpha_{s} C_{1,2,8g}$
& $(5.59^{+3.08}_{-1.61}) $ & $(0.36 \mpm 0.66) $
& & &
%
& $\alpha_s C_{1-10}$ & $0.65^{+0.63}_{-0.21}$ & $0.22^{+0.22}_{-0.20}$ &  
\\
& \ $+i (5.39^{+2.60}_{-1.48})$ &  $+ i (0.28 \mpm 0.49)$
  & & &
  & & & &
\\
\hline
\end{tabular}
\caption{\label{tableII} Numerical predictions for the penguin
  amplitudes, $\hat P_{MM}$, from the factorization theorem. The results were 
  split into terms generated at ${\cal O}(\alpha_s^0)$ and ${\cal O}(\alpha_s)$
  in the short-distance matching coefficients. 
  In each row the theoretical predictions are  broken down by contributions from
  the $\zeta^{BM}$ and $\zeta_J^{BM}$ terms in the leading power amplitude, 
  terms $\zeta^{BM}$, $\zeta_J^{BM}$, $\zeta_\chi^{BM}$ from the chiraly enhanced
  part of the amplitude, and terms from standard penguin annihilation, 
  three-parton annihilation, and chiraly enhanced annihilation. The errors shown are 
  uncertainties propagated from input parameters as described in the text.}
\end{table*}
On the other hand the electroweak coefficients $j^{(1c)}_{00}$ and $\ell^{(2c)}_{00}$
have only $\sim 3\%$ $\mu$-dependence at LO, consistent with our expectations
that the NLO corrections to this term are small. This reflects the fact that the
corresponding $\mu$ dependence occurs in a NLO penguin diagram with photon
exchange, whereas the leading order Wilson coefficients are generated by both
photon and the larger $Z$ exchange. The corrections to the electroweak
coefficient $i^{(1c)}_0$ at LO is larger as a percent ($40$-$50\%$), however the
$i^{(1c)}_0$ is tiny to begin with, since at LO its proportional to the
numerically small combination $C_{10}+C_9/N_c$. Thus we do not expect our
neglect of these one-loop electroweak corrections to have a large effect .

In table~\ref{tableII} we present numbers for the penguin amplitudes in
Eq.~(\ref{PP}), showing separately the tree-level and $\alpha_s$ corrections.
The errors shown in the table include only input parameter uncertainty, and are
computed with Gaussian scans for the errors in the model parameters given in
section~\ref{sect:inputs}.  Despite having a number of hadronic parameters, we
observe a relatively small model parameter dependence in the first four columns
for $\pi\pi$ and $K\pi$. This occurs because there is only a small dependence of
the penguin amplitudes on the shape of $\zeta^{BM}_J(z)$ and $\phi^M(u)$. The
normalization terms, $i^{(4c)}_{0}$ and $j^{(1c,4c)}_{00}$, give the dominant
contribution to the amplitudes and the corresponding model parameters were fit
to independent data to reduce their uncertainty. Furthermore, at LO the only
shape parameter dependence comes from $\langle x^{-1}\rangle_M$ and for the pion
this parameter is quite well known (which in our error analysis is accounted for
by taking into account an important correlation in $a_2^\pi$ and $a_4^\pi$). For
example, we find that further doubling the error bars on the shape parameters
$A_{i=1,2}^M$ only effects the last quoted digit of the error bars on the
penguin amplitudes in table~\ref{tableII}. Since the second and third rows of
$j^{(4t)}_{\alpha\beta}$ are small, the dependence on the shape of
$\zeta_J^{BM}(z)$ is very small.  Though the coefficients
$j^{(4t)}_{01},~j^{(4t)}_{02}$ are similar in size to $j^{(4t)}_{00}$, their
contributions are suppressed by the small $a_{1,2}^M$.  One entry is very
sensitive to the $a_{2}^K$, namely the $\alpha_s C_{1,2,8g}$ contributions to
$\hat P_{K\pi}^{\zeta_J}$. Here the dominant term is
\begin{align}
10^3\hat P_{K\pi}^{\zeta_J}\Big|_{\alpha_sC_i} & \!\! \simeq 
    \hat f_K \zeta_J^{B\pi} \big[(2.9 +i 4.5)-(10.3+i 4.47)a_1^K \nn\\ 
  &\qquad   - (10.8-i 0.15)a_2^K \big] \,,
\end{align}
and exhibits a large cancellation in the real part for the value $a_2^K=0.2$
(explaining its large percent uncertainty for $a_2^K=0.2\pm 0.2$). Some
cancellation is also evident for $\hat P_{\pi\pi}^{\zeta_J}$. However, overall
these are both small contributions to their respective penguin amplitudes. This
type of parameter dependence does not appear in other terms, and we find that it
does not significantly effect the final result. Thus even though our model
parameters vary over a large range we have fairly robust central values for
individual contributions to the leading order penguin amplitudes in
table~\ref{tableII}.  The uncertainty in the 3rd and 4th columns for the chiraly
enhanced amplitudes is also reduced by our knowledge of the normalization of
$\zeta$ and $\zeta_J$, and is a bit bigger than the first two columns due to the
added uncertainty from $\phi_{pp}^M(u)$. The 5th column involves the new form
factor $\zeta_\chi^{BM}(z)$, where we do not have information about the sign,
and hence zero central values.

In the $\pi\pi$ and $K\pi$ entries in table~\ref{tableII} we also observe that
the contributions from $\zeta^{BM}$ and $\zeta_J^{BM}$ are similar in size. This
is a reflection of the fact that there coefficients are similar numerically, and
is in agreement with the power counting $\zeta^{BM}\sim \zeta_J^{BM}$. In
determining the errors associated with these parameters it was quite important
to take into account the correlations, as already described in
section~\ref{sect:zeta}. Also, as mentioned above, the chiraly enhanced penguin
amplitudes compete numerically with the leading power amplitudes due to the
presence of the enhancement by the $1/u\bar u$ momentum fraction factor which
generates a numerical factor of six. For example, we have
\begin{align}
  \hat P_{\pi\pi}^{\zeta_J} \plus \hat P_{\pi\pi}^{\chi\zeta_J} \Big|_{C_{3-10}}
   & \!\!\! 
   \sim \hat f_\pi \zeta_J^{B\pi} \Big( 28 + 215 \frac{\mu_\pi}{3m_B}
\Big)+ \ldots,
\end{align}
where the large numerical value $215$ is generated by this
enhancement.

Examining the annihilation amplitudes we see that $\hat
P_{\pi\pi}^{\rm Lann}+\hat P_{\pi\pi}^{\rm Gann}$ is suppressed by a
factor of $1/5$ relative to $\hat P_{\pi\pi}^{\zeta}+\hat
P_{\pi\pi}^{\zeta_J}$, and so is of the expected size for this power
correction, namely
\begin{align}
 \frac{\hat P_{\pi\pi}^{Lann}+\hat P_{\pi\pi}^{Gann}}{\hat
 P_{\pi\pi}^{\zeta} + \hat P_{\pi\pi}^{\zeta_J}} \sim
\frac{\Lambda_{\rm QCD}}{E}\,\frac{\alpha_s(m_b)}{\alpha_s(\mu_i)} \,.
\end{align}
The same conclusions hold for $K\pi$. On the other hand the chiral enhanced
annihilation terms $\hat P_{\pi\pi}^{\rm Lann\chi}$ and $\hat P_{K\pi}^{\rm
  Lann\chi}$ have much larger parameter uncertainty, and we are not able to draw
definite conclusions about the size of these terms. In fact they provide the
dominant parameter uncertainty for the $\pi\pi$ and $K\pi$ channels.

For $B\to \rho \rho$ decays, our analysis was slightly different from
the pseudoscalars as we used a non-polynomial model for
$\phi^\rho(x)$. Here the errors are dominated by the uncertainty in
$\zeta^{B\rho}$, $\zeta_J^{B\rho}$, and $a_\rho$. The uncertainty from the
shape parameters $A_i$ are negligible in comparison. Since
current data prefers a central value for $\zeta_J^{B\rho}$
significantly smaller than that for $\zeta^{B\rho}$ this same hierarchy is
observed in the penguin amplitudes. The size of $\hat
P_{\rho\rho}^{\zeta}$ is enhanced by $f_\rho$ and the $\rho$-form
factor in comparison to $\hat P_{\pi\pi}^{\zeta}$. Due to the absence
of chiraly enhanced contributions the dominant parameter uncertainty
comes from experimental uncertainties that propagate into the errors
for $\zeta^{B\rho}$ and $\zeta_J^{B\rho}$.

\begin{table*}[t!]
\begin{tabular}{|c|c|c|c|c|c|c|c|}
\hline
  & $\hat P^{\text{LO}} \!\times\! 10^4$
&  $\hat P^{\chi}\!\times\! 10^4$
&  $\hat P^{\text{ann}}\!\times\! 10^4$
& $\hat P^{\text{total}}\!\times\! 10^4$
& $\hat P^{\text{expt}}_{\text{ispin}} \!\times\! 10^4$
& $\hat P^{\text{expt}}_{\text{ispin}} \!\times\! 10^4$
& $\hat P^{\text{expt}}_{\text{TF}} \!\times\! 10^4$
\\
& & & & & $(\gamma=59^\circ)$  & $(\gamma=74^\circ)$  &
 $(\gamma=59^\circ$-$74^\circ)$
  \\
\hline
\raisebox{-6pt}[0pt][-6pt]{$B\to \pi \pi$} &  
 $(8.10 \!\pm\! 0.63)$  &
$(10.2 \!\pm\! 2.9) $ & $-1.31 \pm 5.08 $ & $(16.9 \pm 5.9)$
& $(18\!\pm\! 9)$ &  $(44\!\pm\! 6)$ &
\\
&  \  $+ i (1.61 \!\pm\! 0.21)$  &  \ $+ i (1.10 \!\pm\! 0.39)$
  &
  &$\quad +i(2.71 \pm 0.45)$&
   \ $-i(29\!\pm\! 6)$ & \ $-i(29\!\pm\! 6)$ &
\\
\hline
\raisebox{-6pt}[0pt][-6pt]{$B\to K\pi$} &
 $(9.34 \pm 1.00)$  &
$(13.8 \pm 3.9) $ & $0.46 \pm 8.03 $ &
$(23.6 \pm 9.0)$ & &
& $\pm (48 \pm 4 \pm 10)$\\
& \quad\  $ + i (2.08 \pm 0.25)$  & \ $+ i (1.49 \pm 0.57)$
  &
  &$\quad +i(3.57 \pm 0.62)$& & &  \ $-i(22 \pm 7 \pm 4)$ \\
\hline
\raisebox{-6pt}[0pt][-6pt]{$B\to \rho\rho$} & $22.4^{+3.7}_{-2.3} $
& {\bf ---} & $0.87^{+0.67}_{-0.29}$ & $23.3^{+3.7}_{-2.4}$
& $-(29 \pm 26)$ & $(38 \pm 23)$ &  \\
&  $+i \, 5.68^{+2.45}_{-1.07} $ & {\bf ---}
  &  & $+i \, 5.68^{+2.45}_{-1.07}$
& $\quad\ -i(8 \pm 18)$ & $\quad\ -i(8 \pm 18)$ &  \\
\hline
\end{tabular}
\caption{\label{tableIII} Numerical predictions for the short-distance penguin
  amplitudes at leading power, $\hat P^{LO}$, from chiraly enhanced
  terms $\hat P^{\chi}$, and from the annihilation amplitudes in
  Refs.~\cite{Arnesen:2006vb,Arnesen:2006dc}. The sum of these
  contributions $\hat P^{\text{total}}$, is the total short-distance
  result from the factorization theorems discussed in the text
  (long-distance terms are discussed in the text). The last three columns
  show current experimental data. Comparing them with $\hat P^{\text{total}}$
  shows an order of magnitude short-fall for the imaginary part.}
\end{table*}
In table~\ref{tableIII} we ``sum up'' the individual contributions from the
leading power, chiraly enhanced, and annihilation penguin amplitudes, to obtain
$\hat P^{\rm LO}$, $\hat P^{\chi}$, and $\hat P^{ann}$ respectively. To perform
these sums we do separate Gaussian scans for the total penguin amplitude since
this provides the simplest way of propagating correlated parameter uncertainties.
This also explains why the central values are not precisely the mean from
table~\ref{tableII}, due to small non-linearity effects in the parameter
dependences. The correlation in input parameter uncertainties must be taken into
account to get the errors shown here. The three amplitudes in the first three
columns of table~\ref{tableIII} are then added together to get the total
theoretical contribution, $\hat P^{\rm total}$. These total values can be
compared to the experimental values in the last three columns. The uncertainty
shown only includes the variation of parameters from the Gaussian scans. For the
first column the displayed errors are dominated by the uncertainties in
$a_{2}^\pi+a_4^\pi$, $a_{1,2}^K$, $\zeta^{B\pi}$, $\zeta_J^{B\pi}$, and for $B\to
\rho\rho$ those in $a_\rho$, $\zeta^{B\rho}$, and $\zeta_J^{B\rho}$. The effect
of other parameter uncertainties is quite small. Even the dominant uncertainties
are small due to our proper account of parameter correlations and use of
experimental data. Also due to our fit procedure the errors from $\zeta$ and
$\zeta_J$ will decrease with improved measurements of the tree amplitudes (which
come from improved branching ratios and CP-asymmetries).  In $\hat P^{\rm
  total}$ the uncertainty from the parameters in the chiral enhanced
annihilation by far dominate the errors for $B\to\pi\pi$ and $B\to K\pi$.

In addition we can estimate the uncertainty from
determining the hard coefficients by varying $\mu\in
[m_b/2,2m_b]$. For the real parts this gives an additional
${}^{+7\%}_{-9\%}$ uncertainty for $\hat P_{\pi\pi}^{\rm total}$,
${}^{+15\%}_{-12\%}$ uncertainty for $\hat P_{K\pi}^{\rm total}$, and
${}^{+9\%}_{-10\%}$ uncertainty for $\hat P_{\rho\rho}^{\rm
total}$. For the imaginary parts we find an additional
${}^{+25\%}_{-19\%}$ uncertainty for $\hat P_{\pi\pi}^{\rm total}$,
${}^{+26\%}_{-19\%}$ uncertainty for $\hat P_{K\pi}^{\rm total}$, and
${}^{+30\%}_{-22\%}$ uncertainty for $\hat P_{\rho\rho}^{\rm total}$.
Finally we assign a generic $20\%$ uncertainty to the final $\hat
P^{\rm total}$ results to account for the fact that we have given only
a partial treatment of $1/m_b$ corrections, but do not foresee a reason
why the untreated corrections should be enhanced over the power
counting estimate. Thus with an estimate for all theoretical
uncertainties we find
\begin{align} \label{Pfinal}
  \hat P_{\pi\pi}^{\rm total} &= 
    (16.9 \mpm 5.9 {}^{+1.0}_{-1.7}
  \mpm 2.0
  \mpm 3.4 ) 
  \nn\\
  & \qquad  +i (2.71 \mpm 0.38 {}^{+.68}_{-.51} 
   \mpm 0.33  
   \mpm 0.54 )
  \,,\nn\\
  \hat P_{K\pi}^{\rm total} &= 
    (23.6 \mpm 9.0 {}^{+3.5}_{-2.8}
   \mpm 2.8
   \mpm 4.7 )
   \nn\\
   &\qquad  +i (3.57 \mpm 0.53 {}^{+.93}_{-.68} 
  \mpm 0.43
  \mpm 0.71 )
  \,,\nn\\
  \hat P_{\rho\rho}^{\rm total} &= 
  (23.3 {}^{+3.7}_{-2.4} {}^{+2.1}_{-2.3}
   \mpm 2.8
   \mpm 4.7 )
   \nn\\
  &\qquad   +i (5.68 {}^{+2.81}_{-1.75} {}^{+1.70}_{-1.25}
  \mpm 0.68
  \mpm 1.14 )
  \,.
\end{align}
The first errors are from input parameters and are dominated by chiral-enhanced
annihilation for $B\to\pi\pi,K\pi$. The second errors are our estimates of
higher order perturbative corrections (the $\mu$-variation).  The third terms
are errors from $|V_{ub}|$ which propagate through the form factors and hence
can be added as a $\pm 12\%$ uncertainty.\footnote{We have increased the $7\%$
  error on $|V_{ub}|$ quoted by HFAG~\cite{HFAG}, which we consider to be overly
  optimistic.} Finally the fourth errors are a generic 20\% that we add for
unknown power corrections.

For $\pi\pi$ the real part of the amplitude in Eq.~(\ref{Pfinal})
agrees with the data in table~\ref{tableIII} for
$\gamma=59^\circ$. However, the same is not true for $K\pi$, nor even
for $\pi\pi$ if $\gamma=74^\circ$ (which is the value preferred by
SU(3) and SCET power counting which predicts $\hat P_{\pi\pi}\simeq
\hat P_{K\pi}$~\cite{Bauer:2005kd}). Here the disagreement with data
in the real part is at the level of factor of two.

On the other hand the imaginary part of the short-distance prediction for
$\hat P^{\pi\pi}$ and $\hat P^{K\pi}$ are much smaller than the
corresponding experimental values and have the opposite sign.  Due to a
numerical enhancement $\hat P_{M_1 M_2}^{\chi\zeta}$ and $\hat P_{M_1
M_2}^{\chi\zeta_J}$ are of same size as the leading power
contributions to the amplitude, but as we have demonstrated by
deriving an \SCETa factorization theorem, these terms are real at
zeroth order in $\alpha_s$. After taking into account all theoretical
uncertainties in our analysis, we conclude that it is not possible to
match the $\hat P$ imaginary parts obtained from experimental data.
Therefore the large phase of the penguin relative to tree amplitudes
can only be explained by long distance charm contribution, $\hat
P_{c\bar c}$, within the standard model, or by contributions from new
physics.

If the remainder is generated by long distance charm contributions, then we can
determine what values of $\hat P^{M_1M_2}_{c\bar c}$ reproduce the experimental
data. This gives 
\begin{align}
  \hat P_{c\bar c}^{\pi\pi} &= \bigg\{
  \begin{array}{c} 
     (1\pm 11) -i (32 \pm 6)  \qquad (\gamma=59^\circ) \\
     (27\pm 8) -i (32 \pm 6)  \qquad (\gamma=74^\circ) 
  \end{array}
   \,,\nn\\
   \hat P_{c\bar c}^{K\pi} &= (24\pm 14) -i (26 \pm 8)
   \,,
\end{align}
where we have added the experimental and theoretical errors in
quadrature. Thus a long-distance charm penguin with substantial
imaginary amplitude is one possibility for reproducing the data. This
explanation was favored in Refs.~\cite{Ciuchini, Colangelo,
Bauer:2004tj}, and the analysis here makes the required size of these
long-distance terms fairly precise. In the next section we contrast
this long distance standard model explanation with the more exciting
possibility of a new physics contribution.
An additional test of the penguin amplitudes can be made from studying the
channels $B^-\to \bar K^0 K^-$, $\bar B^0\to K^0\bar K^0$, and $\bar B^0\to
K^+K^-$ which get contributions from penguin and annihilation/exchange type
diagrams. Branching ratios for these channels are available~\cite{KKdata}. Since
for $B\to KK$ we do not have enough experimental information to fix
$\zeta^{B\bar K}$ and $\zeta_J^{B\bar K}$ we resort to SU(3). We can apply SU(3)
directly at the level of SCET amplitudes as discussed in
Ref.~\cite{Bauer:2005kd}, and it implies that $\hat P^{KK}\simeq \hat P^{K\pi}$
up to the small penguin annihilation terms.  Using the experimental value of
$\hat P_{TF}^{\rm expt}(K\pi)$ from table~\ref{tableIII} this value of $\hat
P^{KK}$ reproduces the data for $Br(B^-\to \bar K^0 K^-)= 1.36\pm 0.28$ and
$Br(\bar B^0\to K^0\bar K^0)=0.96\pm 0.20$ from HFAG~\cite{HFAG}.  The channel
$\bar B^0\to K^+K^-$ does not get contributions from the penguin amplitudes
$\hat P^{\zeta}$, $\hat P^{\zeta_J}$, nor the chiraly enhanced penguin amplitudes
as can be seen from Table~\ref{table2a}. It does get contributions from
annihilation, but not from the potentially sizeable chiraly enhanced penguin
annihilation, $\hat P^{Lann\chi}$, as is clear from Table~IV of
Ref.~\cite{Arnesen:2006vb}. Hence the small observed value $Br(\bar B^0\to
K^+K^-)= 0.15 \pm 0.10$ is consistent with the size of the annihilation results
for $\hat P^{Lann}$ and $\hat P^{Gann}$ in table~\ref{tableII}. The size of
these amplitudes is also consistent with the power counting estimate of
$\Lambda/m_b$ suppression relative to leading order terms.

The experimental errors in $\hat P^{\rho\rho}$ are too large at this
time to draw strong conclusions, but it is interesting to note that
the positive sign for the real part of the short-distance standard
model penguin prefers values of $\gamma$ larger than $59^\circ$.

Our numerical results for the penguins can also be compared with earlier
analyses in the BBNS~\cite{BBNS,BBNS2} and KLS~\cite{Keum,Lu} approaches where
light-cone sum-rules are used for the hadronic parameters. The BBNS analysis
also gives numbers where $\hat P^\chi\sim \hat P^{\rm LO}$, and gives small
short-distance imaginary parts in $\hat P^{\rm LO}$. However, individual central
values differ from ours due to their different method for dealing with input
parameters and their use of an expansion in $\alpha_s(\mu_i)$ at the
intermediate scale for the LO penguin and chiral enhanced penguin contributions.
Also a larger (complex) range of annihilation amplitudes was adopted in
Ref.~\cite{BBNS2}, with a non-perturbative strong phase that can be chosen to
fit the data.  In the KLS approach it is more difficult to compare individual
contributions, but generically the penguin amplitudes are somewhat larger, and
have a large strong phase from annihilation graphs.  The most prominent feature
in both comparisons is that our parameter errors in $\hat P^{\rm LO}$ and $\hat
P^{\chi}$ are significantly smaller than earlier results, due to our use of tree
amplitude data to determine the hadronic parameters. From our numerical analysis
of annihilation amplitudes, together with power counting arguments it appears
that nonperturbative charm loops are the most likely culprit for a missing
long-distance contribution to the amplitude.


\section{Penguins Contributions from New Physics} \label{sect:np}

There has been a lot of discussion about the possibility of new physics in
nonleptonic $B$-decays (for
example~\cite{Yoshikawa:2003hb,Buras:2003yc,Grossman:2003qi,Buras:2004ub,Kagan:2004uw,Datta:2004re,Mishima:2004um,Hewett:2004tv,Agashe:2005hk,Botella:2005ks,Imbeault:2006nx,Silvestrini:2007yf}).
The precision achieved for the computation of the standard model penguin
amplitudes in tables~\ref{tableII} and~\ref{tableIII}, and their lack of
concordance with the experimental results, make it interesting to reexamine the
role new physics contributions may play. In this section we aim to look at
general features the new physics contributions should have, and do not attempted
to explore this topic in specific models.

Lets consider adding new physics contributions to the nonleptonic
amplitudes. Since new flavor-changing physics is likely to be heavy we
can suppose that upon integrating out the short-distance new particles
we generate a set of operators whose amplitude is parameterized by a
CP-even matrix element $N$ and a CP-violating phase $\phi$, $\hat
A^{NP}=Ne^{i\phi}$. Here $N$ contains the strong rescattering phase
for the amplitude, and $e^{i\phi}$ has CP violation that need not
follow the CKM paradigm.

In order to fit the data, i.e.~contribute to ${\rm Im}(\hat P)$, we will
demonstrate that $Ne^{i\phi}$ must have a non-zero CP-even strong phase. Given
this we may ask whether a small strong phase in $N$ can be enhanced by a large
new source of CP violation, or by some other new physics effect. We will see
that there is a strict bound that prevents us from enhancing ${\rm Im}(\hat P)$
without having large ${\rm Im}(N)$.

To study these points we follow Ref.~\cite{Botella:2005ks} and use the
fact that we can decompose any new physics amplitude into terms that
simply shift the CP-even standard model amplitudes in
Eq.~(\ref{A}). For example, we can decompose any $Ne^{i\phi}$ to make
it look like terms appearing in $B\to \pi\pi,K\pi,
\rho\rho$:
\begin{equation} \label{NP}
Ne^{i \phi}= N_1+ N_2\, e^{-i \gamma} \,,
\end{equation}
where the first terms acts like the $\lambda_c^{(d,s)}$ term and the second like
$\lambda_u^{(d,s)}$.  Here only $\phi$ and the standard model phase $\gamma$
change sign under CP, while $N_{1}=N_{1}^R+i N_{1}^I$ and $N_2=N_2^R+i N_2^I$
are CP-even. Adding $Ne^{i\phi}$ terms to the SM amplitudes, we see that
$N_{1,2}$ simply shift the SM amplitude parameters.  Eq.~(\ref{NP}) was used in
Ref.~\cite{Imbeault:2006nx} to point out that it is not possible to observe new
physics in penguin amplitudes in decays like $B\to \pi\pi, K\pi$ without having
information about the SM penguins that goes beyond isospin. Given the
computations of the SM penguins in the previous section, we can use
Eq.~(\ref{NP}) to explore how new physics effects can appear. To generate large
${\rm Im}(\hat P)$ in our phase convention we need large ${\rm Im}(N_1)$ and/or
large ${\rm Im}(N_2)$.

Being CP-even the parameters $N_{1,2}$ act like strong interaction amplitudes,
despite the fact that they contain short-distance CP-violating parameters.
Solving Eq.~(\ref{NP}) gives
\begin{align}
N_{1}^I&= \frac{{\rm Im}(N)}{\sin\gamma}\, \sin(\gamma+\phi)
  \,,
 & N_{2}^I &= -\frac{{\rm Im}(N)}{\sin\gamma}\, 
    \sin(\phi)\,, \nn \\
N_1^R &= \frac{{\rm Re}(N)}{\sin\gamma} \,  \sin(\gamma+\phi)
 \,, 
 & N_2^R &= -\frac{{\rm Re}(N)}{\sin\gamma}\, \sin(\phi) \,.
\end{align}
Hence the shift to the imaginary part of the standard model amplitudes is zero if ${\rm
  Im}(N)=0$. Furthermore we have the bounds
\begin{align} \label{NPbnd}
  \big| N_{1}^I \big| \le \bigg| \frac{ { \rm Im }(N) }{\sin\gamma }  \bigg| \,,
  \qquad
  \big| N_{2}^I \big| \le \bigg| \frac{ { \rm Im }(N) }{\sin\gamma }  \bigg|
  \,.
\end{align}
The SM value of $\gamma$ is not small ($\sin\gamma\sim 0.9$), so these
bounds imply that enhancement in $N_1^I$ or $N_2^I$ requires large
${\rm Im}(N)$, and hence a large strong phase for this new physics
amplitude. Thus given $\sin\gamma$, no enhancement of the effective
strong phase can occur due simply to new sources of CP-violation. This
conclusion does not appear to be changed if one or more new physics
amplitudes are added in the various standard model decay channels.

The CP-even phase in $N$ will be generated by strong rescattering, and
it is useful to consider $N$ as an amplitude generated by new
dimension-6 four-quark operators not present in the SM. Our analysis
of SM four-quark operators gave power suppressed non-perturbative
strong phases and small strong phases from hard penguin loops, so we
might speculate that the same would be true for four-quark operators
with non-SM symmetry properties. In this case the imaginary part of
$N$ will be small, and Eq.~(\ref{NPbnd}) implies that adding new
physics will not significantly improve the situation with ${\rm
Im}(\hat P)$.  One might think that the inclusion of new physics into
the process of extracting a value for $\hat P$ could mollify the need
for a large imaginary part.  However, a simple analysis, say in the
$\pi \pi$ modes, shows that the existence of an $N$ with a small
imaginary part can not lead to penguin completion. It will simply
shift the meaning of the real parts of the tree and penguin amplitudes
in the fit, with only a small change to the meaning of the imaginary
parts.

Thus for new physics to play a significant role in the observed ${\rm Im}(\hat
P)$ we need to find a large imaginary part for an $N$ from analyzing an operator
not generated by the standard model. Though there is no reason to expect an
enhanced short-distance contribution, this is a logical possibility which
deserves further study.  Significant new physics contributions could in fact be
obtained by modifying the coefficient of four-quark operators with charm quarks,
since then a large long distance charming penguin amplitude could provide the
necessary contribution in ${\rm Im}(N)$.  It might be interesting to attempt to
construct explicit new physics models of this type which are not ruled out by
other constraints on flavor changing neutral currents. Thus it seems to be quite
a challenge to complete the penguin without the aid of a long-distance
contribution.


\section{Discussion and Conclusion} \label{sect:conclusion}

Let us now address the question raised in the title.  The results in
table III show a lack of concordance between the theoretical
prediction for the short-distance standard model penguin contributions
and the extracted value for the penguin amplitude.\footnote{ As shown
in Eq. (\ref{z0}) for this conclusion uncertainties in the weak phase
$\gamma$ are irrelevant for ${\rm Im}(\hat P_{\pi\pi})$, but not for
${\rm Re}(\hat P_{\pi\pi})$.}  Chiraly enhanced operators
substantially increase the penguin contributions, but they are not
able to generate the necessary imaginary pieces.  Thus it would seem
that the shortfall must be due to either the long distance charm or
new physics.

Before addressing these possibilities we must be assured that the
assumptions leading to this conclusion are justified.  Our theory
predictions for SM penguins assume that the expansion in powers of
$\Lambda/m_b$ is trustworthy, since the convergence of this series is
a necessary criteria for factorization to apply.  The experimental
extraction of $\hat P_{\pi \pi}$ and $\hat P_{\rho \rho}$ relies on
isospin, and hence is quite robust.  The penguin extraction for the
$K\pi$ system relies slightly more on the factorization (the
$\Lambda/m_b$ expansion) since we use factorization for the tree
amplitude $T^{K^+\pi^-}$.

What evidence do we have that the large mass expansion is indeed converging?
The factorization theorem for color allowed $B\to D^{(*)}M^-$ decays (proven
in~\cite{Bauer:2001cu}) agrees with data with the expected accuracy.  For color
suppressed charmed decays, $B\to D^{(*)}M^0$ the SCET prediction for the strong
phases~\cite{Mantry:2003uz,Blechman:2004vc} is in good accord with the data for
many channels, which provides a non-trivial test of the large energy expansion.
The same expansion is also used in analyzing the photon cut dependence of $B\to
X_s\gamma$~\cite{Becher:2006pu} and for the analysis of $|V_{ub}|$ from $B\to
X_u\ell\bar\nu$~\cite{Lange:2005yw}, where power corrections appear with the
expected size. One might object that these last two examples are inclusive,
summing over states up to $\mu^2\sim m_b\Lambda$. However, our analysis is quite
similar, since the factorization theorems we use do not attempt to factorize
physics below $\mu^2\sim m_b\Lambda$, and instead retain it as form factors. Due
to experimental cuts an analysis of $B\to X_s\ell^+\ell^-$ data will also rely
on this type of expansion~\cite{Lee:2005pw}.

More direct evidence for our methodology for analyzing charmless nonleptonic
decays comes from successes in the exclusive modes themselves.
In~\cite{Bauer:2005kd,JureAlex}, a complete list of the predictions for
branching ratios and CP asymmetries was given, by using the data to fit the
unknown hadronic parameters (including long-distance charm penguin amplitudes).
The theory fits the data quite well, with all of the theory points falling
within 1-2 $\sigma$ of the data. (The only significant exception is the ratio of
$A_{CP}(K^-\pi^0)/A_{CP}(K^-\pi^+)$ where the sign disagrees with the data.) It
is interesting to note that SCET predicts certain asymmetries to be negative
while the current experimental central values are positive.  The factorization
theorem for charmless nonleptonic decays also gives a prediction for $|V_{ub}|
f_+(0)$ given in Eq.~(\ref{flzJ}), which is in good agreement with the recent
extractions based upon dispersion relations \cite{iain,Flynn:2007ii} utilizing
lattice data \cite{Okamoto:2004xg,Mackenzie:2005wu,fermilab, deWater}. (Using
Hill's $\delta$ parameter~\cite{Hill:2005ju} an analogous test will be possible
for $\zeta_J^{B\pi}$ with future experimental improvements on the
$B\to\pi\ell\bar\nu$ spectrum.)  Note that for all of these successes the
penguins were fit to the data and any deviation from the short distance
prediction was absorbed into the long distance charm piece $A_{cc}$.  Thus these
successes do not directly imply convergence of the large energy expansion for
the penguin amplitudes. However, from the point of view of QCD there is not much
distinction between short distance tree and penguin contributions. Although the
pattern of contributions to each of their amplitudes differs, the $\Lambda/m_b$
expansion for each type of contribution involves very similar hadronic physics.
An exception occurs for charm quarks, where the non-relativistic region and
poorer convergence of the $\Lambda/m_c$ expansion may play a role.

For the penguin amplitudes we can see from table III, that the chiraly enhanced
power correction is of the same order as the leading order penguin contribution.
It is interesting to understand the origin of the enhancement for these power
corrections.  First off, the chiral condensate gives an enhancement of a factor
of three~\cite{BBNS2}. As we discussed the chiraly enhanced contribution also
has a Wilson coefficient which gives an added numerical enhancement compared to
leading order penguins by $\sim 6$, coming from a factor of $1/(u \bar u)$.  One
should worry that there could be higher dimensional operators which are chiraly
enhanced as well. However, for these operators to be as large as the leading
chiraly enhanced contribution they would have to have additional enhancement
from their coefficient function. At present no such subleading operators are
known to exist, but further investigation is warranted.

From the power counting and form of the factorization formulas for the
nonleptonic amplitudes, terms with a long distance phase (outside of
long-distance charm amplitudes $P_{cc}$) arise from contributions which are down
by $\alpha(\mu_i)/\pi$ relative to the corrections considered in this paper,
see~\cite{Arnesen:2005ez}.  In principle these corrections, which come in at
order $\Lambda/m_b \alpha(\mu_i)/\pi$, as well as $\Lambda\mu_M/m^2_{b}
\alpha(\mu_i)/\pi$ (the chiraly enhanced pieces), could account for the penguin
deficit. However, this would be in gross violation of the power counting. Even
if the expansion in $\alpha(\mu_i)$ were very poorly behaved, which seems not to
be the case in the calculations performed to date \cite{Becher:2004kk}, these
contributions could still not make up the deficit, as they would be expected to
be the same size as the chiraly enhanced penguin annihilation (at best), shown
in the last column in table II. Of course if the chiraly enhanced annihilation
were truly as big as the lower order terms in the power expansion, which our
error in the table allows, then we would question the whole power expansion in
the penguin sector. However, to push the penguin annihilation to the limits of
our errors one needs large deviations from naturalness.

Two possible resolutions are, new physics and long distance charm. Let
us consider the former possibility first.  As we have shown in the
previous section, introducing a large CP violating phase from new
physics, does indeed have the effect of mimicking a CP conserving
imaginary penguin, however its size is bounded by the strong phase
induced by QCD. Thus, given that we have shown that such imaginary
pieces (modulo $A_{cc}$) are small, it would seem to be a challenge to
complete the penguin using new physics. One open possibility is that
the new physics generates new operators not present in the standard
model electro-weak Hamiltonian, that generate large imaginary parts
when matching onto \SCETa.  However, given our experience matching the
standard model operators, there is no compelling reason to believe
that such a scenario is likely.

In addition, generically, the new physics is constrained to only arise in
certain operators.  In particular, we note that the new physics would not
fall under the rubric of Minimally Flavor
Violation~\cite{D'Ambrosio:2002ex,Buras:2000dm}, since there are strong
constraints on $\epsilon^\prime$ in the kaon system. This in itself is not a
problem as one might expect the new physics to couple differently to the third
generation, given the top quark mass. Furthermore the new physics should leave
the $\Delta B=2$ operators responsible for $B-\bar B$ mixing essentially
unscathed.  It would seem to be an interesting challenge to build a model which
accomplishes these goal without fine-tunings.

Long distance charm contributions are perhaps the most compelling explanation
for the penguin deficit.  As was shown in section~\ref{sect:np} current data
appears to require a sizeable long-distance strong phase, such as the
long-distance charm amplitude described in section~\ref{sect:long}.  Moreover,
the long distance charm has the potential to explain another discrepancy with
the data~\cite{Poldata}, namely the deficit of transversely polarized vectors in
the decay $\phi K^\star$ channel~\cite{Bauer:2004tj}.  In SCET one does not
generate any leading operators which produce transversely polarized vectors.
This suppression follows from simple chirality arguments~\cite{Kagan:2004uw}.
To derive an amplitude factorization formula for the long distance charming
penguins which generates transverse polarization was beyond the scope of this
paper. However, it is simple to see that we would expect transverse polarization
by noting that the helicity arguments mentioned above no longer apply because
the valence quarks which make up the mesons are no longer produced on the light
cone.  Moreover, a post-diction of our SCET analysis method would be that one
would expect a large transverse polarization fraction in the $\phi K^\star$
channel, but not in the $\rho \rho$, simply because the latter is tree dominated
while the former is penguin, and hence $A_{cc}$ dominated. Thus it would seem that
the long distance charm contribution can explain both the penguin dearth as well
as the transverse polarization in $\phi K^\star$.  Whereas a new physics
scenario would seem to need some organizing principle which would lead to an
enhanced $C_{3,4}$ coupling, the generation of a set of new operators to explain
the polarization \cite{Kagan:2004uw}, and at the same time not disturb all the
successes of the standard model in the B and K sectors. Recent work on the
polarization question was done in Ref.~\cite{Datta:2007qb}
and~\cite{Beneke:2006hg}.

Note: While this manuscript was in preparation
Ref.~\cite{Beneke:2006mk} appeared where $\Delta b_4^{(f)}$ was also
computed in the NDR scheme. We have verified that our result for
$\Delta b_4^{(f)}$ in the NDR agrees with theirs as a function of $u$
and $z$ (they use a different basis, and the relevant comparison is
for $2\Delta b_4^{(f)}-2\Delta c_4^{(f)}$). Unlike
Ref.~\cite{Beneke:2006mk}, for the up and charm loops we demonstrated
the simplicity of using the offshell UV subtraction procedure, and
also presented the computation in the HV scheme for $\gamma_5$.
Ref.~\cite{Beneke:2006mk} includes small $C_{i\ge 3}\alpha_s$ terms in
$\Delta b_4^{(f)}$, which we neglected because they are expected to
compete with other terms of similar numerical size (such as the
complete two-loop corrections with $C_1$) which remain unknown.  We
also derived a new factorization theorem for chiraly enhanced penguins,
and demonstrated that only the short-distance perturbative
coefficients give imaginary parts to the corresponding amplitudes. In
contrast Ref.~\cite{Beneke:2006mk} includes a complex hadronic
parameter in their modeling of the analogous terms which they obtain
with an additional expansion in $\alpha_s(\mu_i)$, where
$\mu_i^2\simeq E\Lambda$ is the intermediate scale.  Finally our
phenomenological analysis differs from Ref.~\cite{Beneke:2006mk}.  Our
strategy was to avoid expanding in $\alpha_s(\mu_i)$, and to use data
on the tree-amplitudes to determine the most important
non-perturbative parameters.  This allowed us to reduce the model
uncertainty considerably while still predicting the penguin
amplitudes. In contrast Ref.~\cite{Beneke:2006mk} models all
non-perturbative parameters, and hence has larger parameter
uncertainty in their final result.

We thank C.~Arnesen, Z.~Ligeti and J.~Zupan for their comments on the manuscript. This work was supported in part by the DOE under DE-FG03-91-ER40683,
and by the DOE Office of Nuclear Science under the grant
DE-FG02-94ER40818. I.S. was also supported in part by the DOE
Outstanding Junior Investigator program and Sloan Foundation.

\appendix 

\section{One-Loop Functions} \label{appA}

\begin{widetext}
  In this appendix we quote some of the basic loop integrals used in the text.
  We work in $d=4-2\epsilon$ dimensions in the $\overline {\rm MS}$ scheme, and
  the $(4\pi)^{-1} e^{\gamma_E}$ constant is absorbed in momentum subtraction
  scale $\mu^{2}$. The basic loop integrals $J_0$ and $I_0(q^2)$ are defined as,
\begin{eqnarray}
\label{loopint}
J_0(p,q)&=&\left ( \frac{16\pi^2}{i} \right )\cdot \int \frac{d^dk}{(2 \pi)^d}
\frac{1}{[k^2-m_c^2]}\frac{1}{[(k+p)^2-m_c^2]}\frac{1}{[(k-q)^2-m_c^2]}\nonumber
\\
&=&\frac{1}{2 p\cdot q}\bigg[
  \text{Li}_2\bigg(\frac{2}{1-\sqrt{1-4m_c^2/q^2}}\bigg)
  +\text{Li}_2\bigg(\frac{2}{1+\sqrt{1-4m_c^2/q^2}}\bigg) \nonumber \\
&&  \qquad\qquad\quad
  - \text{Li}_2\bigg(\frac{2}{1+\sqrt{1-4m_c^2/(p+q)^2}} \bigg)
  - \text{Li}_2\bigg(\frac{2}{1-\sqrt{1-4m_c^2/(p+q)^2}} \bigg) \bigg]  \, ,
  \nonumber \\
I_0(q^2)&=&\left ( \frac{16\pi^2}{i} \right )\cdot \int \frac{d^dk}{(2 \pi)^d}
\frac{1}{(k^2-m_c^2)}\frac{1}{((k-q)^2-m_c^2)}
  = \frac{1}{\epsilon} + \overline I_0(q^2) \nonumber\\
  \overline I_0(q^2) &=&
  2 +\ln\Big( \frac{\mu^2}{m_{c}^2}\Big) 
   - \theta(q^2 \!-\! 4m_c^2) \sqrt{1-{4m_c^2}/{q^2}}\,\,
   \bigg\{  \ln\Big(\frac{1+\sqrt{1-4m_c^2/q^2}}
      {1-\sqrt{1-4m_c^2/q^2}}\Big) - i\pi \bigg\}
    \nonumber \\
  && \qquad \quad -2\,\theta(4m_c^2\!-\! q^2) \sqrt{{4m_c^2}/{q^2}-1}\,\, 
   \cot^{-1}\Big(\sqrt{{4m_c^2}/{q^2}-1}\Big) \, ,
\end{eqnarray}
\end{widetext}
Some limits for these functions are 
\begin{eqnarray}\label{divfree2}
I_0^{(u)}(q^2)&\equiv&\lim_{m_c\to 0} I_0(q^2) =   \frac{1}{\epsilon} +
  2 +\ln\Big( \frac{\mu^2}{q^2}\Big) + i\pi  \, ,
 \nonumber\\
  I_0&\equiv& \lim_{q^2 \to 0}I_0(q^2) =   \frac{1}{\epsilon} +
    \ln\Big(\frac{\mu^2}{m_c^2} \Big),\nn\\
\overline I_0^{(u)}(q^2)&=& 
  2 +\ln\Big( \frac{\mu^2}{q^2}\Big) + i\pi  \, , \nn \\
  \overline I_0 &=& \, \ln\Big(\frac{\mu^2}{m_c^2} \Big) \, .
\end{eqnarray}
The combinations of these integrals that appear in the body of the paper are
\begin{align}
h_0^c(u,z,\rho) &= m_c^2 J_0(p,q) \,,
 \nonumber \\
 \bar I_0(q^2) &= G \left ( \frac{q^2}{m_b^2}\, , \rho \right ),\\
 \bar I_0^{(u)}(q^2) &= G_0 \left ( \frac{q^2}{m_b^2} \right ), \nn
\end{align}
where $u$ and $z$ are momentum fractions of quarks in $M_1$ and $M_2$
respectively.  Notice that functions $h_i^p(u,z)$ are finite and dimensionless.
Change of the functional dependence from square of momenta to momentum fractions
comes from the relations $q^2 = m_b^2(1-u)z$ and $(p+q)^2 = m_b^2(1-u)$.

We also list some useful Fierz relations. For convenience we define,
\begin{eqnarray}
P^{\lambda a}= \big [ \bar q_n \gamma_\perp^\lambda T^a P_L b_v \big] 
   \big [ \bar d_{\bar n} \nslash P_L q_{\bar n} \big ] 
   \nonumber \\
V^{\lambda a}=\big [ \bar q_n  T^a P_L b_v \big] 
   \big [ \bar d_{\bar n} \nslash \gamma_\perp^\lambda  P_R q_{\bar n} \big ].
    \nonumber
\end{eqnarray}
Feirzing gives following formulas useful for simplification,
\begin{align}
 \epsilon_\perp^{\lambda \mu}\big[\bar{d}_{\bar{n}} \nslash P_{L}b_{v}\big]
  \big[\bar{q}_{n}\gamma_{\perp \mu}T^{a}q_{\bar{n}}\big]
  &=-\frac{i}{3}(P^{\lambda a}-V^{\lambda a}) ,
  \nonumber \\
 d^{abc} \epsilon_\perp^{\lambda \mu}\big[\bar{d}_{\bar{n}} \nslash P_{L} T^c
 b_{v}\big]\big[\bar{q}_{n}\gamma_{\perp \mu}T^{b}q_{\bar{n}}\big]
  &=-\frac{i5}{18}(P^{\lambda a}-V^{\lambda a}) ,
  \nonumber \\
 f^{abc} \big[\bar{d}_{\bar{n}} \nslash P_{L}T^c
 b_{v}\big]\big[\bar{q}_{n}\gamma_{\perp}^\lambda T^{b}q_{\bar{n}}\big]
   &=-\frac{i}{2}(P^{\lambda a}+V^{\lambda a}) ,
   \nonumber \\
\big[\bar{d}_{\bar{n}} \nslash P_{L} \gamma_{\perp}^\lambda \gamma_{\perp}^\mu
     b_{v}\big]\big[\bar{q}_{n}\gamma_{\perp \mu}T^{a}q_{\bar{n}}\big]
  &=-\frac{2}{3}V^{\lambda a} ,
   \nonumber  \\
 \big[\bar{d}_{\bar{n}} \nslash P_{L} \gamma_{\perp}^\mu \gamma_{\perp}^\lambda
    b_{v}\big]
  \big[\bar{q}_{n}\gamma_{\perp \mu}T^{a}q_{\bar{n}}\big]
  &=-\frac{2}{3}P^{\lambda a} ,
  \nonumber \\
 \big[\bar{d}_{\bar{n}} \nslash P_{L} b_{v}\big]
  \big[\bar{q}_{n}\gamma_{\perp}^\lambda T^{a} q_{\bar{n}}\big]
  &=-\frac{1}{3}(P^{\lambda a}+V^{\lambda a}).
\end{align}
These results include the minus sign from permuting fermion fields.

\section{Chiraly Enhanced terms in \SCETb }  \label{appB}

In section~\ref{sect:chiral} we made the statement that the operators
$Q_i^{(2\chi)*}$ in Eq.~(\ref{Bstar}) only contribute for $B^*$
decays. The simplest way to verify this statement is to consider the
Dirac structures generated by matching the $T_2$ time-ordered product
of these operators onto operators in \SCETb. This can be done working
to all orders in $\alpha_s$. From Ref.~\cite{Bauer:2004tj} the most
general perturbative matching generates Wilson coefficients given by
jet functions $J$ and $J_\perp$ whose form is constrained by RPI,
chirality, power counting and dimensional analysis [\,$\omega_1 =
z\omega$, $\omega_4=(1\!-\!z)\omega$, $\bar x=1\!-\!x$\,,
$\chi_{n,\omega}=(W^\dagger\xi_n)_{\omega}$],
\begin{eqnarray}\label{Jdef}
&& \hspace{-.4cm}
 T\,\big[ (\bar \xi_n W)_{\omega_1}
   i g\, {\cal B}^{\perp\alpha}_{n,\omega_4}
   P_{R,L} \big]^{ia}(0)\:
   \big[ i g\, \slash\!\!\!\!{\cal B}^\perp_{n}
   W^\dagger \xi_n \big]_0^{jb}(y) \nn
   \\
&=&\!\!\!
   i\, \delta^{ab} {\delta(y^+) \delta^{(2)}(y_\perp)}\,
    \frac{1}{\omega}
   \int_0^1\!\!\! dx \int\! \frac{dk^+}{2\pi}\: e^{+i  k^+ y^-/2}
   \nn\\
&\times&\!\!\!\! \Big\{\! -\!J_\perp(z,x,k_+)   \Big( \frac{\nslash}{2} P_{R,L}
  \gamma_\perp^\alpha \gamma_\perp^\beta \Big)_{ji}
  \big[\bar\chi_{n,x\omega}^{L,R}
   {\bnslash\gamma^\perp_\beta }  \chi_{n,-\bar x\omega}^{R,L}
   \big] \nn\\
&+&\!\!\!
  J(z,x,k_+) \Big({\nslash} P_{L,R} \gamma_\perp^\alpha\Big)_{ji}
  [\bar\chi_{n,x\omega}^{L,R} \bnslash \chi_{n,-\bar
    x\omega}^{L,R} ] \Big\},
\end{eqnarray}
where $\{i,j\}$ and $\{a,b\}$ are spin and color indices. This result
on the RHS includes the sign from antipermuting the fermion fields
that are contracted with the spin indices $ij$. To use this formula
other Dirac structures occurring in the $T_2$ time-ordered product
should be grouped with the heavy-quark field $h_v$ and soft-quark
field $\bar q_s$. Using Eq.~(\ref{Jdef}) we find that all the
operators $Q_{i}^{(2\chi)*}$ give the structure $\bar q_s \nslash
\gamma_\perp^\alpha h_v$ which has a vanishing $B$-meson matrix
element, but would be nonzero for $B^*$ initial states.

In section~\ref{sect:chiral} we used $\zeta_\chi^{BM}(z)\sim z$ which
followed from the power counting in \SCETa that indicates that it can
not be more singular than $\zeta_J^{BM}(z)$. We also stated that this
scaling could be checked by factorizing the \SCETa time-ordered
product that defines this form factor using \SCETb. To do so we again
use Eq.~(\ref{Jdef}) and take the matrix element of the resulting
operators to find
\begin{align} \label{zetaJchifactor}
\zeta_{\chi}^{BM}(z) &= \frac{f_B f_M}{ m_b}\int_0^1\!\!\!\!\! dx \!\!
\int_0^\infty \!\!\!\!\!\! dk^+ \, \frac{J_\perp(z,k^+,x)}{1-z} 
  \phi_B^+(k^+)\phi_{pp}^M(x) .
\end{align}
Here $J_\perp(z,x,k_+) = {\delta(x -z) \pi \alpha_s(\mu) C_F}/{(N_c\,
\bar x k_+)}$ at lowest order, so the behavior of $\zeta_\chi^{BM}(z)$
as $z\to 0$ is inherited from $\phi_{pp}^M(x)$ as $x\to 0$, giving
linear scaling $\propto z$. The one-loop result for $J_\perp$ is also
known~\cite{Becher:2004kk,Beneke:2005gs} and the linear scaling is
reproduced at this order. The limit $x\to 0$ corresponds to the
collinear quark in the form factor becoming soft, and for this matrix
element there is no corresponding diagram with a soft quark at this
order. Hence, because we do not expect an overlap with a soft diagram,
we do not expect there is a need for any zero-bin subtractions, and
hence no endpoint divergences which would result from constant scaling
as $z\to 0$.  The other interesting limit is $z\to 1$, where we expect
$\zeta_\chi^{BM}(z)\sim 1$.  From the point of view of \SCETb this
limit is more interesting because there are diagrams with soft
antiquarks in the form factor, and we must avoid double counting
them. Indeed the tree level jet function appears to give
$\zeta_\chi^{BM}(z)\sim
\phi_{pp}^M(z)/(1-z)^2$, which would imply singular behavior as $z\to
1$. However in \SCETb to avoid double counting the region where this
quark is soft we must make zerobin subtractions in defining this
singular moment~\cite{Manohar:2006nz}. These subtractions modify the
distribution, causing scaling behavior of $\phi_{pp}^{0-bin}(z)\sim
(1-z)^2$ in the endpoint region (and also generate dependence of this
distribution on an additional rapidity parameter). The result is that
as $z\to 1$, $\zeta_\chi^{BM}(z)\sim 1$ as expected from power
counting in \SCETa.


\section{Beta's at $\mu=m_b/2$ and $\mu=2m_b$} \label{appC}

In Eq.~(\ref{betas}) we quoted values for the annihilation moments at
$\mu=m_b=4.7\,{\rm GeV}$. For our error analysis we also required the values at
$\mu=m_b/2=2.3\,{\rm GeV}$
\begin{widetext}
\begin{align} \label{betas2}
\beta_{1c}^{\pi\pi} &= (-5.5\pm 3.0)\times 10^{-2}
   \,, 
 & \beta_{3c}^{\pi\pi} &= 1.15 \mpm 0.58
   \,,
 & \beta_{4c}^{\pi\pi} &=  -0.25 \mpm 0.15
   \,,  \\
\beta_{hc 1}^{\pi\pi} &= -2.35 \pm 0.78 
   \,, 
 & \beta_{hc 2}^{\pi\pi} &= 0.61 \pm 0.34 
   \,,
 & \beta_{hc 3}^{\pi\pi} &= (4.4\pm 5.3)\times 10^{-3} 
   \,,
 & \beta_{hc 4}^{\pi\pi} &= (-8.6\pm 2.8)\times 10^{-2} 
   \,,\nn\\
 \beta_{\chi 1}^{\pi\pi} &=  0.0 \pm 9.7
   \,, 
 & \beta_{\chi 2}^{\pi\pi} &= 0.0 \pm 9.1 
   \,,
 & \beta_{\chi 5}^{\pi\pi} &= 0.0 \pm 0.13
   \,,
 & \beta_{\chi 6}^{\pi\pi} &= 0.0 \pm 0.16
   \,,\nn\\
\beta_{4c}^{\pi K} &= -0.27 \pm 0.11
   \,, \nn\\
\beta_{hc 1}^{\pi K} &= -2.45 \pm 0.80 
   \,, 
 & \beta_{hc 2}^{\pi K} &= 0.61 \pm 0.34 
   \,,
 & \beta_{hc 3}^{\pi K} &= (0.8\pm 5.5) \times 10^{-3}  
   \,,
 & \beta_{hc 4}^{\pi K} &= (-8.6\pm 2.7) \times 10^{-2}  
   \,,\nn\\
 \beta_{\chi 1}^{\pi K} &= 0.0 \pm 12.3 
   \,, 
 & \beta_{\chi 2}^{\pi K} &= 0.0 \pm 11.2
   \,,
 & \beta_{\chi 5}^{\pi K} &= 0.0 \pm 0.17
   \,,
 & \beta_{\chi 6}^{\pi K} &= 0.0 \pm 0.20 
   \,,\nn\\
\beta_{1c}^{\rho\rho} &= (9.2 {}^{+7.6}_{-2.5})\times 10^{-3}
   \,, 
 & \beta_{3c}^{\rho\rho} &= -0.19 {}^{+.16}_{-.05}
   \,,
 & \beta_{4c}^{\rho\rho} &=  (4.3 {}^{+3.5}_{-1.2})\times 10^{-2}
   \,, \nn\\
\beta_{hc 1}^{\rho\rho} &= (-7.0 {}^{+6.1}_{-5.3})\times 10^{-2}
   \,, 
 & \beta_{hc 2}^{\rho\rho} &= (-1.4 {}^{+1.4}_{-1.1})\times 10^{-2}
   \,,
 & \beta_{hc 3}^{\rho\rho} &= (-1.4 {}^{+1.2}_{-1.0})\times 10^{-4}
   \,,
 & \beta_{hc 4}^{\rho\rho} &= (2.6 {}^{+2.2}_{-2.0})\times 10^{-3}
   \,, \nn
\end{align}
and at $\mu=2m_b$
\begin{align} \label{betas3}
\beta_{1c}^{\pi\pi} &= (-1.6\pm 0.9)\times 10^{-2}
   \,, 
 & \beta_{3c}^{\pi\pi} &= 0.35 \mpm 0.18
   \,,
 & \beta_{4c}^{\pi\pi} &=  -0.089 \mpm 0.054
   \,,  \\
\beta_{hc 1}^{\pi\pi} &= -0.73 \pm 0.23 
   \,, 
 & \beta_{hc 2}^{\pi\pi} &= -0.015 \pm 0.040 
   \,,
 & \beta_{hc 3}^{\pi\pi} &= (-4.9\pm 0.8)\times 10^{-3} 
   \,,
 & \beta_{hc 4}^{\pi\pi} &= (-3.7\pm 1.1)\times 10^{-2} 
   \,,\nn\\
 \beta_{\chi 1}^{\pi\pi} &=  0.0 \pm 2.7
   \,, 
 & \beta_{\chi 2}^{\pi\pi} &= 0.0 \pm 2.5 
   \,,
 & \beta_{\chi 5}^{\pi\pi} &= 0.0 \pm 0.040
   \,,
 & \beta_{\chi 6}^{\pi\pi} &= 0.0 \pm 0.053
   \,,\nn\\
\beta_{4c}^{\pi K} &= -0.095 \pm 0.069
   \,, \nn\\
\beta_{hc 1}^{\pi K} &= -0.76 \pm 0.24 
   \,, 
 & \beta_{hc 2}^{\pi K} &= -0.015 \pm 0.040 
   \,,
 & \beta_{hc 3}^{\pi K} &= (-4.9\pm 0.8) \times 10^{-3}  
   \,,
 & \beta_{hc 4}^{\pi K} &= (-3.7\pm 1.1) \times 10^{-2}  
   \,,\nn\\
 \beta_{\chi 1}^{\pi K} &= 0.0 \pm 3.4 
   \,, 
 & \beta_{\chi 2}^{\pi K} &= 0.0 \pm 3.0
   \,,
 & \beta_{\chi 5}^{\pi K} &= 0.0 \pm 0.051
   \,,
 & \beta_{\chi 6}^{\pi K} &= 0.0 \pm 0.065 
   \,,\nn\\
\beta_{1c}^{\rho\rho} &= (2.7 {}^{+2.2}_{-0.7})\times 10^{-3}
   \,, 
 & \beta_{3c}^{\rho\rho} &= -0.06 {}^{+.05}_{-.02}
   \,,
 & \beta_{4c}^{\rho\rho} &=  (1.5 {}^{+1.2}_{-0.4})\times 10^{-2}
   \,, \nn\\
\beta_{hc 1}^{\rho\rho} &= (-2.2 {}^{+1.9}_{-1.7})\times 10^{-2}
   \,, 
 & \beta_{hc 2}^{\rho\rho} &= (1.7 {}^{+1.3}_{-1.3})\times 10^{-3}
   \,,
 & \beta_{hc 3}^{\rho\rho} &= (-2.1 \pm 1.7)\times 10^{-4}
   \,,
 & \beta_{hc 4}^{\rho\rho} &= (1.1 {}^{+1.0}_{-0.9})\times 10^{-3}
   \,. \nn
\end{align}
\end{widetext}

\vspace{-.4cm}

\newpage

\end{document}